\RequirePackage[l2tabu, orthodox]{nag}
\RequirePackage{snapshot}

\documentclass[9pt,onecolumn]{extarticle}

\makeatletter
\if@twocolumn
  \usepackage[dvips,letterpaper,top=0.5in, bottom=0.5in, left=0.75in, right=0.5in,includefoot,heightrounded]{geometry}
\else
  \usepackage[dvips,letterpaper,margin=1in,includefoot,heightrounded]{geometry}
\fi

\usepackage{srcltx}

\usepackage[russian,portuges,english]{babel}

\iflanguage{portuges}
    {\newcommand{\keywordname}{Palavras-chaves}}
    {\newcommand{\keywordname}{Keywords}}

\usepackage{amsmath}
\usepackage{amssymb}
\usepackage{amsfonts}

\usepackage{abstract}

\usepackage{graphicx}
\usepackage[usenames,dvipsnames,svgnames,x11names]{xcolor}
\usepackage{subfigure}
\usepackage{psfrag}

\usepackage{booktabs}

\usepackage{setspace}
\usepackage{flushend}

\usepackage{multicol}

\usepackage{cite}

\usepackage{hyperref}
\usepackage[normalem]{ulem}

\usepackage{enumerate}

\usepackage[noend]{algpseudocode}

\usepackage{listings}

\usepackage[short,12hr]{datetime}
       \usepackage{fouriernc}

\newtheorem{definition}{Definition}
\newtheorem{theorem}{Theorem}

\newtheorem{example}{Example}

\makeatletter

\newenvironment{rsmallmatrix}{\null\,\vcenter\bgroup
  \Let@\restore@math@cr\default@tag
  \baselineskip6\ex@ \lineskip1.5\ex@ \lineskiplimit\lineskip
  \ialign\bgroup\hfil$\m@th\scriptstyle##$&&\thickspace\hfil
  $\m@th\scriptstyle##$\crcr
}{%
  \crcr\egroup\egroup\,%
}

\newcommand{\printtitle}{%
\makeatletter
\if@twocolumn

\twocolumn[%
  \maketitle
  \begin{onecolabstract}
    \myabstract
  \end{onecolabstract}
  \begin{center}
    \small
    \textbf{\keywordname}
    \\\medskip
    \mykeywords
  \end{center}
  \bigskip
]
\saythanks
\else
  \maketitle
  \begin{onecolabstract}
    \myabstract
  \end{onecolabstract}
  \begin{center}
    \small
    \textbf{\keywordname}
    \\\medskip
    \mykeywords
  \end{center}
  \bigskip
  \onehalfspacing
\fi
\makeatother
}

\author{%
D.~F.~G.~Coelho%
\thanks{%
D.~F.~G.~Coelho
was with
the
Signal Processing Group,
Departamento de Estat\'{\i}stica;
the Graduate Program in Electrical Engineering
and
the Graduate Program in Statistics,
Universidade Federal de Pernambuco,
Recife, Brazil;
currently
he is an independent researcher.}
\qquad
R.~J.~Cintra%
\thanks{%
R. J. Cintra
was with
the
Signal Processing Group,
Departamento de Estat\'{\i}stica,
Universidade Federal de Pernambuco;
Equipe Cairn,
IRISA-INRIA,
Universit\'e de Rennes,
Rennes, France;
LIRIS,
Institut National des Sciences Appliqu\'ees (INSA),
Lyon, France;
currently
he is with the Department of Technology,
UFPE, Caruaru, Brazil.
E-mail: \mbox{rjdsc@de.ufpe.br}
}
\thanks{This report was elaborated in October, 2016,
as part of the first author's doctorate research.
The second author acknowledges the Conselho Nacional de Desenolvimento Científico
for the partial support during his sabbatical year in France.}
}

\title{Discrete Fourier Transform Approximations Based on the Cooley-Tukey Radix-2 Algorithm}

\newcommand{\myabstract}{%
This report elaborates on approximations for the discrete Fourier transform
by means of replacing the exact Cooley-Tukey algorithm twiddle-factors
by
low-complexity integers, such as $0, \pm \frac{1}{2}, \pm 1$.
}

\newcommand{\mykeywords}{%
DFT Approximation,
Cooley-Tukey algorithm,
Matrix factorization,
Low-complexity methods.
}

\date{}

\begin{document}

\printtitle

\section{Cooley-Tukey Algorithm Review and Matrix Formalism}
\label{review}
A large number of algorithms has been developed for the exact computation of DFT.
The Cooley-Tukey algorithm radix-2 is among the most prominent techniques for computing the exact DFT.
The decimation in time (DIT) version of Cooley-Tukey algorithm is given as follows.

Considering a power-of-two $N>2$, let~$W_N = \exp(-2\pi j/N)$ be the~$N$th unity root and%
\begin{equation*}
\mathbf{A}_N =  \left[
\begin{rsmallmatrix}
1 & 1\\
1 & -1
\end{rsmallmatrix}\right] \otimes
\mathbf{I}_{N/2}
=
\left[
\begin{rsmallmatrix}
\mathbf{I}_{N/2} & \mathbf{I}_{N/2}\\
\mathbf{I}_{N/2} & -\mathbf{I}_{N/2}
\end{rsmallmatrix}\right],
\end{equation*}
where~$\mathbf{I}_{N/2}$ is the identity matrix of order~$N/2$ and~$\otimes$ denotes the Kronecker product~\cite{Manassah2001}.

Let~$\mathbf{B}_{N}$ represent the decimation in time matrix whose non-zero elements are given by
\begin{equation*}
\mathbf{B}_{N}(i,j) = \mathbf{B}_{N}(i+\frac{N}{2},j+1) = 1,
\end{equation*}
for $i = 1, 2, \ldots, \frac{N}{2}$ and $j = 2 i-1$.

The Cooley-Tukey algorithm in DIT version can be represented as
\begin{equation}
\label{CT}
\mathbf{F}_{N} = \mathbf{A}_N \mathbf{W}_{N}
\left( \mathbf{I}_2 \otimes \mathbf{F}_{N/2} \right)
\mathbf{B}_{N},
\end{equation}
where~$\mathbf{W}_{N} = \operatorname{diag}\left(\begin{bmatrix}\mathbf{1}_{\mathrm{N}\times1}, 1, W_N, W_N^2, \ldots, W_N^{(\frac{N}{2}-1)}\end{bmatrix}^\top\right)$ and the operator~$\operatorname{diag}(\cdot)$ returns a diagonal matrix with its vector argument in the main diagonal.

The above expression  is recursive and depends on the~$N/2$-point DFT.
Note also that~$\mathbf{F}_2$ is a butterfly matrix~\cite{Blahut2010}.

In particular, for $N = 4$ we have:
\begin{equation*}
\mathbf{F}_\mathrm{4} =
\left[
\begin{rsmallmatrix}
\mathbf{I}_{\mathrm{2}} & \mathbf{I}_{\mathrm{2}}\\
\mathbf{I}_{\mathrm{2}} & -\mathbf{I}_{\mathrm{2}}
\end{rsmallmatrix}\right]
\left[
\begin{rsmallmatrix}
1 & \phantom{\:\:\:\:}0 & \phantom{\:\:\:\:}0 & 0\\
0 & 1 & 0 & 0\\
0 & 0 & 1 & 0\\
0 & 0 & 0 & W_4
\end{rsmallmatrix}\right]
\left[
\begin{rsmallmatrix}
\mathbf{F}_\mathrm{2} & \mathbf{0}_{\mathrm{2}}\\
\mathbf{0}_{\mathrm{2}} & \mathbf{F}_\mathrm{2}
\end{rsmallmatrix}\right]
\left[
\begin{rsmallmatrix}
1 & 0 & 0 & 0\\
0 & 0 & 1 & 0\\
0 & 1 & 0 & 0\\
0 & 0 & 0 & 1
\end{rsmallmatrix}\right].
\end{equation*}

\section{Discrete Fourier Transform Approximation}
\label{approximation}

In the following the scaled rounding operator is defined.
This operator is analyzed throughout this section.

\begin{definition}[Scaled Rounding Operator]
\label{approx-twiddle-factor-def}
Let~$\alpha>0$ and~$x$ be real numbers.
The scaled rounding operator is the result of scaling and rounding operation followed by a normalization as given by:
\begin{align*}
\frac{1}{\alpha}\operatorname{round}\left(\alpha \cdot x\right),
\end{align*}
where~$\operatorname{round}(\cdot)$ is the usual rounding function.
\end{definition}

Let~$2^\mathbb{N}$ be the set of all integers power of two.
For a fixed~$N$,  we define
an~$N$-point complex valued vector according the mapping:
\begin{align}
\label{matrixmapping-dft}
L: 2^\mathbb{N} & \longrightarrow \mathbb{C}^N \nonumber \\
\alpha & \longmapsto \frac{1}{\alpha}\operatorname{round}\left(\alpha \cdot \mathbf{W}_{N}\right),
\end{align}
where~$\operatorname{round}(\cdot)$ returns
the componentwise rounding function~\cite{Cintra2011a}.

Because the entries of~$\mathbf{W}_{N}$ are complex numbers, $\operatorname{round}(\cdot)$ operates both real and imaginary parts~\cite{Cintra2011a}.
Therefore,~\eqref{matrixmapping-dft} introduces an approximation for the twiddle factor vector,
which is denoted by
\begin{equation*}
\tilde{\mathbf{W}}_{N} = \frac{1}{\alpha}\operatorname{round}\left(\alpha \cdot \mathbf{W}_{N}\right).
\end{equation*}
The above approximation for the twiddle factor is called approximate twiddle factor.

\subsection{The Scaled Rounding Operator Properties}
The proposed scheme for approximated DFT is based on the approximate twiddle factor.
The approximate twiddle factor is obtained through the scaled rounding operator as mentioned in the previous section.
This sub-section analyzes  the scaled rounding operator and its properties.
In particular, we aim at  deriving  bounds to the scaled rounding operator applied to the exact twiddle factors and the error due its application.

The atom element of this investigation is the~$\operatorname{round}(\cdot)$ operator.
The rounding-off operator is primarily defined for real arguments.
If~$x$ is an arbitrary real number, $\operatorname{round}(x)$ is the nearest integer to its argument.
Besides that, we have that
\begin{equation*}
x-\frac{1}{2} \leq \operatorname{round}(x) \leq x+\frac{1}{2}.
\end{equation*}
Thus, the absolute error of the rounding operator is bounded by~$1/2$.
Standard manipulations yields the following inequalities:
\begin{equation}
\label{x-round-ineq}
x-\frac{1}{2\alpha} \leq \frac{1}{\alpha} \operatorname{round}(\alpha x) \leq x+\frac{1}{2\alpha}.
\end{equation}
The introduction of a scale factor reduces the overall maximum error due the rounding operation from~$1/2$ to~$1/2\alpha$.
Thus, the consideration of scale factor in the rounding operation furnishes closer representation to its argument.
Using above inequality, we have that
\begin{equation*}
-\frac{1}{2\alpha} \leq x -\frac{1}{\alpha} \operatorname{round}(\alpha x) \leq \frac{1}{2\alpha}.
\end{equation*}
As consequence, we obtain
\begin{equation}
\label{x-error-mag-bound}
\Big|x -\frac{1}{\alpha} \operatorname{round}(\alpha x) \Big| \leq \frac{1}{2\alpha},
\end{equation}
where the operator~$|\cdot|$ returns the absolute value of its argument~\cite{Avila2000, Avila1974}.
The  above inequality gives the sense of approximating~$x$ by  the  scaled rounding operator.
As far as~$\alpha$ increases, the magnitude error bound decreases.
In consequence, the approximation is thus improved by providing closer representation.

Using~\eqref{x-round-ineq}, we can obtain the following lower and upper bounds for the absolute value of scaled rounding operator
\begin{equation}
\label{x-round-mag-bound}
\operatorname{min}\left(\Big|x+\frac{1}{2\alpha}\Big|, \Big|x-\frac{1}{2\alpha}\Big|\right) \leq \Big|\frac{1}{\alpha} \operatorname{round}(\alpha x) \Big| \leq \operatorname{max}\left(\Big|x+\frac{1}{2\alpha}\Big|, \Big|x-\frac{1}{2\alpha}\Big|\right),
\end{equation}
where~$\operatorname{min}(\cdot, \cdot)$ and~$\operatorname{max}(\cdot, \cdot)$ returns the minimum and maximum value between its arguments.

Let the argument of  the  rounding operator be a complex number~$z=x+jy$ and~$\mathfrak{Re}(\cdot)$ and~$\mathfrak{Im}(\cdot)$ returns the real and imaginary parts of its complex arguments.
Thus, we have that
\begin{equation}
\label{z-round-re-ineq}
x-\frac{1}{2\alpha} \leq \mathfrak{Re}\left( \frac{1}{\alpha} \operatorname{round}(\alpha z) \right)\leq x+\frac{1}{2\alpha}
\end{equation}
and
\begin{equation}
\label{z-round-im-ineq}
y-\frac{1}{2\alpha} \leq \mathfrak{Im}\left( \frac{1}{\alpha} \operatorname{round}(\alpha z) \right)\leq y+\frac{1}{2\alpha}.
\end{equation}
In essence, the scaled rounding operator for complex arguments acts separately in its real and imaginary  parts.
Applying~\eqref{x-error-mag-bound} to above inequalities, we obtain
\begin{equation*}
\Big|x -\mathfrak{Re}\left(\frac{1}{\alpha} \operatorname{round}(\alpha z)\right) \Big| \leq \frac{1}{2\alpha}
\end{equation*}
and
\begin{equation*}
\Big|y -\mathfrak{Im}\left(\frac{1}{\alpha} \operatorname{round}(\alpha z)\right) \Big| \leq \frac{1}{2\alpha}.
\end{equation*}
Summing up the squared magnitude of above quantities and taking the square root, we obtain
\begin{equation}
\label{z-error-mag-bound}
\Big|z -\frac{1}{\alpha} \operatorname{round}(\alpha z) \Big| \leq \frac{1}{\sqrt{2}\alpha}.
\end{equation}

Note that inequalities furnished by~\eqref{z-round-re-ineq} and~\eqref{z-round-im-ineq} can be given a different  format  as
\begin{equation*}
\mathfrak{Re}\left(z-\frac{1}{2\alpha}(1+j)\right) \leq \mathfrak{Re}\left( \frac{1}{\alpha} \operatorname{round}(\alpha z) \right)\leq \mathfrak{Re}\left(z+\frac{1}{2\alpha}(1+j)\right)
\end{equation*}
and
\begin{equation*}
\mathfrak{Im}\left(z-\frac{1}{2\alpha}(1+j)\right) \leq \mathfrak{Im}\left( \frac{1}{\alpha} \operatorname{round}(\alpha z) \right)\leq \mathfrak{Im}\left(z+\frac{1}{2\alpha}(1+j)\right).
\end{equation*}
Applying~\eqref{x-round-mag-bound} for both real and imaginary parts of~$\operatorname{round}(\alpha z)/\alpha$, we have that
\begin{equation}
\label{z-round-mag-bound}
S_{\operatorname{min}}(x,y)\leq \Big|\frac{1}{\alpha} \operatorname{round}(\alpha z) \Big| \leq S_{\operatorname{max}}(x,y),
\end{equation}
where
\begin{equation*}
S_{\operatorname{min}}(x,y)= \operatorname{min}\left(\Big|z-\frac{1}{2\alpha}(1+j)\Big|, \Big|z+\frac{1}{2\alpha}(1+j)\Big|\right)
\end{equation*}
and
\begin{equation*}
S_{\operatorname{max}}(x,y)= \operatorname{max}\left(\Big|z-\frac{1}{2\alpha}(1+j)\Big|, \Big|z+\frac{1}{2\alpha}(1+j)\Big|\right).
\end{equation*}

\begin{figure*}
\centering
\psfrag{pi}{$\pi$}
\psfrag{pi/2}{$\frac{\pi}{2}$}
\psfrag{3pi/2}{$\frac{3\pi}{2}$}
\psfrag{2pi}{$2\pi$}
\subfigure[Scaled rounded cosine wave for precision parameter~$\alpha = 2$.]{\includegraphics{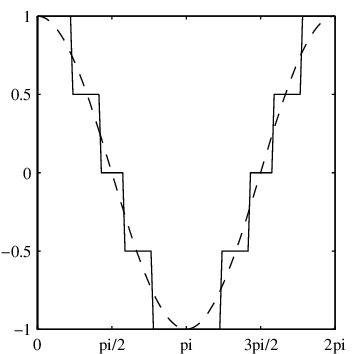}}
\subfigure[Scaled rounded sine wave for precision parameter~$\alpha = 2$.]{\includegraphics{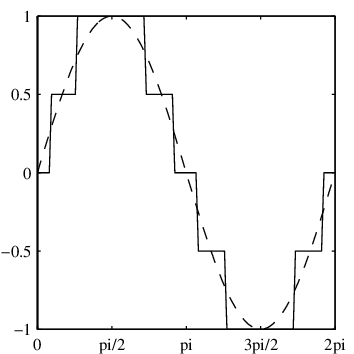}}\\
\subfigure[Scaled rounded cosine wave for precision parameter~$\alpha = 4$.]{\includegraphics{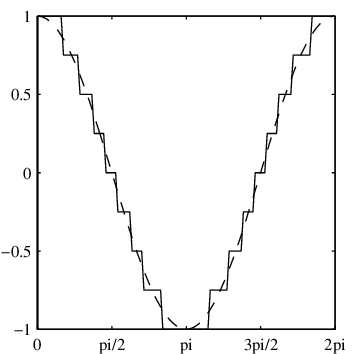}}
\subfigure[Scaled rounded sine wave for precision parameter~$\alpha = 4$.]{\includegraphics{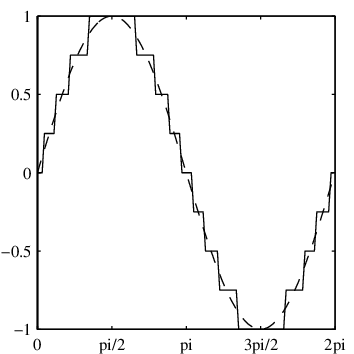}}
\caption{Curves for scaled rounded sine and cosine waves for scale parameters~$\alpha = 2$ and~$4$. Dashed lines represent original sine and cosine waves.}
\label{SclRnd}
\end{figure*}

 Since the quantities in the DFT matrix are complex numbers with unity norm, we analyze in the following~\eqref{z-error-mag-bound} and~\eqref{z-round-mag-bound} considering~$z = e^{j\theta}$ for~$\theta \in [0, 2\pi)$.
For sake of clarity, Figure~\ref{SclRnd} shows scaled rounded sine and cosine waves for scale parameters~$\alpha = 2$ and~$4$.
We have that
\begin{align*}
\Big|e^{j\theta}-\frac{1}{2\alpha}(1+j)\Big| & = \sqrt{\left(\cos(\theta)-\frac{1}{2\alpha}\right)^2+\left(\sin(\theta)-\frac{1}{2\alpha}\right)^2}\\
& = \sqrt{1-\frac{1}{\alpha}\left(\cos(\theta)+\sin(\theta)\right)+\frac{1}{2\alpha^2}}.
\end{align*}

The minimum and maximum of the above expression are~$\sqrt{1-\sqrt{2}/\alpha+1/2\alpha^2}$ for~$\theta = \pi/4$ and~$\sqrt{1+\sqrt{2}/\alpha+1/2\alpha^2}$ for~$\theta = 5\pi/4$, respectively.

Thus, we have that
\begin{equation*}
\sqrt{1-\frac{\sqrt{2}}{\alpha}+\frac{1}{2\alpha^2}} \leq \Big|e^{j\theta}-\frac{1}{2\alpha}(1+j)\Big| \leq \sqrt{1+\frac{\sqrt{2}}{\alpha}+\frac{1}{2\alpha^2}}.
\end{equation*}
Similarly, we have that
\begin{align*}
\Big|e^{j\theta}+\frac{1}{2\alpha}(1+j)\Big| & = \sqrt{\left(\cos(\theta)+\frac{1}{2\alpha}\right)^2+\left(\sin(\theta)+\frac{1}{2\alpha}\right)^2}\\
& = \sqrt{1+\frac{1}{\alpha}\left(\cos(\theta)+\sin(\theta)\right)+\frac{1}{2\alpha^2}}.
\end{align*}

The minimum and maximum of the above expression are~$\sqrt{1-\sqrt{2}/\alpha+1/2\alpha^2}$ for~$\theta = 5\pi/4$ and~$\sqrt{1+\sqrt{2}/\alpha+1/2\alpha^2}$ for~$\theta = \pi/4$, respectively.

Thus, we have that
\begin{equation*}
\sqrt{1-\frac{\sqrt{2}}{\alpha}+\frac{1}{2\alpha^2}} \leq \Big|e^{j\theta}+\frac{1}{2\alpha}(1+j)\Big| \leq \sqrt{1+\frac{\sqrt{2}}{\alpha}+\frac{1}{2\alpha^2}}.
\end{equation*}

Since the maximum and minimum of~$$\Big|e^{j\theta}-\frac{1}{2\alpha}(1+j)\Big|$$ and~$$\Big|e^{j\theta}+\frac{1}{2\alpha}(1+j)\Big|$$  are the same  for~$\theta \in [0,2\pi)$, we have that
\begin{equation*}
S_{\operatorname{min}}(x,y)\geq  \sqrt{1-\frac{\sqrt{2}}{\alpha}+\frac{1}{2\alpha^2}}
\end{equation*}
and
\begin{equation*}
S_{\operatorname{max}}(x,y)\leq  \sqrt{1+\frac{\sqrt{2}}{\alpha}+\frac{1}{2\alpha^2}}.
\end{equation*}
Since~$\alpha$ is a positive power-of-two and greater than one, we note that
\begin{equation*}
\sqrt{1\pm\frac{\sqrt{2}}{\alpha}+\frac{1}{2\alpha^2}} = 1\pm\frac{1}{\sqrt{2}\alpha}.
\end{equation*}
Then, using~\eqref{z-round-mag-bound}, we can establish the following inequality,
\begin{equation}
\label{ejtheta-round-mag-bound}
1-\frac{1}{\sqrt{2}\alpha}\leq \Big|\frac{1}{\alpha} \operatorname{round}(\alpha e^{j\theta}) \Big| \leq 1+\frac{1}{\sqrt{2}\alpha},
\end{equation}
where~$\theta \in [0, 2\pi)$.

Let us now consider the case of interest where~$z$ is a root of unity.
Thus, we need only to consider~$\theta = 2k\pi/N$ for above development, where~$N$ is a integer power-of-two and~$k = 0, 1, \ldots, N/2-1$.
Then, Euler trigonometric equation~\cite{Avila2000} furnish us with~$z = e^{j2\pi k/N} = \cos(2\pi k/N)+j\sin(2\pi k/N)$.
Note that the way~$z$ is defined  correspond to  the known  $k$th  power of the root of unity~$W_N = e^{j2\pi/N}$.
 Thus, applying ~$z = W_N^k$ to~\eqref{z-error-mag-bound}, we obtain
\begin{equation}
\label{W-error-bound}
\Big|W_N^k -\frac{1}{\alpha} \operatorname{round}(\alpha W_N^k) \Big| \leq \frac{1}{\sqrt{2}\alpha}.
\end{equation}
Above inequality shows that the error bound for the scaled rounding operator is decreasing.
In particular, the error due the scaled rounding operator goes to zero as~$\alpha$ increase arbitrarily.
It means the the exact twiddle factor can be approximated with arbitrary precision~\cite{Cintra2011a}.
 Also,  it can be noted on Figure~\ref{SclRnd}  that  error magnitude of scaled rounded sine and cosine waves decay as precision parameter~$\alpha$ increases.
The precision on the representation of each twiddle factor is dependent only on the value of~$\alpha$.
That is  why we can call~$\alpha$ a precision parameter.

Also, applying~$z = W_N^k$ to~\eqref{ejtheta-round-mag-bound}, it yields
\begin{equation}
\label{W-round-bound}
1-\frac{1}{\sqrt{2}\alpha}\leq \Big|\frac{1}{\alpha} \operatorname{round}(\alpha W_N^k) \Big| \leq 1+\frac{1}{\sqrt{2}\alpha}.
\end{equation}
Above lower and upper bound for the approximate twiddle factor paves the way for important results in the remaining paper.
For instance, the lower bound is the bases for proving that all DFT approximation of proposed class allows perfect signal reconstruction.
In others words, it implies that the approximated DFT matrix is not singular.
 Also,  the upper bound is used for showing that the approximated DFT is asymptotically convergent to the exact DFT when~$\alpha$ is arbitrarily increased.

\subsection{The Approximate Twiddle Factor Properties}

In this sub-section we examine the approximate twiddle factor defined in~\eqref{matrixmapping-dft}.
The adequate representation of twiddle factor has direct impact on the quality of DFT approximation.
This calls for clear understanding of approximate twiddle factor.

As shown in the proceeding sub-section, the approximate twiddle factor has an upper and lower bound.
In particular,~\eqref{W-round-bound} shows that approximate twiddle factor has upper and lower bounds as a function of precision parameter~$\alpha$.
Furthermore, the upper and lower bounds are, respectively,  monotonically decreasing  and increasing  for precision parameter~$\alpha > 1$.
Next, we state and demonstrate a theorem on the upper and lower bound for all twiddle factor.

\begin{theorem}
\label{theo-W-mag-bound}
The norm of all twiddle factor has upper and lower bounds given by the  following  inequality:
\begin{equation*}
1-\frac{1}{\sqrt{2}} \leq \Big|\frac{1}{\alpha}\operatorname{round}\left(\alpha\cdot  W_N^k\right) \Big| \leq 1+\frac{1}{\sqrt{2}},
\end{equation*}
where~$k = 1, 2, \ldots, N/2-1$.

\begin{proof}
From~\eqref{W-round-bound}, we have that the upper and lower bounds for the approximate twiddle factor are, respectively,  decreasing  and increasing  functions of~$\alpha$.
Note that for any fixed precision parameter~$\alpha$, we have
\begin{equation*}
\inf_{\beta}\left(1-\frac{1}{\sqrt{2}\beta}\right)  \leq 1-\frac{1}{\sqrt{2}\beta},
\end{equation*}
where~$\inf_{\beta}(\cdot)$ represents the~\emph{infimum} of  the  sequence generated by setting~the parameter~$\beta \in 2^{\mathbb{N}}$ in its argument~\cite{James1996}.
Above inequality stems from the defintion of~$\inf_{\beta}(\cdot)$.
Also, we have that
\begin{equation*}
\sup_{\beta}\left(1+\frac{1}{\sqrt{2}\beta}\right)  \geq 1+\frac{1}{\sqrt{2}\beta},
\end{equation*}
where~$\sup_{\beta}(\cdot)$ represents the~\emph{supremum} of  the  sequence generated by setting~the parameter~$\beta \in 2^{\mathbb{N}}$ in its argument~\cite{James1996}.
Since
\begin{equation*}
\inf_{\beta}\left(1-\frac{1}{\sqrt{2}\beta}\right)  = 1-\frac{1}{\sqrt{2}}
\end{equation*}
and
\begin{equation*}
\sup_{\beta}\left(1+\frac{1}{\sqrt{2}\beta}\right) = 1+\frac{1}{\sqrt{2}\beta},
\end{equation*}
we have that
\begin{equation*}
1-\frac{1}{\sqrt{2}} \leq \Big|\frac{1}{\alpha}\operatorname{round}\left(\alpha\cdot  W_N^k\right) \Big| \leq 1+\frac{1}{\sqrt{2}},
\end{equation*}
where~$k = 1, 2, \ldots, N/2-1$.
\end{proof}
\end{theorem}

Now, above theorem allows us to have lower and upper bounds for approximate twiddle factor not depending on the parameter precision~$\alpha$.
Theorem~\ref{theo-W-mag-bound}  will be useful for demonstrating that all DFT approximations in the proposed class are reversible transforms, thus allowing signal perfect reconstruction~\cite{Britanak2007, Oppenheim2009}.

Next, we state and demonstrate a theorem about approximate twiddle factor convergence to the exact twiddle factor.
\begin{theorem}[Asymptotic Convergence of Approximate Twiddle Factor]
\label{twiddle-approx-theorem}
The approximate twiddle factor matrix given by
\begin{equation*}
\tilde{\mathbf{W}}_{N} = \frac{1}{\alpha}\operatorname{round}\left(\alpha \cdot \mathbf{W}_{N}\right),
\end{equation*}
that is induced by the mapping in~\eqref{matrixmapping-dft} is asymptotically convergent to the exact twiddle factor matrix
\begin{equation*}
\mathbf{W}_{N} = \operatorname{diag}\left(\begin{bmatrix}\mathbf{1}_{\mathrm{N}\times1}, 1, W_N, W_N^2, \ldots, W_N^{(\frac{N}{2}-1)}\end{bmatrix}^\top\right),
\end{equation*}
in the sense of Frobenius norm.

\begin{proof}
Let~$\tilde{\mathbf{W}}_{N}$ be given by~\eqref{matrixmapping-dft}. We want to show that~$\|\mathbf{W}_{N}-\tilde{\mathbf{W}}_{N}\|_{\mathrm{F}} \to 0$ as~$\alpha \to \infty$.
Thus, we have
\begin{align*}
\|\mathbf{W}_{N}-\tilde{\mathbf{W}}_{N}\|_{\mathrm{F}}
& = \Bigg\|\operatorname{diag}\left(\begin{bmatrix}\mathbf{1}_{\mathrm{N}\times1}, 1, W_N, W_N^2, \ldots, W_N^{(\frac{N}{2}-1)}\end{bmatrix}^\top\right)\\
& \phantom{=\Bigg\|\:\:} - \frac{1}{\alpha}\operatorname{round}\left(\alpha\cdot \operatorname{diag}\left(\begin{bmatrix}\mathbf{1}_{\mathrm{N}\times1}, 1, W_N, W_N^2, \ldots, W_N^{(\frac{N}{2}-1)}\end{bmatrix}^\top\right)\right)\Bigg\|_{\mathrm{F}}.
\end{align*}
Let us denote by~$d_{ii}$ each element of principal diagonal of~$\mathbf{W}_{N}-\tilde{\mathbf{W}}_{N}$.
Since~$\alpha$ is  an  integer, we have that~$d_{ii} = 0$ for~$i = 1, 2, \ldots, N/2+1$.
The remaining elements of diagonal principal are given by
\begin{equation*}
d_{ii} = W_N^{i-(\frac{N}{2}-1)} - \frac{1}{\alpha}\operatorname{round}\left(\alpha\cdot  W_N^{i-(\frac{N}{2}-1)}\right),
\end{equation*}
where~$i = N/2+2, N/2+3, \ldots, N$.

Note that the magnitude of~$d_{ii}$ is the expression in~\eqref{W-error-bound}.
Thus, we have that
\begin{align*}
|d_{ii}| \leq \frac{1}{\sqrt{2}\alpha}.
\end{align*}
Above expression shows that the error due the scaled rounding operator can be arbitrarily small as far~$\alpha$ increases.
Therefore, we have that
\begin{align*}
|d_{ii}| \to 0,
\end{align*}
as~$\alpha \to \infty$ for~$i = N/2+2, N/2+3, \ldots, N$.

Since both the exact and approximate twiddle factor matrix are diagonal matrices, calculating the Frobenius norm of~$\mathbf{W}_{N}-\tilde{\mathbf{W}}_{N}$ yields
\begin{align*}
\|\mathbf{W}_{N}-\tilde{\mathbf{W}}_{N}\|_{\mathrm{F}} & = \sqrt{\sum_{i = 1}^{N}|d_{ii}|^2}\\
& = \sqrt{\sum_{i = N/2+2}^{N}|d_{ii}|^2},
\end{align*}
since~$d_{ii} = 0$ for~$i = 1, 2, \ldots, N/2+1$.
Also, since~$|d_{ii}| \to 0 $ as~$\alpha \to \infty$ for~$i = 1, 2, \ldots, N/2+1$, we have %
\begin{align*}
\sum_{i = N/2+2}^{N}|d_{ii}|^2 \to 0
\end{align*}
as~$\alpha \to \infty$.
Nevertheless,~$\sqrt{\cdot}$ is a continuous function, thus we have that
\begin{align*}
\|\mathbf{W}_{N}-\tilde{\mathbf{W}}_{N}\|_{\mathrm{F}} \to 0,
\end{align*}
as~$\alpha \to \infty$.

\end{proof}
\end{theorem}

Since approximate twiddle factors  are  asymptotically convergent to the exact twiddle factor  for arbitrary power-of-two lengths~$N$, we   expect that  the DFT approximations induced by the use of such approximate twiddle factors  are  also asymptotically convergent to the exact DFT.
Thus, above theorem sets the bases for demonstrate that DFT approximations are asymptotically convergent to the exact DFT.

Above condition of asymptotic convergence of approximate twiddle factor to the exact twiddle factor is necessary, but not sufficient.
In the following, we state and demonstrate a theorem setting magnitude constraints to the Frobenius norm of approximate twiddle factor matrix.

\begin{theorem}
\label{twiddle-ineq}
Let~$\tilde{\mathbf{W}}_{N}$ be the approximate twiddle factor to~$\mathbf{W}_{N}$ as induced by~\eqref{matrixmapping-dft}.
Thus, we have that
\begin{align*}
\left(1-\frac{1}{\sqrt{2}}\right)\sqrt{N} \leq \|\tilde{\mathbf{W}}_{N}\|_\mathrm{F} & \leq \left(1+\frac{1}{\sqrt{2}}\right)\sqrt{N}.
\end{align*}

\begin{proof}

Let us denote each element of main diagonal of~$\tilde{\mathbf{W}}_{N}$ by~$\tilde{w}_{ii}$ for~$i = 1, 2, \ldots, N$.
Note that~$\tilde{w}_{ii} = 1$ for~$i = 1, 2, \ldots, N/2+1$.
The remaining elements of main diagonal are given by
\begin{equation*}
\tilde{w}_{ii} = \frac{1}{\alpha}\operatorname{round}\left(\alpha\cdot  W_N^{i-(\frac{N}{2}-1)}\right),
\end{equation*}
where~$i = N/2+2, N/2+3, \ldots, N$.
Applying  Theorem~\ref{theo-W-mag-bound} for the upper and lower bounds for the magnitude of~$\tilde{w}_{ii}$, we have that
\begin{align*}
1-\frac{1}{\sqrt{2}} \leq |\tilde{w}_{ii}| & \leq 1+\frac{1}{\sqrt{2}},
\end{align*}
where~$i = N/2+2, N/2+3, \ldots, N$.
Since~$\tilde{w}_{ii} = 1$ for~$i = 1, 2, \ldots, N/2+1$, we can generalize the above result for all~$i = 1, 2, \ldots, N$.
Since the approximate twiddle factor matrix is diagonal, summing up the squared value of~$|\tilde{w}_{ii}|$ in above inequality for~$i = 1, 2, \ldots, N$ and taking the square root  returns  the Frobenius norm of~$\tilde{\mathbf{W}}_{N}$.
Therefore, we have that
\begin{align*}
\left(1-\frac{1}{\sqrt{2}}\right)\sqrt{N} \leq \|\tilde{\mathbf{W}}_{N}\|_\mathrm{F} & \leq \left(1+\frac{1}{\sqrt{2}}\right)\sqrt{N}.
\end{align*}
\end{proof}
\end{theorem}

Above theorem goes beyond initial demonstration that there  are  lower and upper bounds for the magnitude of each approximate twiddle factor.
Instead, it paves the way for showing that DFT approximations in the proposed class are asymptotically convergent to the exact DFT.
Furthermore, it allows us demonstrate that each DFT approximation is invertible, i.e., there is a inverse transformation that furnish exact signal reconstruction.

\subsection{The DFT Approximation and Its Properties}

First, let us define the DFT approximation according the mapping in~\eqref{matrixmapping-dft}.

\begin{definition}
\label{DFT-approx-def}
Let~$N$ be a power-of-two such that~$\operatorname{log}_2(N) > 2$ and~$\tilde{\mathbf{W}}_{N}$ be the approximate twiddle factor obtained by the mapping~\eqref{matrixmapping-dft} for~$N$.
The $N$-point DFT approximation associated to the exact $N$-point DFT with precision parameter~$\alpha$ is given by
\begin{equation*}
\tilde{\mathbf{F}}_{N} = \mathbf{A}_N \tilde{\mathbf{W}}_{N}
\left( \mathbf{I}_2 \otimes \tilde{\mathbf{F}}_{N/2} \right)
\mathbf{B}_{N},
\end{equation*}
where~$\tilde{\mathbf{F}}_{N/2}$ is the associated $N/2$-point DFT approximation  and the initial condition is
\begin{equation*}
\tilde{\mathbf{F}}_4 = \mathbf{F}_4 =
\left[
\begin{rsmallmatrix}
1 & 1& 1 & 1\\
1 & -j&-1&j\\
1 & -1&1 & -1\\
1 & j&-1&-j
\end{rsmallmatrix}\right].
\end{equation*}

\end{definition}

Note that the above approximation for $N$-point DFT is constructed by just replacing the exact twiddle factor matrix in~\eqref{CT} by the approximate twiddle factor matrix~$\tilde{\mathbf{W}}_N$, since it is given the $N/2$-point DFT approximation.
Since~$\alpha$ now is chosen as a power-of-two, the division required by~$\tilde{\mathbf{W}}_N$ is computationally not expensive.
Indeed, it is performed just by a bit-shifting operations~\cite{Blahut2010}.
In the following we have an example that shows how to construct a DFT approximation based on the above definition.

\begin{example}
Take the case where~$N = 8$.
Then~\eqref{matrixmapping-dft} defines~$\tilde{\mathbf{W}}_\mathrm{8} = \frac{1}{\alpha}\operatorname{round}\left(\alpha \cdot \mathbf{W}_\mathrm{8}\right)$.
For the fixed~$N = 8$, consider~\eqref{CT} then substitute~$\mathbf{W}_\mathrm{8}$ by~$\tilde{\mathbf{W}}_\mathrm{8}$.
Thus, we have
\begin{equation*}
\tilde{\mathbf{F}}_\mathrm{8} = \mathbf{A}_\mathrm{8} \tilde{\mathbf{W}}_\mathrm{8}
\left( \mathbf{I}_2 \otimes \tilde{\mathbf{F}}_\mathrm{4} \right)
\mathbf{B}_\mathrm{8},
\end{equation*}
where~$\tilde{\mathbf{F}}_\mathrm{4} = \mathbf{F}_4$ as in the initial conditions in~Definition~\ref{DFT-approx-def}.

If we explicitly compute~$\tilde{\mathbf{F}}_\mathrm{8}$ for~$\alpha = 2$, we obtain
\begin{equation*}
\tilde{\mathbf{F}}_\mathrm{8} = \left[
\begin{rsmallmatrix}
1 & 1 & 1 & 1 & 1 & 1 & 1 & 1  \\
1 & \bar{a} & -j & -a & -1 & -\bar{a} & j &  a\\
1 &  -j & -1 &  j &  1 & -j & -1 &  j\\
1 & -a & j &  \bar{a} & -1 & a & -j & -\bar{a}\\
1 & -1 & 1 & -1 & 1 &    -1 & 1 & -1 \\
1 & -\bar{a} & -j  & a & -1 & \bar{a} & j & -a\\
1 &  j & -1 &  -j  & 1 & j & -1 &  -j\\
1 & a & j & -\bar{a} & -1 & -a & -j  & \bar{a}
\end{rsmallmatrix}
\right],
\end{equation*}
where~$a = (1+j)/2$.

\end{example}

In the following theorem we demonstrate that~$8$-point DFT approximation induced by~\eqref{matrixmapping-dft} is also asymptotically convergent to the exact~$8$-point DFT in the sense of Frobenius norm.
\begin{theorem}
\label{DFTA8-conv}
The approximation for the~$8$-point DFT  as
\begin{equation*}
\tilde{\mathbf{F}}_\mathrm{8} = \mathbf{A}_\mathrm{8} \tilde{\mathbf{W}}_\mathrm{8}
\left( \mathbf{I}_2 \otimes \tilde{\mathbf{F}}_\mathrm{4} \right)
\mathbf{B}_\mathrm{8},
\end{equation*}
is asymptotically convergent to the exact DFT as~$\alpha \to \infty$ in the sense of Frobenius norm~\cite{Britanak2007, Manassah2001},
where~$\tilde{\mathbf{W}}_\mathrm{8}$ is given by
\begin{equation*}
\tilde{\mathbf{W}}_\mathrm{8} = \frac{1}{\alpha}\operatorname{round}\left(\alpha \cdot \mathbf{W}_\mathrm{8}\right),
\end{equation*}
and~$\tilde{\mathbf{F}}_\mathrm{4}$ is the $4$-point DFT approximation for accuracy parameter~$\alpha$.

\begin{proof}
Let us calculate the Frobenius norm of difference between the exact and approximate~$8$-point DFT. It yields,
\begin{align*}
\|\mathbf{F}_\mathrm{8}-\tilde{\mathbf{F}}_\mathrm{8}\|_{\mathrm{F}} & = \|\mathbf{A}_\mathrm{8} \mathbf{W}_\mathrm{8}
\left( \mathbf{I}_2 \otimes \mathbf{F}_\mathrm{4} \right)\mathbf{B}_\mathrm{8} - \mathbf{A}_\mathrm{8} \tilde{\mathbf{W}}_\mathrm{8}\left( \mathbf{I}_2 \otimes \tilde{\mathbf{F}}_\mathrm{4} \right) \mathbf{B}_\mathrm{8}\|_{\mathrm{F}}\\
& = \|\mathbf{A}_\mathrm{8} \left[\mathbf{W}_\mathrm{8} \left( \mathbf{I}_2 \otimes \mathbf{F}_\mathrm{4} \right) - \tilde{\mathbf{W}}_\mathrm{8}\left( \mathbf{I}_2 \otimes \tilde{\mathbf{F}}_\mathrm{4} \right)\right] \mathbf{B}_\mathrm{8}\|_{\mathrm{F}}.
\end{align*}
Using the sub-multiplicative property of Frobenius norm~\cite{Britanak2007}, we have that
\begin{align*}
\|\mathbf{F}_\mathrm{8}-\tilde{\mathbf{F}}_\mathrm{8}\|_{\mathrm{F}} & \leq \|\mathbf{A}_\mathrm{8}\|_\mathrm{F} \|\left[\mathbf{W}_\mathrm{4} \left( \mathbf{I}_2 \otimes \mathbf{F}_\mathrm{4} \right) - \tilde{\mathbf{W}}_\mathrm{8}\left( \mathbf{I}_2 \otimes \tilde{\mathbf{F}}_\mathrm{4} \right)\right]\|_\mathrm{F} \|\mathbf{B}_\mathrm{8}\|_{\mathrm{F}}\\
& \leq \sqrt{16} \|\left[\mathbf{W}_\mathrm{8} \left( \mathbf{I}_2 \otimes \mathbf{F}_\mathrm{4} \right) - \tilde{\mathbf{W}}_\mathrm{8}\left( \mathbf{I}_2 \otimes \tilde{\mathbf{F}}_\mathrm{4} \right)\right]\|_\mathrm{F} \cdot \sqrt{8}\\
& \leq 8\sqrt{2} \|\left[\mathbf{W}_\mathrm{8} \left( \mathbf{I}_2 \otimes \mathbf{F}_\mathrm{4} \right) - \tilde{\mathbf{W}}_\mathrm{8}\left( \mathbf{I}_2 \otimes \tilde{\mathbf{F}}_\mathrm{4} \right)\right]\|_\mathrm{F}.
\end{align*}
The Theorem~\ref{twiddle-approx-theorem} states that~$\|\mathbf{W}_\mathrm{8}-\tilde{\mathbf{W}}_\mathrm{8}\|_\mathrm{F} \to 0$ as~$\alpha \to \infty$.
Also, note that~$\mathbf{I}_2 \otimes \tilde{\mathbf{F}}_\mathrm{4}$ is convergent in Frobenius norm to~$\mathbf{I}_2 \otimes \mathbf{F}_\mathrm{4}$ as~$\alpha \to \infty$. Let~$\mathbf{0}_4$ be the~$4\times 4$ null matrix.
We have that
\begin{align*}
\|\mathbf{I}_2 \otimes \mathbf{F}_\mathrm{4}- \mathbf{I}_2 \otimes \tilde{\mathbf{F}}_\mathrm{4}\|_\mathrm{F}
& = \|\mathbf{I}_2 \otimes \left(\mathbf{F}_\mathrm{4}- \tilde{\mathbf{F}}_\mathrm{4}\right)\|_\mathrm{F}\\
& =
\Big\|\left[ \begin{rsmallmatrix}
\mathbf{F}_\mathrm{4}- \tilde{\mathbf{F}}_\mathrm{4} & \mathbf{0}_4\\
\mathbf{0}_4 & \phantom{--}\mathbf{F}_\mathrm{4}- \tilde{\mathbf{F}}_\mathrm{4}
\end{rsmallmatrix}\right]\Big\|_\mathrm{F}\\
& = \sqrt{2}\cdot \|\mathbf{F}_\mathrm{4}- \tilde{\mathbf{F}}_\mathrm{4}\|_\mathrm{F}.
\end{align*}
The initial condition~$\tilde{\mathbf{F}}_4 = \mathbf{F}_4$ in Definition~\ref{DFT-approx-def} implies in $\|\mathbf{F}_\mathrm{4}- \tilde{\mathbf{F}}_\mathrm{4}\|_\mathrm{F} = 0$, what yields~$\|\mathbf{I}_2 \otimes \mathbf{F}_\mathrm{4}- \mathbf{I}_2 \otimes \tilde{\mathbf{F}}_\mathrm{4}\|_\mathrm{F} = 0$.
In particular, we have that
\begin{align*}
\|\mathbf{I}_2 \otimes \mathbf{F}_\mathrm{4}- \mathbf{I}_2 \otimes \tilde{\mathbf{F}}_\mathrm{4}\|_\mathrm{F} \to 0,
\end{align*}
as~$\alpha \to \infty$.

Now we have that both~$\|\mathbf{W}_\mathrm{8}-\tilde{\mathbf{W}}_\mathrm{8}\|_\mathrm{F} \to 0$  and~$\|\mathbf{I}_2 \otimes \mathbf{F}_\mathrm{4}- \mathbf{I}_2 \otimes \tilde{\mathbf{F}}_\mathrm{4}\|_\mathrm{F} \to 0$ as~$\alpha \to \infty$.
Also, note that~$\|\tilde{\mathbf{W}}_\mathrm{8}\|_\mathrm{F} \leq (1+1/\sqrt{2})\sqrt{8}$ according Theorem~\ref{twiddle-ineq}.
We invoke~Appendix~\ref{prod-conv}, what yields
\begin{equation*}
\|\left[\mathbf{W}_\mathrm{8} \left( \mathbf{I}_2 \otimes \mathbf{F}_\mathrm{4} \right) - \tilde{\mathbf{W}}_\mathrm{8}\left( \mathbf{I}_2 \otimes \tilde{\mathbf{F}}_\mathrm{4} \right)\right]\|_\mathrm{F} \to 0,
\end{equation*}
as~$\alpha \to \infty$.
Therefore, we have that
\begin{equation*}
\|\mathbf{F}_\mathrm{8}-\tilde{\mathbf{F}}_\mathrm{8}\|_{\mathrm{F}} \to 0,
\end{equation*}
as~$\alpha \to \infty$.
\end{proof}
\end{theorem}

The above demonstrated convergence in Frobenius norm for~$N = 8$ is just a example of the framework followed by the next general theorem.
In the following we demonstrate that for any~$N$ power-of-two such that~$\operatorname{log}_2(N) > 2$, the associated DFT approximation is asymptotically convergent to the exact DFT.

\begin{theorem}[Asymptotic Convergence of DFT Approximations]
\label{DFTAN-conv}
The approximation for the~$N$-point DFT  as in Definition~\ref{DFT-approx-def} is asymptotically convergent to the exact DFT as~$\alpha \to \infty$ in the sense of Frobenius norm~\cite{Britanak2007, Manassah2001}, since~$\tilde{\mathbf{F}}_{N/2}$ is the $N/2$-point DFT approximation for precision parameter~$\alpha$.

\begin{proof}
In order to prove the above statement, we use finite induction.
Thus, we have:
\begin{itemize}
\item[\textbf{Base Case:}] For the case~$N = 4$, DFT approximation is the exact DFT as given by  the initial condition.
\\
\item[\textbf{Hypothesis:}] Consider that the~$N/2$-point DFT approximation is asymptotically convergent to the correspondent exact $N/2$-point DFT.
\end{itemize}

Let us prove that the~$N$-point DFT approximation as in Definition~\ref{DFT-approx-def} is asymptotically convergent to the exact $N$-point DFT as~$\alpha \to \infty$ with above considerations.
The calculation the Frobenius norm of difference between the exact and approximate~$N$-point DFT yields:
\begin{align*}
\|\mathbf{F}_{N}-\tilde{\mathbf{F}}_{N}\|_{\mathrm{F}} & = \|\mathbf{A}_N \mathbf{W}_{N}
\left( \mathbf{I}_2 \otimes \mathbf{F}_{N/2} \right)\mathbf{B}_{N} - \mathbf{A}_N \tilde{\mathbf{W}}_{N}\left( \mathbf{I}_2 \otimes \tilde{\mathbf{F}}_{N/2} \right) \mathbf{B}_{N}\|_{\mathrm{F}}\\
& = \|\mathbf{A}_N \left[\mathbf{W}_{N} \left( \mathbf{I}_2 \otimes \mathbf{F}_{N} \right) - \tilde{\mathbf{W}}_{N}\left( \mathbf{I}_2 \otimes \tilde{\mathbf{F}}_{N} \right)\right] \mathbf{B}_{N}\|_{\mathrm{F}}.
\end{align*}
Using the sub-multiplicative property of Frobenius norm~\cite{Britanak2007}, we have that
\begin{align*}
\|\mathbf{F}_{N}-\tilde{\mathbf{F}}_{N}\|_{\mathrm{F}} & \leq \|\mathbf{A}_N\|_\mathrm{F} \|\left[\mathbf{W}_{N} \left( \mathbf{I}_2 \otimes \mathbf{F}_{N/2} \right) - \tilde{\mathbf{W}}_{N}\left( \mathbf{I}_2 \otimes \tilde{\mathbf{F}}_{N/2} \right)\right]\|_\mathrm{F} \|\mathbf{B}_{N}\|_{\mathrm{F}}\\
& \leq \sqrt{2N} \|\left[\mathbf{W}_{N} \left( \mathbf{I}_2 \otimes \mathbf{F}_{N/2} \right) - \tilde{\mathbf{W}}_{N}\left( \mathbf{I}_2 \otimes \tilde{\mathbf{F}}_{N/2} \right)\right]\|_\mathrm{F} \cdot \sqrt{N}\\
& \leq \sqrt{2}N \|\left[\mathbf{W}_{N} \left( \mathbf{I}_2 \otimes \mathbf{F}_{N/2} \right) - \tilde{\mathbf{W}}_{N}\left( \mathbf{I}_2 \otimes \tilde{\mathbf{F}}_{N/2} \right)\right]\|_\mathrm{F}.
\end{align*}
Theorem~\ref{twiddle-approx-theorem} states that~$\|\mathbf{W}_{N}-\tilde{\mathbf{W}}_{N}\|_\mathrm{F} \to 0$ as~$\alpha \to \infty$.
Also, note that~$\mathbf{I}_2 \otimes \tilde{\mathbf{F}}_{N/2}$ is convergent in Frobenius norm to~$\mathbf{I}_2 \otimes \mathbf{F}_{N/2}$ as~$\alpha \to \infty$. Let~$\mathbf{0}_{N/2}$ be the~$(N/2)\times (N/2)$ null matrix.
We have that
\begin{align*}
\|\mathbf{I}_2 \otimes \mathbf{F}_{N/2}- \mathbf{I}_2 \otimes \tilde{\mathbf{F}}_{N/2}\|_\mathrm{F}
& = \|\mathbf{I}_2 \otimes \left(\mathbf{F}_{N/2}- \tilde{\mathbf{F}}_{N/2}\right)\|_\mathrm{F}\\
& =
\Big\|\left[ \begin{rsmallmatrix}
\mathbf{F}_{N/2}- \tilde{\mathbf{F}}_{N/2} & \mathbf{0}_{N/2}\\
\mathbf{0}_{N/2} & \phantom{--}\mathbf{F}_{N/2}- \tilde{\mathbf{F}}_{N/2}
\end{rsmallmatrix}\right]\Big\|_\mathrm{F}\\
& = \sqrt{2}\cdot \|\mathbf{F}_{N/2}- \tilde{\mathbf{F}}_{N/2}\|_\mathrm{F}.
\end{align*}
According the inductive hypothesis, we have that~$\|\mathbf{F}_{N/2}- \tilde{\mathbf{F}}_{N/2}\|_\mathrm{F} \to 0$ as~$\alpha \to \infty$.
Thus, we obtain
\begin{align*}
\|\mathbf{I}_2 \otimes \mathbf{F}_{N/2}- \mathbf{I}_2 \otimes \tilde{\mathbf{F}}_{N/2}\|_\mathrm{F} \to 0,
\end{align*}
as~$\alpha \to \infty$.

Now we have that both~$\|\mathbf{W}_{N}-\tilde{\mathbf{W}}_{N}\|_\mathrm{F} \to 0$  and~$\|\mathbf{I}_2 \otimes \mathbf{F}_{N/2}- \mathbf{I}_2 \otimes \tilde{\mathbf{F}}_{N/2}\|_\mathrm{F} \to 0$ as~$\alpha \to \infty$.
Also, note that~$\|\tilde{\mathbf{W}}_{N}\|_\mathrm{F} \leq (1+1/\sqrt{2})\sqrt{N}$ according Theorem~\ref{twiddle-ineq}.
We invoke~Appendix~\ref{prod-conv}, what yields
\begin{equation*}
\|\left[\mathbf{W}_{N} \left( \mathbf{I}_2 \otimes \mathbf{F}_{N/2} \right) - \tilde{\mathbf{W}}_{N}\left( \mathbf{I}_2 \otimes \tilde{\mathbf{F}}_{N/2} \right)\right]\|_\mathrm{F} \to 0,
\end{equation*}
as~$\alpha \to \infty$.
Therefore, we have that
\begin{equation*}
\|\mathbf{F}_{N}-\tilde{\mathbf{F}}_{N}\|_{\mathrm{F}} \to 0,
\end{equation*}
as~$\alpha \to \infty$.
\end{proof}
\end{theorem}

The above theorem proves that for any power-of-two~$N$, the $N$-point DFT approximation is asymptotically convergent to the exact~$N$-point DFT.
In signal processing problems, we are usually interested in represent back to time domain signals after being processed in transformed domain.
It is possible only if the employed linear transformation is invertible.
Following we consider the invertibility of the proposed class of approximations.

\subsection{Invertibility of DFT Approximations}
In this section, we investigate the invertibility condition for DFT approximations.
Invertibility is of main concern for it assures exact signal reconstruction~\cite{Britanak2007}.

Consider the case where~$N = 8$.
We obtain
\begin{align}
\label{det-DFTA8}
\operatorname{det}\left(\tilde{\mathbf{F}}_\mathrm{8}\right) & = \operatorname{det}\left(\mathbf{A}_\mathrm{8}\right) \operatorname{det}\left(\tilde{\mathbf{W}}_\mathrm{8}\right) \operatorname{det}\left( \mathbf{I}_2 \otimes \mathbf{F}_\mathrm{4} \right)\operatorname{det}\left(\mathbf{B}_\mathrm{8}\right)\nonumber\\
& = \operatorname{det}\left(\mathbf{A}_\mathrm{8}\right) \operatorname{det}\left(\tilde{\mathbf{W}}_\mathrm{8}\right) \operatorname{det}\left(\mathbf{I}_2\right)^4 \operatorname{det}\left(\mathbf{F}_\mathrm{4}\right)^2 \operatorname{det}\left(\mathbf{B}_\mathrm{8}\right)\nonumber\\
& = \operatorname{det}\left(\mathbf{A}_\mathrm{8}\right) \operatorname{det}\left(\tilde{\mathbf{W}}_\mathrm{8}\right)  \operatorname{det}\left(\mathbf{F}_\mathrm{4}\right)^2 \operatorname{det}\left(\mathbf{B}_\mathrm{8}\right).
\end{align}

Above result  serve as bases for the demonstration of a general expression for determinant of a~$N$-point DFT approximation.
In the following, we present and prove a general expression for determinant of a~$N$-point DFT approximation.

\begin{theorem}
\label{DFT-approx-det-theorem}
Let~$\tilde{\mathbf{F}}_{N}$ be the~$N$-point DFT approximation with a precision parameter~$\alpha$ as in~Definition~\ref{DFT-approx-def}.
A general expression for determinant of an~$N$-point DFT approximation is given by
\begin{equation*}
\operatorname{det}\left(\tilde{\mathbf{F}}_{N}\right) = \prod_{i = 0}^{\operatorname{log}(N)- 3 } \left( \operatorname{det}\left(\mathbf{A}_{\mathrm{N}/2^i}\right) \operatorname{det}\left(\tilde{\mathbf{W}}_{N/2^i}\right) \operatorname{det}\left(\mathbf{B}_{N/2^i}\right) \right)^{2^i}\operatorname{det}\left(\mathbf{F}_\mathrm{4}\right)^{N/4}.
\end{equation*}

\begin{proof}
In order to prove the above statement, we use finite induction.
Thus, we have:
\begin{itemize}
\item[\textbf{Base Case:}] The determinant of~$8$-point DFT approximation matrix is given by~\eqref{det-DFTA8},  which  can be obtained by making $N = 8$ in above equation for the general case.\\
\item[\textbf{Hypothesis:}] Suppose the equation for the general case is valid for~$N/2$, thus~$\operatorname{det}\left(\tilde{\mathbf{F}}_{N/2}\right)$ equals to
\begin{equation*}
\prod_{i = 0}^{\operatorname{log}(N/2)-3} \left( \operatorname{det}\left(\mathbf{A}_{\mathrm{N}/2^{i+1}}\right) \operatorname{det}\left(\tilde{\mathbf{W}}_{\mathrm{N}/2^{i+1}}\right) \operatorname{det}\left(\mathbf{B}_{\mathrm{N}/2^{i+1}}\right) \right)^{2^i}\operatorname{det}\left(\mathbf{F}_\mathrm{4}\right)^{N/8}.
\end{equation*}
\end{itemize}
Let us prove the formula for the general case~$N$ assuming the above considerations.
Using Definition~\ref{DFT-approx-def} and  Kronecker product properties~\cite{Zhang2013, Britanak2007}, we have that
\begin{align*}
\operatorname{det}\left(\tilde{\mathbf{F}}_{N}\right) & = \operatorname{det}\left(\mathbf{A}_N\right) \operatorname{det}\left(\tilde{\mathbf{W}}_{N}\right) \operatorname{det}\left( \mathbf{I}_2 \otimes \tilde{\mathbf{F}}_{N/2} \right)\operatorname{det}\left(\mathbf{B}_{N}\right)\\
& = \operatorname{det}\left(\mathbf{A}_N\right) \operatorname{det}\left(\tilde{\mathbf{W}}_{N}\right) \operatorname{det}\left(\mathbf{I}_2\right)^{N/2} \operatorname{det}\left(\tilde{\mathbf{F}}_{N/2}\right)^2 \operatorname{det}\left(\mathbf{B}_{N}\right)\\
& = \operatorname{det}\left(\mathbf{A}_N\right) \operatorname{det}\left(\tilde{\mathbf{W}}_{N}\right)  \operatorname{det}\left(\tilde{\mathbf{F}}_{N/2}\right)^2 \operatorname{det}\left(\mathbf{B}_{N}\right).
\end{align*}
Using the inductive hypothesis, we have that
\begin{align*}
\operatorname{det}\left(\tilde{\mathbf{F}}_{N}\right) & = \operatorname{det}\left(\mathbf{A}_N\right) \operatorname{det}\left(\tilde{\mathbf{W}}_{N}\right)  \operatorname{det}\left(\tilde{\mathbf{F}}_{N/2}\right)^2 \operatorname{det}\left(\mathbf{B}_{N}\right)\\
&=  \left(\prod_{i = 0}^{\operatorname{log}(N/2)-3} \left( \operatorname{det}\left(\mathbf{A}_{\mathrm{N}/2^{i+1}}\right) \operatorname{det}\left(\tilde{\mathbf{W}}_{\mathrm{N}/2^{i+1}}\right) \operatorname{det}\left(\mathbf{B}_{\mathrm{N}/2^{i+1}}\right) \right)^{2^i}\operatorname{det}\left(\mathbf{F}_\mathrm{4}\right)^{N/8}\right)^2 \\
&  \phantom{---}\cdot \operatorname{det}\left(\mathbf{A}_N\right) \operatorname{det}\left(\tilde{\mathbf{W}}_{N}\right)\operatorname{det}\left(\mathbf{B}_{N}\right)\\
&=  \prod_{i = 0}^{\operatorname{log}(N/2)-3} \left( \operatorname{det}\left(\mathbf{A}_{\mathrm{N}/2^{i+1}}\right) \operatorname{det}\left(\tilde{\mathbf{W}}_{\mathrm{N}/2^{i+1}}\right) \operatorname{det}\left(\mathbf{B}_{\mathrm{N}/2^{i+1}}\right) \right)^{2^{i+1}}\operatorname{det}\left(\mathbf{F}_\mathrm{4}\right)^{N/4} \\
&  \phantom{---}\cdot \operatorname{det}\left(\mathbf{A}_N\right) \operatorname{det}\left(\tilde{\mathbf{W}}_{N}\right)\operatorname{det}\left(\mathbf{B}_{N}\right).
\end{align*}
Making the variable change in above product of~$j = i+1$, where~$j = 1, 2, \ldots, \operatorname{log}(N/2)-1$, we obtain
\begin{align*}
\operatorname{det}\left(\tilde{\mathbf{F}}_{N}\right) &=  \prod_{j = 1}^{\operatorname{log}(N/2)-2} \left( \operatorname{det}\left(\mathbf{A}_{\mathrm{N}/2^{j}}\right) \operatorname{det}\left(\tilde{\mathbf{W}}_{N/2^{j}}\right) \operatorname{det}\left(\mathbf{B}_{\mathrm{N}/2^{j}}\right) \right)^{2^{j}}\operatorname{det}\left(\mathbf{F}_\mathrm{4}\right)^{N/4} \\
&  \phantom{---}\cdot \operatorname{det}\left(\mathbf{A}_N\right) \operatorname{det}\left(\tilde{\mathbf{W}}_{N}\right)\operatorname{det}\left(\mathbf{B}_{N}\right).
\end{align*}
Using above equation, we can include the case~$j=0$ and use that~$\operatorname{log}(N/2) = \operatorname{log}(N)-1$, what yields
\begin{align*}
\operatorname{det}\left(\tilde{\mathbf{F}}_{N}\right) &=  \prod_{j = 0}^{\operatorname{log}(N)-3} \left( \operatorname{det}\left(\mathbf{A}_{\mathrm{N}/2^{j}}\right) \operatorname{det}\left(\tilde{\mathbf{W}}_{N/2^{j}}\right) \operatorname{det}\left(\mathbf{B}_{\mathrm{N}/2^{j}}\right) \right)^{2^{j}}\operatorname{det}\left(\mathbf{F}_\mathrm{4}\right)^{N/4}.
\end{align*}
\end{proof}
\end{theorem}

Above theorem is relevant to shows that the invertibility of a~$N$-point DFT approximation is dependent only on approximation of twiddle factor for~$N/2, N/4, \ldots, 8$.
First, note that by Definition~\ref{DFT-approx-def} we have that~$\tilde{\mathbf{F}}_\mathrm{4} = \mathbf{F}_\mathrm{4}$.
Also,~$\operatorname{det}\left(\mathbf{F}_4\right) = 16j$, where~$j = \sqrt{-1}$  and the decimation-in-time matrix~$\mathbf{B}_{N}$ is just a permutation matrix, what results in~$\operatorname{det}\left(\mathbf{B}_{N}\right) = \pm 1$.
For the post-addition matrix~$\mathbf{A}_N$, we have that~$\operatorname{det}\left(\mathbf{A}_N\right) = 2^{N/2}$.
The following theorem sets a result for invertibility.

\begin{theorem}
\label{invertibility-condition-theorem}
Let~$\tilde{\mathbf{F}}_{N}$ be a $N$-point DFT approximation for the exact~$N$-point DFT.
The $N$-point DFT approximation is invertible if and only if
\begin{equation*}
\prod_{i = 0}^{\operatorname{log}(N)-3} \big|\operatorname{det}\left(\tilde{\mathbf{W}}_{N/2^{j}}\right)\big| \neq 0,
\end{equation*}
where~$|\cdot|$ returns the norm of its complex argument.

\begin{proof}
Note that~$\big|\operatorname{det}\left(\mathbf{F}_\mathrm{4}\right)\big| = 16$,~$\big|\operatorname{det}\left(\mathbf{B}_{N}\right)\big| = 1$ and~$\big|\operatorname{det}\left(\mathbf{A}_N\right)\big| = 2^{N/2}$. Using the expression in Theorem~\ref{DFT-approx-det-theorem}, we have that
\begin{align*}
\big|\operatorname{det}\left(\tilde{\mathbf{F}}_{N}\right)\big| & = \Big|\prod_{i = 0}^{\operatorname{log}(N)-3} \left( \operatorname{det}\left(\mathbf{A}_{\mathrm{N}/2^i}\right) \operatorname{det}\left(\tilde{\mathbf{W}}_{N/2^i}\right) \operatorname{det}\left(\mathbf{B}_{N/2^i}\right) \right)^{2^i}\operatorname{det}\left(\mathbf{F}_\mathrm{4}\right)^{N/4}\Big|\\
& = \prod_{i = 0}^{\operatorname{log}(N)-3} \left( \big|\operatorname{det}\left(\mathbf{A}_{\mathrm{N}/2^i}\right)\big| \big|\operatorname{det}\left(\tilde{\mathbf{W}}_{N/2^i}\right)\big| \big|\operatorname{det}\left(\mathbf{B}_{N/2^i}\right)\big| \right)^{2^i}\big|\operatorname{det}\left(\mathbf{F}_\mathrm{4}\right)^{N/4}\big|\\
& = \prod_{i = 0}^{\operatorname{log}(N)-3} \left( 2^{N/2^i} \big|\operatorname{det}\left(\tilde{\mathbf{W}}_{N/2^i}\right)\big|\right)^{2^i}16^{N/4}\\
& = \prod_{i = 0}^{\operatorname{log}(N)-3} 2^{N}  \big|\operatorname{det}\left(\tilde{\mathbf{W}}_{N/2^i}\right)\big|^{2^i}2^N\\
& = 2^{N(\operatorname{log}(N)-1)}\prod_{i = 0}^{\operatorname{log}(N)-3} \big|\operatorname{det}\left(\tilde{\mathbf{W}}_{N/2^i}\right)\big|^{2^i}.%
\end{align*}
Since~$2^{N(\operatorname{log}(N)-1)}$ is always non-zero, invertibility concerns remain in the product operator on the right side.
Thus, $\big|\operatorname{det}\left(\tilde{\mathbf{F}}_{N}\right)\big| \neq 0$ if and only if the product operator on the right side is non-zero.
Also, note that
\begin{align*}
\prod_{i = 0}^{\operatorname{log}(N)-3} \big|\operatorname{det}\left(\tilde{\mathbf{W}}_{N/2^i}\right)\big|^{2^i} \neq 0 \Longleftrightarrow
\prod_{i = 0}^{\operatorname{log}(N)-3} \big|\operatorname{det}\left(\tilde{\mathbf{W}}_{N/2^i}\right)\big| \neq 0.
\end{align*}

Therefore, the~$N$-point DFT approximation is invertible if and only if
\begin{align*}
\prod_{i = 0}^{\operatorname{log}(N)-3} \big|\operatorname{det}\left(\tilde{\mathbf{W}}_{N/2^i}\right)\big| \neq 0.
\end{align*}

\end{proof}

\end{theorem}

Remember that for a arbitrary diagonal matrix, the determinant is given by the product of its principal diagonal elements~\cite{Britanak2007}.
In particular, for any approximate twiddle factor matrix for a~$N$-point DFT approximation, we have that
\begin{equation}
\label{twiddle-factor-det}
\operatorname{det}\left(\tilde{\mathbf{W}}_{N}\right) = \prod_{k = N/2+1}^{N} \frac{1}{\alpha}\operatorname{round}\left( \alpha W_N^k \right).
\end{equation}

In the following, we use the above expression in~\eqref{twiddle-factor-det} with Theorem~\ref{invertibility-condition-theorem} to show that all approximations included in this class of DFT approximations are invertible for arbitrary positive power-of-two~$N$.

\begin{theorem}
Any $N$-point DFT approximation with a non-zero precision parameter~$\alpha$ is invertible.

\begin{proof}
Consider a $N$-point DFT approximation with non-zero precision parameter~$\alpha$.
Consider the bounds for approximate twiddle factor as in Theorem~\ref{theo-W-mag-bound}.
We have that
\begin{equation*}
\left|\frac{1}{\alpha}\operatorname{round}\left(\alpha\cdot  W_N^k\right)\right| \geq 1-\frac{1}{\sqrt{2}} ,
\end{equation*}
where~$k = 1, 2, \ldots, N/2-1$.
Since~$\tilde{\mathbf{W}}_{N}$ is a diagonal matrix, its determinant is given by~\eqref{twiddle-factor-det}.
Then, using the inequality stated above, we have that
\begin{equation*}
\operatorname{det}\left(\tilde{\mathbf{W}}_{N}\right) \geq \left(1-\frac{1}{\sqrt{2}}\right)^{N/2-1}.
\end{equation*}
Above quantity is non-zero and always positive for all~$N$.
Thus, we can state that
\begin{equation*}
\operatorname{det}\left(\tilde{\mathbf{W}}_{N}\right) > 0,
\end{equation*}
for an arbitrary positive power-of-two integer~$N$.
Invoking Theorem~\ref{invertibility-condition-theorem}, we have that
\begin{align*}
\prod_{i = 0}^{\operatorname{log}(N)-3} \big|\operatorname{det}\left(\tilde{\mathbf{W}}_{N/2^i}\right)\big| & > 0.
\end{align*}
Thus, we have that
\begin{align*}
\prod_{i = 0}^{\operatorname{log}(N)-3} \big|\operatorname{det}\left(\tilde{\mathbf{W}}_{N/2^i}\right)\big| & \neq 0,
\end{align*}
for any positive power-of-two~$N$.
Therefore, any $N$-point DFT approximation with a non-zero precision parameter~$\alpha$ is invertible.

\end{proof}
\end{theorem}

\section{Evaluating DFT Approximations According the Total Error Energy and Orthogonality Deviation}
\label{dft-approx-evaluation}
In this section we summarize the definition of total error energy defined in~\cite{Cintra2011}.
Suppose the transfer function defined by each row of a linear transformation~$\mathbf{T}_N$ of order~$N$ defined by
\begin{equation*}
\mathbf{H}(\omega,\mathbf{T}_N) =
\left[
\begin{array}{c}
H_0(\omega,\mathbf{T}_N)\\ H_1(\omega,\mathbf{T}_N)\\ H_2(\omega,\mathbf{T}_N)\\ \vdots \\ H_{N-1}(\omega,\mathbf{T}_N)
\end{array}
\right]
 = \mathbf{T}_N\cdot \pmb{\omega},
\end{equation*}
where~$\omega \in [-\pi, \pi]$ and
\begin{equation*}
\pmb{\omega} = \left[
\begin{array}{c}
1\\ e^{-j\omega}\\ e^{-j2\omega}\\ \vdots \\ e^{-j(N-1)\omega}
\end{array}
\right].
\end{equation*}

A way to evaluate an approximation to the exact DFT is by the squared difference between transfer functions of exact and approximated DFT.
Thus, we define the error related quantity
\begin{equation*}
D_i(\omega,\tilde{\mathbf{F}}_N) = | H_i(\omega,\mathbf{F}_N)- H_i(\omega,\tilde{\mathbf{F}}_N)|^2,
\end{equation*}
where~$i = 0, 1, \ldots, N-1$ and~$D_i(\omega,\tilde{\mathbf{F}}_N)$ represents the difference between transfer functions in each $i$th row. Each row in the DFT approximation contributes for a departure from the exact DFT. It is measured by the total error energy
\begin{equation*}
\epsilon_i(\tilde{\mathbf{F}}_N) = \int_{-\pi}^{\pi}D_i(\omega,\tilde{\mathbf{F}}_N)d \omega.
\end{equation*}

Note that the definition of total error energy is a different definition of that given in~\cite{Suarez2014}.
Here we does not consider the normalizing factor as defined in~\cite{Suarez2014}.
Also, the integral interval for obtaining~$D_i(\omega,\tilde{\mathbf{F}}_N)$ is slightly different.

Since the the exact DFT is a orthogonal transform, an way to evaluate the proposed DFT approximations is by measuring the deviation from orthogonality.
Due the complex nature of matrices entries in the proposed DFT approximations, we use the orthogonality definition in~\cite{Britanak2007}.
If a square complex matrix~$\mathbf{M}$ is orthogonal, then~$\mathbf{M}\mathbf{M}^\mathsf{H}$ returns a diagonal matrix, where~$\mathbf{M}^\mathsf{H}$ is the Hermitian conjugate matrix associated with~$\mathbf{M}$~\cite{Britanak2007}.
Thus, we define the measure of deviation of orthogonality by
\begin{equation*}
\delta(\mathbf{M}) = 1-\frac{\|\operatorname{diag}(\mathbf{M}\cdot \mathbf{M}^{\mathsf{H}})\|_{\mathsf{F}}^2}{\|\mathbf{M}\cdot \mathbf{M}^{\mathsf{H}}\|_{\mathsf{F}}^2}.
\end{equation*}
As a threshold for near-orthogonal approximations we consider the value assumed to DCT approximations of maximum deviation from orthogonality of~$0.20$~\cite{Cintra2014, Haweel2001, Potluri2012}.

Several DFT approximations for different lengths and precision parameters ranging from~$\alpha = 2$ to~$\alpha = 8$ are shown in Tables~\ref{measuresALPHA2},~\ref{measuresALPHA4},~\ref{measuresALPHA8}, and~\ref{measuresALPHA16}.
Also, it is shown the deviation from orthogonality for each DFT approximation.
Note that all approximations possess a very low deviation from orthogonality.

The total error measures for the considered DFT approximations are decreasing with the precision parameter~$\alpha$.
It can be seen across the tables that for a fixed length~$N$, the total error energy reduces as the precision parameter increases.
It is in accordance to what is expected since larger the precision parameter, better is the approximation.
Thus, for high values of precision parameter~$\alpha$ the total error energy is essentially null.
The same behavior is observed for the orthogonality deviation measure.
For a fixed length~$N$, the orthogonality deviation decreases as~$\alpha$ increases.

\begin{table}
\centering
\caption{Total error energy for DFT approximations for~\mbox{$\alpha = 2$}}
\label{measuresALPHA2}
\begin{tabular}[c]{c@{\quad}c@{\quad}c@{\quad}}
\toprule
$N$&$\sum_{i=0}^{N-1} \epsilon_i(\tilde{\mathbf{F}}_N)$& $\delta(\tilde{\mathbf{F}}_N)$\\
\midrule
$4$ &$0$ &$0$ \\\midrule
$8$ &$4.86\cdot 10^{-1}$ &$3.85\cdot 10^{-2}$ \\\midrule
$16$ &$5.08\cdot 10^{-1}$ &$1.48\cdot 10^{-2}$ \\\midrule
$32$ &$6.49\cdot 10^{-1}$ &$2.12\cdot 10^{-2}$ \\\midrule
$64$ &$7.08\cdot 10^{-1}$ &$5.85\cdot 10^{-2}$ \\\midrule
$128$ &$7.58\cdot 10^{-1}$ &$8.04\cdot 10^{-2}$ \\\midrule
$256$ &$7.89\cdot 10^{-1}$ &$9.98\cdot 10^{-2}$ \\\midrule
$512$ &$8.12\cdot 10^{-1}$ &$1.14\cdot 10^{-1}$ \\\midrule
$1024$ &$8.23\cdot 10^{-1}$ &$1.28\cdot 10^{-1}$ \\\bottomrule
\end{tabular}
\end{table}

\begin{table}
\centering
\caption{Total error energy for DFT approximations for~\mbox{$\alpha = 4$}}
\label{measuresALPHA4}
\begin{tabular}[c]{c@{\quad}c@{\quad}c@{\quad}}
\toprule
$N$&$\sum_{i=0}^{N-1} \epsilon_i(\tilde{\mathbf{F}}_N)$& $\delta(\tilde{\mathbf{F}}_N)$\\
\midrule
$4$ &$0$ &$0$ \\\midrule
$8$ &$1.01\cdot 10^{-1}$ &$1.83\cdot 10^{-3}$ \\\midrule
$16$ &$3.00\cdot 10^{-1}$ &$7.36\cdot 10^{-3}$ \\\midrule
$32$ &$3.75\cdot 10^{-1}$ &$5.56\cdot 10^{-3}$ \\\midrule
$64$ &$3.81\cdot 10^{-1}$ &$3.93\cdot 10^{-4}$ \\\midrule
$128$ &$3.93\cdot 10^{-1}$ &$5.47\cdot 10^{-3}$ \\\midrule
$256$ &$4.03\cdot 10^{-1}$ &$1.01\cdot 10^{-2}$ \\\midrule
$512$ &$4.07\cdot 10^{-1}$ &$1.47\cdot 10^{-2}$ \\\midrule
$1024$ &$4.12\cdot 10^{-1}$ &$1.93\cdot 10^{-2}$ \\\bottomrule
\end{tabular}
\end{table}

\begin{table}
\centering
\caption{Total error energy for DFT approximations for~\mbox{$\alpha = 8$}}
\label{measuresALPHA8}
\begin{tabular}[c]{c@{\quad}c@{\quad}c@{\quad}}
\toprule
$N$&$\sum_{i=0}^{N-1} \epsilon_i(\tilde{\mathbf{F}}_N)$& $\delta(\tilde{\mathbf{F}}_N)$\\
\midrule
$4$ &$0$ &$0$ \\\midrule
$8$ &$1.01\cdot 10^{-1}$ &$1.83\cdot 10^{-3}$ \\\midrule
$16$ &$3\cdot 10^{-1}$ &$7.36\cdot 10^{-3}$ \\\midrule
$32$ &$3.75\cdot 10^{-1}$ &$5.56\cdot 10^{-3}$ \\\midrule
$64$ &$3.81\cdot 10^{-1}$ &$3.93\cdot 10^{-4}$ \\\midrule
$128$ &$3.93\cdot 10^{-1}$ &$5.47\cdot 10^{-3}$ \\\midrule
$256$ &$4.03\cdot 10^{-1}$ &$1.01\cdot 10^{-2}$ \\\midrule
$512$ &$4.07\cdot 10^{-1}$ &$1.47\cdot 10^{-2}$ \\\midrule
$1024$ &$4.12\cdot 10^{-1}$ &$1.93\cdot 10^{-2}$ \\ \bottomrule
\end{tabular}
\end{table}

\begin{table}
\centering
\caption{Total error energy for DFT approximations for~\mbox{$\alpha = 16$}}
\label{measuresALPHA16}
\begin{tabular}[c]{c@{\quad}c@{\quad}c@{\quad}}
\toprule
$N$&$\sum_{i=0}^{N-1} \epsilon_i(\tilde{\mathbf{F}}_N)$& $\delta(\tilde{\mathbf{F}}_N)$\\
\midrule
$4$ &$0$ &$0$ \\\midrule
$8$ &$4.60\cdot 10^{-2}$ &$3.84\cdot 10^{-4}$ \\\midrule
$16$ &$4.98\cdot 10^{-2}$ &$2.32\cdot 10^{-4}$ \\\midrule
$32$ &$5.74\cdot 10^{-2}$ &$2.41\cdot 10^{-5}$ \\\midrule
$64$ &$6.18\cdot 10^{-2}$ &$2.02\cdot 10^{-4}$ \\\midrule
$128$ &$6.74\cdot 10^{-2}$ &$3.75\cdot 10^{-4}$ \\\midrule
$256$ &$7.57\cdot 10^{-2}$ &$5.46\cdot 10^{-4}$ \\\midrule
$512$ &$8.18\cdot 10^{-2}$ &$7.98\cdot 10^{-4}$ \\\midrule
$1024$ &$8.64\cdot 10^{-2}$ &$1.10\cdot 10^{-3}$ \\\bottomrule
\end{tabular}
\end{table}

\section{Arithmetic Complexity}
\label{complexity}
In this subsection, we evaluate the arithmetic complexity associated with the proposed DFT approximation.
The number of complex additions due to the proposed transform is inherited from radix-2 Cooley-Tukey algorithm for exact DFT.
The direct method for evaluating a complex multiplication is employing~$4$ real multiplications and $2$ real additions~\cite{Oppenheim2009}.
However, usual methods for fast computation recast usual complex product complexity into~$3$ real multiplication and~$3$ real additions~\cite{Blahut2010}.
Since real multiplication demands much more hardware resources than real addition during calculation in a digital computer, above modification in order to reduce the overall number of real multiplication is largely employed.
In the proposed method, the multiplications required by the exact DFT computation are replaced by~\emph{small} integers multiplications.
Since the product by small integers can be efficiently implemented with only bit-shifting operations and additions~\cite{Cintra2014}, the multiplications required by exact DFT are thus replaced by bit-shifting operations and additions.
That is why we does not employ the aforementioned recast in order to reduce the overall number of real multiplications in a complex multiplication.
Thus, the arithmetic complexity in a DFT approximation is highly dependent by the quantity of complex additions as required by Cooley-Tukey radix-2 decimation-in-time algorithm.
Smaller the value of employed precision parameter~$\alpha$, smaller the resulting integers used for approximating each twiddle factor.
In consequence, less additions are required.
In particular, if one uses~$\alpha = 2$ in producing DFT approximations by the method proposed here, there is no increasing in the number of real additions.
This is due the fact that possible outcomes of mapping introduced in~\eqref{matrixmapping-dft} are~$\pm p/2 \pm q/2j$, where~$p,q = 0, 1, 2$.
The product between any quantity and~$p$ or~$q$ is just a trivial multiplication.
Thus,~\emph{any}~$N$-point DFT approximation with precision parameter~$\alpha = 2$ does not require further addition than that required by the original Cooley-Tukey radix-2 decimation-in-time fast algorithm.
Therefore, the arithmetic complexity associated with an arbitrary~$N$-point DFT approximation with precision parameter~$\alpha = 2$ is~$A_c(N) = N\operatorname{log}_2(N)$ complex additions and~$0$ complex multiplications.
The same happens to DFT approximations generated by precision parameter~$\alpha = 1$.
The real and imaginary parts of approximate twiddle factor are either~$0, \pm1$.
Therefore, an arbitrary~$N$-point DFT with precision parameter~$\alpha = 1$ requires only~$A_c(N) = N\operatorname{log}_2(N)$.
Evaluate the arithmetic complexity for higher precision parameters requires further investigation.

\section{Computational Evaluation of DFT Approximations Error}
\label{dft-approx-err-ana}
The $N$-point DFT approximations in the introduced class is asymptotically convergent to the exact $N$-point DFT as shown in Theorem~\ref{DFTAN-conv}.
Since digital computers are able to represent only finite quantities, the situation described in Theorem~\ref{DFTAN-conv} is  hypothetical, where the precision parameter~$\alpha$ is arbitrarily increased.

As consequence of cited limitations on digital computers, we aim in this section to investigate the error due the approximation procedure described in this paper.
First of all, let us define the approximate twiddle factor error matrix as
\begin{equation*}
\Delta \tilde{\mathbf{W}}_{N} \triangleq \mathbf{W}_{N}- \tilde{\mathbf{W}}_{N}
\end{equation*}
and approximate DFT error matrix as
\begin{equation*}
\Delta \tilde{\mathbf{F}}_{N} \triangleq \mathbf{F}_{N}- \tilde{\mathbf{F}}_{N}.
\end{equation*}

Applying above definitions for the approximate $N$-point DFT in Definition~\ref{DFT-approx-def}, we obtain
\begin{align*}
\tilde{\mathbf{F}}_{N} = & \mathbf{A}_N \left(\mathbf{W}_{N}-\Delta\tilde{\mathbf{W}}_{N}\right)
\left[ \mathbf{I}_2 \otimes \left(\mathbf{F}_{N/2}-\Delta\tilde{\mathbf{F}}_{N/2}\right) \right]
\mathbf{B}_{N},
\end{align*}
what yields
\begin{align*}
\Delta\tilde{\mathbf{F}}_{N} = & \mathbf{A}_N \left[ \mathbf{W}_{N}(\mathbf{I}_2 \otimes \Delta\tilde{\mathbf{F}}_{N/2}) + \Delta\tilde{\mathbf{W}}_{N} (\mathbf{I}_2 \otimes \Delta\mathbf{F}_{N/2}) - \Delta\tilde{\mathbf{W}}_{N}(\mathbf{I}_2 \otimes \Delta\tilde{\mathbf{F}}_{N/2})\right] \mathbf{B}_{N}.
\end{align*}

If we open above expression in the format of Kronecker product, we would obtain the following
\begin{equation*}
\Delta\tilde{\mathbf{F}}_{N} = \mathbf{A}_N
\left[
\begin{rsmallmatrix}
\Delta\tilde{\mathbf{F}}_{N/2} & \\
& \left[\mathbf{W}_{N}\right]_\mathrm{L}\Delta \tilde{\mathbf{F}}_{N/2}+ \left[\Delta\tilde{\mathbf{W}}_{N}\right]_\mathrm{L}\mathbf{F}_{N/2}-\left[\Delta \tilde{\mathbf{W}}_{N}\right]_\mathrm{L}\Delta \tilde{\mathbf{F}}_{N/2}
\end{rsmallmatrix}
\right]\mathbf{B}_{N},
\end{equation*}
where the operator~$\left[\cdot \right]_\mathrm{L}$ returns the low half  diagonal block of its matrix argument.
Using Appendix~\ref{prod-frob}, we have that
\begin{equation}
\label{FN-general-recursion}
\|\Delta\tilde{\mathbf{F}}_{N}\|_\mathrm{F}^2 =  2\|\Delta\tilde{\mathbf{F}}_{N/2}\|_\mathrm{F}^2+2\|\left[\mathbf{W}_{N}\right]_\mathrm{L}\Delta \tilde{\mathbf{F}}_{N/2}+ \left[\Delta\tilde{\mathbf{W}}_{N}\right]_\mathrm{L}\mathbf{F}_{N/2}-\left[\Delta \tilde{\mathbf{W}}_{N}\right]_\mathrm{L}\Delta \tilde{\mathbf{F}}_{N/2}\|_\mathrm{F}^2.
\end{equation}
Note that the elements of~$\left[\mathbf{W}_{N}\right]_\mathrm{L}$ are all complex number with unity norm.
Thus, the following identity is verified~$\|\left[\mathbf{W}_{N}\right]_\mathrm{L}\Delta \tilde{\mathbf{F}}_{N/2}\|_\mathrm{F} = \|\Delta \tilde{\mathbf{F}}_{N/2}\|$.
Concerns remains on the Frobenius norm of the products~$\left[\Delta\tilde{\mathbf{W}}_{N}\right]_\mathrm{L}\mathbf{F}_{N/2}$ and~$\left[\Delta \tilde{\mathbf{W}}_{N}\right]_\mathrm{L}\Delta \tilde{\mathbf{F}}_{N/2}$.
In the following, we provide three different error analysis based on the equation in~\eqref{FN-general-recursion}.

\subsection{Error Analysis I}
\label{sec:ERR_ANA_I}
Due to the difficult for evaluating~$\left[\Delta\tilde{\mathbf{W}}_{N}\right]_\mathrm{L}\mathbf{F}_{N/2}$ and~$\left[\Delta \tilde{\mathbf{W}}_{N}\right]_\mathrm{L}\Delta \tilde{\mathbf{F}}_{N/2}$ we need to provide an alternative or approximation for these products.
The Frobenius norm of~$\left[\Delta \tilde{\mathbf{W}}_{N}\right]_\mathrm{L}$ is dependent on the precision parameter~$\alpha$ and the DFT length~$N$, i.e.,~$\|\left[\Delta \tilde{\mathbf{W}}_{N}\right]_\mathrm{L}\|_\mathrm{F} = \epsilon(\alpha, N)$.
In order to approximately quantify the error due the introduced  approximation method, we resort to assume that
\begin{equation*}
\epsilon(\alpha, N) = \frac{1}{\sqrt{2}\alpha}\frac{1}{N/2}.
\end{equation*}
Note that~$1/\sqrt{2}\alpha$ is the upper bound for the error due the scaled rounding operator to twiddle factors as in~\eqref{z-error-mag-bound}.
Since the lower half block matrix~$\left[\Delta \tilde{\mathbf{W}}_{N}\right]_\mathrm{L}$ possess exactly~$N/2$ elements, we scale the approximate twiddle factor upper bound, thus  representing the amount of error that each twiddle factor contribute to the considered approximated $N$-point DFT.
Thus, we have that
\begin{align*}
\|\Delta\tilde{\mathbf{F}}_{N}\|_\mathrm{F}^2 \approx  &  2\|\Delta\tilde{\mathbf{F}}_{N/2}\|_\mathrm{F}^2+ 2\left[ \left( \frac{N}{2}+\|\Delta \tilde{\mathbf{F}}_{N/2}\|_\mathrm{F}\right)\epsilon(\alpha, N) + \|\Delta \tilde{\mathbf{F}}_{N/2}\|_\mathrm{F}\right]^2\\
\approx & 2\|\Delta\tilde{\mathbf{F}}_{N/2}\|_\mathrm{F}^2+ 2\left[ \left( \frac{N}{2}+\|\Delta \tilde{\mathbf{F}}_{N/2}\|_\mathrm{F}\right)\frac{\sqrt{2}}{\alpha N} + \|\Delta \tilde{\mathbf{F}}_{N/2}\|_\mathrm{F}\right]^2.
\end{align*}
Above approximation furnish a way to estimate the $N$-point DFT approximation error in terms of previously computed $N/2$-point DFT approximation error.
 Since absolute error sometimes cannot indicate what is happening in background, we resort to relative error that is~$\|\Delta\tilde{\mathbf{F}}_N\|_\text{rel} = \|\Delta\tilde{\mathbf{F}}_N\|_\mathrm{F}/\|\mathbf{F}_N\|_\mathrm{F}$.
Using that~$\|\mathbf{F}_N\|_\mathrm{F} = N$, we have that
\begin{align*}
\|\Delta\tilde{\mathbf{F}}_{N}\|_\text{rel}^2 \approx  &  2\|\Delta\tilde{\mathbf{F}}_{N/2}\|_\text{rel}^2+ 2\left[ \left( \frac{1}{2}+\|\Delta \tilde{\mathbf{F}}_{N/2}\|_\text{rel}\right)\frac{\sqrt{2}}{\alpha N} + \|\Delta \tilde{\mathbf{F}}_{N/2}\|_\text{rel}\right]^2.
\end{align*}

\begin{figure*}
\centering
\psfrag{ErrorFN}{$\|\Delta \tilde{\mathbf{F}}_{N}\|_\mathrm{F}$}
\psfrag{logN}{$\operatorname{log}(N)$}
\subfigure[Precision parameter~$\alpha = 2$.]{\includegraphics{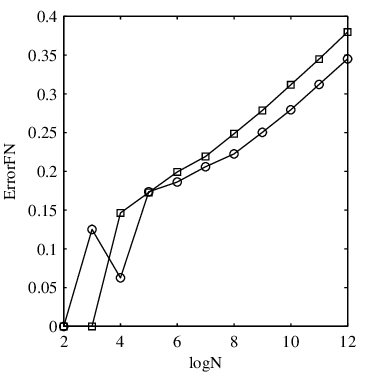}}
\subfigure[Precision parameter~$\alpha = 4$.]{\includegraphics{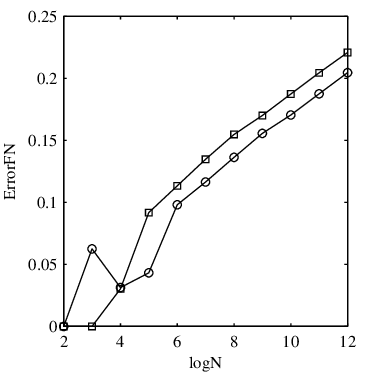}}\\
\subfigure[Precision parameter~$\alpha = 8$.]{\includegraphics{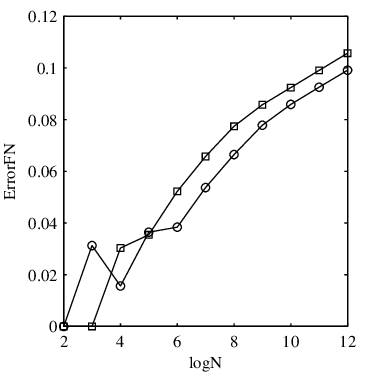}}
\subfigure[Precision parameter~$\alpha = 16$.]{\includegraphics{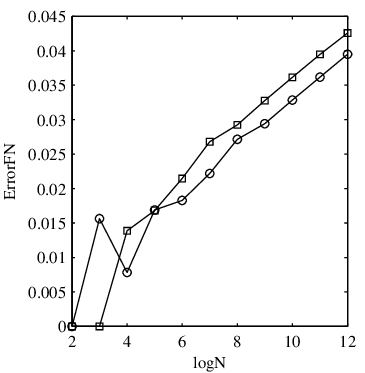}}
\caption{Curves for estimated and computed  relative  $N$-point DFT approximation error in terms of Frobenius norm for precision parameters~$\alpha = 2, 4, 8$ and~$16$.
Circle mark represents estimated  relative  $N$-point DFT approximation error and squared mark represents computed  relative  $N$-point DFT approximation error.}
\label{fig:DFT-error-study}
\end{figure*}

Figure~\ref{fig:DFT-error-study} shows curves for estimated  $N$-point DFT approximation  relative  error from above approximation and computed error, where the precision parameter is~$\alpha = 8$.
The computed error was obtained by direct computation of Frobenius norm for each $N$-point DFT approximation.

Note that errors plot presented in Figure~\ref{fig:DFT-error-study} are lower for greater precision parameters~$\alpha$.
In particular, the approximation errors for~$\alpha = 16$ are lower when compared to the others small precision parameters~$\alpha = 2, 4$ and~$8$.
It is in accordance to what was stated through this text that greater the precision parameter~$\alpha$, better the is the considered DFT approximation.

\subsection{Error Analysis II}
\label{sec:ERR_ANA_II}
Another way to estimate the DFT approximation error is by making use of series expansion of approximate twiddle factors.
In particular, a useful series expansion is the Fourier series~\cite{Folland2000}.
Fourier series expansion can furnish good approximate twiddle factor  representation.
In particular, using Fourier series we can~\emph{nearly} represent the approximate twiddle factor with good precision with few harmonics.
It is due the fact that Fourier series coefficients decay with~$1/n$, where~$n$ is the harmonic order~\cite{Papoulis1997}.
Doing so, we represent the approximate twiddle factor by its first harmonics in real and imaginary parts.

Appendix~\ref{sf-round} develop the Fourier series coefficients for the approximate twiddle factor.
The expressions for~$a_n$ and~$b_n$, respectively, are given in~\eqref{sf-an} and~\eqref{sf-bn}.
For the first harmonics we have
\begin{align*}
a_1 = & b_1 = \frac{4}{\pi\alpha}  \sum_{i=1}^{\alpha} T_1\left(\sqrt{1-\left(\frac{2i-1}{2\alpha}\right)^2}\right) .
\end{align*}
where~$T_1(\cdot)$ is the Chebyshev polynomials of first kind with degree~$1$.
Here at this point we use Chebyshev polynomials and recognize that for any real number we have that~$T_1(x)= x$.
Therefore, above expressions for~$a_1$ and~$b_1$ takes the simpler forms as
\begin{align}
\label{a1b1}
a_1=b_1= & \frac{4}{\pi\alpha}  \sum_{i=1}^{\alpha} \sqrt{1-\left(\frac{2i-1}{2\alpha}\right)^2}.
\end{align}

Note above formulation for the first harmonics for the approximate twiddle factor does not depends on the approximation length, but only on the precision parameter~$\alpha$.
For the~$N$-point DFT approximation, we can provide the following approximation to the real and imaginary parts of exact twiddle factor~$W_N^k$ for~$k = 1, 2, \ldots, N/2-1$,
\begin{equation*}
\frac{1}{\alpha}\operatorname{round}\left(\alpha \cos\left(\frac{2\pi k}{N}\right)\right) \approx  a_1 \cos\left(\frac{2\pi k}{N}\right)
\end{equation*}
and
\begin{equation*}
\frac{1}{\alpha}\operatorname{round}\left(\alpha \sin\left(\frac{2\pi k}{N}\right)\right) \approx  b_1 \sin\left(\frac{2\pi k}{N}\right).
\end{equation*}
Therefore, the scaled rounded twiddle factor can be approximated by
\begin{equation}
\label{twiddle-approx-approx}
\frac{1}{\alpha}\operatorname{round}\left(\alpha W_N^k\right) \approx  a_1 W_N^k,
\end{equation}
where we use that~$a_1 = b_1$ as in~\eqref{a1b1}.
Also, we can establish an approximation to the approximate twiddle factor error.
Above approximation yields
\begin{equation*}
W_N^k - \frac{1}{\alpha}\operatorname{round}\left(\alpha W_N^k\right) \approx  (1-a_1) W_N^k.
\end{equation*}
In consequence, we have that
\begin{equation*}
\Big|W_N^k - \frac{1}{\alpha}\operatorname{round}\left(\alpha W_N^k\right)\Big| \approx  |1-a_1|.
\end{equation*}

Using above approximation for the approximate twiddle factor, we can propose that each element of~$\left[\Delta \tilde{\mathbf{W}}_{N}\right]_\mathrm{L}$ is nearly the approximation in above equation.
Thus, we have that
\begin{equation*}
\left[\Delta \tilde{\mathbf{W}}_{N}\right]_\mathrm{L} \approx |1-a_1|\cdot \mathbf{I}_{\mathrm{N}/2},
\end{equation*}
where~$\approx$ denotes elementwise approximation.
In order to compensate the approximating error for each component of~$\left[\Delta \tilde{\mathbf{W}}_{N}\right]_\mathrm{L}$, we normalize above approximation by the factor~$N/2$.
Thus, we apply above approximation to~\eqref{FN-general-recursion} in the following manner.
Using Frobenius norm properties, we have
\begin{align*}
\|\Delta\tilde{\mathbf{F}}_{N}\|_\mathrm{F}^2 \approx & 2\|\Delta\tilde{\mathbf{F}}_{N/2}\|_\mathrm{F}^2+2\left\|\left[\mathbf{W}_{N}\right]_\mathrm{L}\Delta \tilde{\mathbf{F}}_{N/2}+ \frac{|1-a_1|}{N/2}\cdot \mathbf{I}_{\mathrm{N}/2}\mathbf{F}_{N/2}-\frac{|1-a_1|}{N/2}\cdot \mathbf{I}_{\mathrm{N}/2}\Delta \tilde{\mathbf{F}}_{N/2}\right\|_\mathrm{F}^2\\
\approx & 2\|\Delta\tilde{\mathbf{F}}_{N/2}\|_\mathrm{F}^2+2\left\|\left[\mathbf{W}_{N}\right]_\mathrm{L}\Delta \tilde{\mathbf{F}}_{N/2}+ \frac{|1-a_1|}{N/2}\mathbf{F}_{N/2}-\frac{|1-a_1|}{N/2}\Delta \tilde{\mathbf{F}}_{N/2}\right\|_\mathrm{F}^2\\
\approx & 2\|\Delta\tilde{\mathbf{F}}_{N/2}\|_\mathrm{F}^2+ 2\left[ \left( \frac{N}{2}+\|\Delta \tilde{\mathbf{F}}_{N/2}\|_\mathrm{F}\right)\frac{|1-a_1|}{N/2} + \|\Delta \tilde{\mathbf{F}}_{N/2}\|_\mathrm{F}\right]^2.
\end{align*}

Again we resort to relative error that is~$\|\Delta\tilde{\mathbf{F}}_N\|_\text{rel} = \|\Delta\tilde{\mathbf{F}}_N\|_\mathrm{F}/\|\mathbf{F}_N\|_\mathrm{F}$.
Using that~$\|\mathbf{F}_N\|_\mathrm{F} = N$, we have that
\begin{align*}
\|\Delta\tilde{\mathbf{F}}_{N}\|_\text{rel}^2 \approx  &  2\|\Delta\tilde{\mathbf{F}}_{N/2}\|_\text{rel}^2+ 2\left[ \left( \frac{1}{2}+\|\Delta \tilde{\mathbf{F}}_{N/2}\|_\text{rel}\right)\frac{|1-a_1|}{N/2} + \|\Delta \tilde{\mathbf{F}}_{N/2}\|_\text{rel}\right]^2.
\end{align*}

\begin{figure*}
\centering
\psfrag{ErrorFN}{$\|\Delta \tilde{\mathbf{F}}_{N}\|_\text{rel}$}
\psfrag{logN}{$\operatorname{log}(N)$}
\subfigure[Precision parameter~$\alpha = 2$.]{\includegraphics{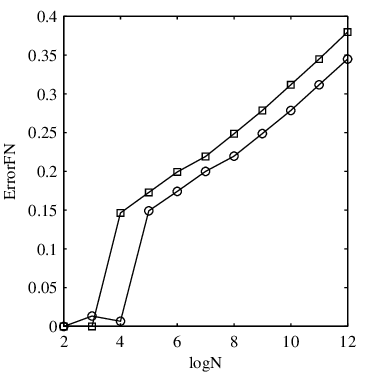}}
\subfigure[Precision parameter~$\alpha = 4$.]{\includegraphics{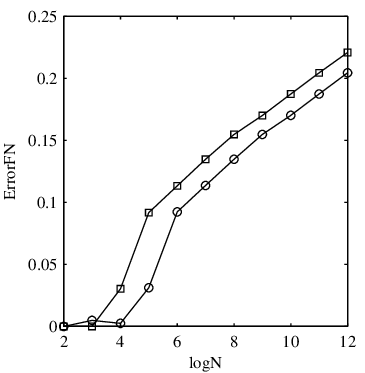}}\\
\subfigure[Precision parameter~$\alpha = 8$.]{\includegraphics{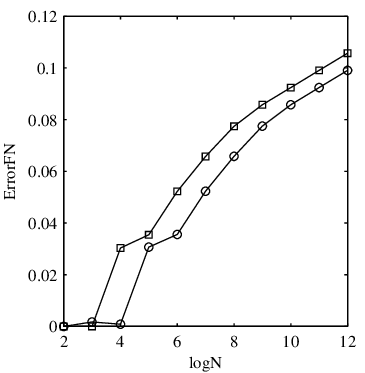}}
\subfigure[Precision parameter~$\alpha = 16$.]{\includegraphics{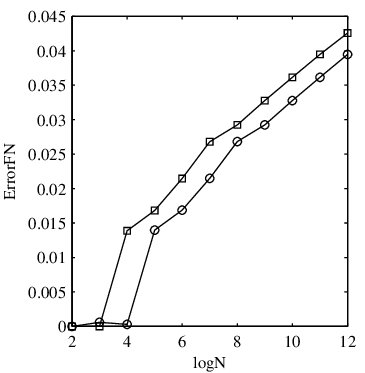}}
\caption{Curves for estimated and computed $N$-point DFT approximation  relative  error in terms of Frobenius norm for precision parameters~$\alpha = 2, 4, 8$ and~$16$.
Circle mark represents estimated $N$-point DFT approximation error and squared mark represents computed $N$-point DFT approximation  relative  error.}
\label{fig:ERR_ANA_II}
\end{figure*}

Figure~\ref{fig:ERR_ANA_II} shows the computed and estimated Frobenius norm for~$N$-point DFT approximation  relative  error matrix.
The estimate for the Frobenius norm in Figure~\ref{fig:ERR_ANA_II} is carried out based on above recursive formulation for the~$N$-point DFT approximation  relative  error matrix.
As observed in the previous error analysis in the subsection~\ref{sec:ERR_ANA_I}, the Frobenius norm for the $N$-point DFT approximation error matrix is decreasing as the precision parameter~$\alpha$ increases.

\subsection{Error Analysis III}
\label{sec:ERR_ANA_III}
Here we continue our analysis based on~\eqref{twiddle-approx-approx}, that states
\begin{equation*}
\frac{1}{\alpha}\operatorname{round}\left(\alpha W_N^k\right) \approx  a_1 W_N^k.
\end{equation*}

Now we can apply above approximation for the scaled rounded twiddle factor in DFT approximation definition.
Here we use Definition~\eqref{DFT-approx-def}.
Note that use above approximation for the scaled rounded twiddle factor is simple carried out by substituting the twiddle factor multiplied by the term~$a_1$.
Since all twiddle factors are multiplied by the term~$a_1$, the approximation corresponding to the~$N$-point DFT approximation is thus multiplied by the term~$a_1$.
This process is repeated in each step of recursive definition of~$N$-point DFT approximation.
It leads to the following result
\begin{equation}
\label{DFTA-approx}
\tilde{\mathbf{F}}_{N} \approx a_1^{\operatorname{log}_2(N/4)}\mathbf{F}_{N},
\end{equation}
where the sign~$\approx$ represents an elementwise approximation.
Thus, we obtain
\begin{align*}
\|\tilde{\mathbf{F}}_{N}\|_\mathrm{F} & \approx  a_1^{\operatorname{log}_2(N/4)}\|\mathbf{F}_{N}\|_\mathrm{F}\\
& \approx  a_1^{\operatorname{log}_2(N/4)}N.
\end{align*}
\begin{figure*}
\centering
\psfrag{FN}{$\|\tilde{\mathbf{F}}_{N}\|_\mathrm{F}$}
\psfrag{logN}{$\operatorname{log}(N)$}
\subfigure[Precision parameter~$\alpha = 1$.]{\label{fig:APPROX-FRO-ALPHA1}\includegraphics{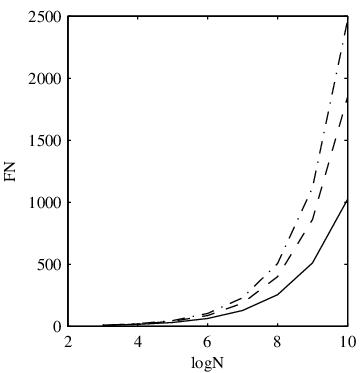}}
\subfigure[Precision parameter~$\alpha = 2$.]{\label{fig:APPROX-FRO-ALPHA2}\includegraphics{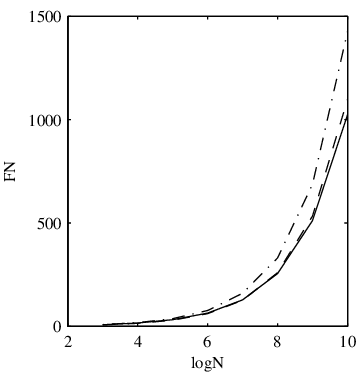}}\\
\subfigure[Precision parameter~$\alpha = 4$.]{\label{fig:APPROX-FRO-ALPHA4}\includegraphics{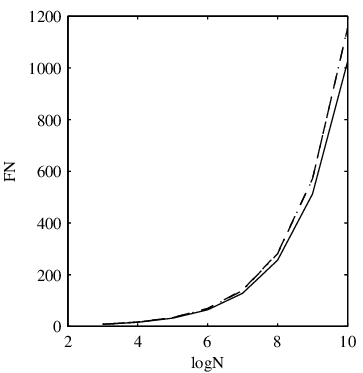}}
\subfigure[Precision parameter~$\alpha = 8$.]{\label{fig:APPROX-FRO-ALPHA8}\includegraphics{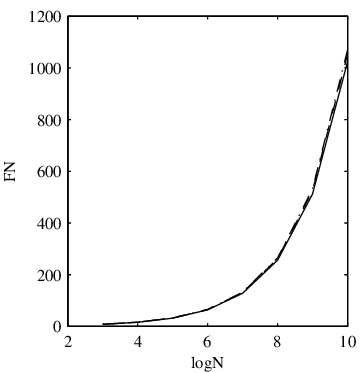}}
\caption{Curves for Frobenius norm of exact and approximated $N$-point DFT and its Frobenius norm approximation.
Solid lines represents the curve for  exact $N$-point DFT Frobenius norm, dashed lines represents the Frobenius norm for the $N$-point DFT approximation and dashed-dotted lines represents the approximating curve for Frobenius norm of $N$-point DFT approximation.}
\label{fig:APPROX-FRO-ALPHA}
\end{figure*}

Figure~\ref{fig:APPROX-FRO-ALPHA} shows the plot for Frobenius norm of approximated~$N$-point DFT, the exact~$N$-point DFT and the above approximation for the~$N$-point DFT approximation.
The behavior of Frobenius norm for approximate~$N$-point DFT (dashed-dotted lines) is very close to the curve~$a_1^{\operatorname{log}_2(N/4)}N$ (dashed lines).
For small precision parameters, the curve~$a_1^{\operatorname{log}_2(N/4)}N$ representing the approximation for Frobenius norm of~$N$-point DFT possess a small deviation from the Frobenius norm of exact~$N$-point DFT, that is~$N$ (solid lines).
As much as precision parameter is increased, the curve~$a_1^{\operatorname{log}_2(N/4)}N$ gets close to the Frobenius norm of exact~$N$-point DFT.
In such a situation, the curve~$a_1^{\operatorname{log}_2(N/4)}N$ tends to the linear curve~$N$.
This is due to the fact that  as much as~$\alpha$ increases, the first harmonic for the scaled rounded twiddle factor decreases, thus approximating the unity.
This indicates that $N$-point DFT approximation better approximates the exact~$N$-point DFT as the precision parameter~$\alpha$ increases.

Also, as the precision parameter~$\alpha$ is increased, the curve for the Frobenius norm of $N$-point DFT approximation gets close to the exact $N$-point DFT.
This reinforce the fact that~$N$-point DFT approximation tends to the exact~$N$-point DFT.
In Figure~\ref{fig:APPROX-FRO-ALPHA4}, the curves for Frobenius norm of exact~$N$-point DFT and curve~$a_1^{\operatorname{log}_2(N/4)}N$ are so close that they are indistinguishable.
It is more evident in  Figure~\ref{fig:APPROX-FRO-ALPHA8}, where Frobenius norm for exact~$N$-point DFT, approximated $N$-point DFT Frobenius norm and curve~$a_1^{\operatorname{log}_2(N/4)}N$ are much more closer.

Invoking the result in Appendix~\ref{a1-prop}, we use~\eqref{lim-a1}.
Since~$a_1^{\operatorname{log}_2(N/4)}$ is a continuous function of~$a_1$, we have that
\begin{align*}
\lim_{\alpha \to \infty} \|\tilde{\mathbf{F}}_{N}\|_\mathrm{F} & \approx \lim_{\alpha \to \infty} a_1^{\operatorname{log}_2(N/4)}N\\
& \approx \left(\lim_{\alpha \to \infty} a_1\right)^{\operatorname{log}_2(N/2)}N\\
& \approx N.
\end{align*}
Therefore, above approximation is an algebraic explanation for the behavior in Figure~\ref{fig:APPROX-FRO-ALPHA}, where curves for the Frobenius norm of~$N$-point DFT approximation gets closer to the Frobenius norm of exact~$N$-point DFT.

Besides that, we can use~\eqref{DFTA-approx} in order to build an estimate for the Frobenius norm of~$\Delta\tilde{\mathbf{F}}_{N}$.
We plug~\eqref{DFTA-approx} into the definition of~$\Delta\tilde{\mathbf{F}}_{N}$, what yields
\begin{align}
\label{DFTA-approx-error}
\Delta\tilde{\mathbf{F}}_{N} &\approx  \mathbf{F}_{N} - a_1^{\operatorname{log}_2(N/4)}\mathbf{F}_{N}\nonumber\\
&\approx  \left(1- a_1^{\operatorname{log}_2(N/4)}\right)\mathbf{F}_{N}.
\end{align}
Note that above approximation for the approximate DFT error matrix is reasonable, since its Frobenius norm tends to zero as demonstrated in Theorem~\ref{DFTAN-conv} as~$\alpha$ goes to infinity.
The Frobenius norm properties allows us to state that
\begin{align*}
\|\Delta\tilde{\mathbf{F}}_{N}\|_\mathrm{F} & \approx \left|1- a_1^{\operatorname{log}_2(N/4)}\right|N.
\end{align*}
Here we take the limit for~$\alpha \to \infty$ and use the fact that modular~$|\cdot|$ function is continuous in~$1$, what yields
\begin{align*}
\lim_{\alpha \to \infty} \|\Delta\tilde{\mathbf{F}}_{N}\|_\mathrm{F} & \approx \left|1- \left(\lim_{\alpha \to \infty} a_1\right)^{\operatorname{log}_2(N/2)}\right|N.
\end{align*}
Therefore, invoking again~\eqref{lim-a1}, we obtain
\begin{align*}
\lim_{\alpha \to \infty} \|\Delta\tilde{\mathbf{F}}_{N}\|_\mathrm{F} & \approx  0.
\end{align*}

\begin{figure*}
\centering
\psfrag{FN}{$\|\Delta \tilde{\mathbf{F}}_{N}\|_\text{rel}$}
\psfrag{logN}{$\operatorname{log}(N)$}
\subfigure[Precision parameter~$\alpha = 1$.]{\label{fig:DFTAN-APPROX-FRO-ALPHA1}\includegraphics{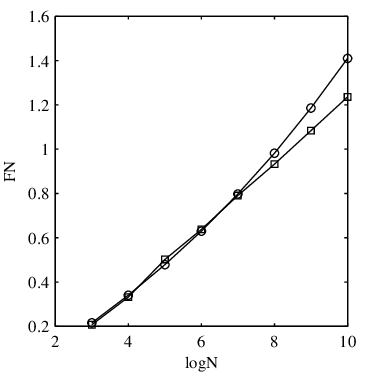}}
\subfigure[Precision parameter~$\alpha = 2$.]{\label{fig:DFTAN-APPROX-FRO-ALPHA2}\includegraphics{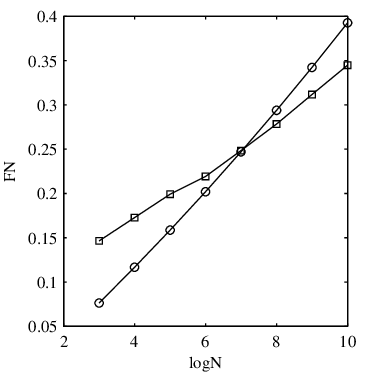}}\\
\subfigure[Precision parameter~$\alpha = 4$.]{\label{fig:DFTAN-APPROX-FRO-ALPHA4}\includegraphics{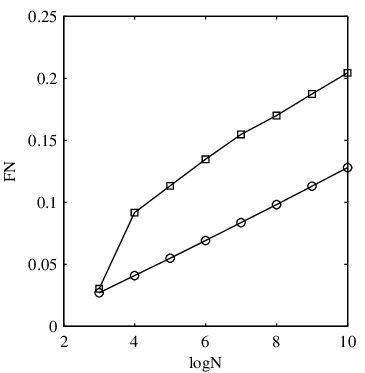}}
\subfigure[Precision parameter~$\alpha = 8$.]{\label{fig:DFTAN-APPROX-FRO-ALPHA8}\includegraphics{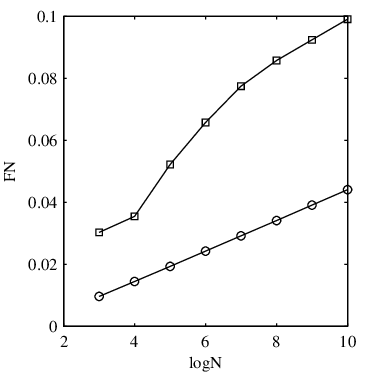}}
\caption{
Solid lines represents the curve for  exact $N$-point DFT Frobenius norm, dashed lines represents the estimated Frobenius norm for the $N$-point DFT approximation error matrix and dashed-dotted lines represents the curve for Frobenius norm of computed $N$-point DFT approximation error matrix.}
\label{fig:DFTAN-APPROX-FRO-ALPHA}
\end{figure*}

If we consider relative error and use that~$\|\mathbf{F}_N\|_\mathrm{F} = N$, we have that
\begin{align*}
\|\Delta\tilde{\mathbf{F}}_{N}\|_\text{rel} & \approx \left|1- a_1^{\operatorname{log}_2(N/4)}\right|.
\end{align*}

Figure~\ref{fig:DFTAN-APPROX-FRO-ALPHA} shows the Frobenius norm for the computed $N$-point DFT approximation  relative  error matrix and the estimated $N$-point DFT approximation  relative  error matrix obtained by~\eqref{DFTA-approx-error}.
Note that both computed and estimated Frobenius based on~\eqref{DFTA-approx-error} decreases as the precision parameter~$\alpha$ increases.
It is in accordance with above algebraic result for the Frobenius norm of $N$-point DFT approximation error matrix that tends to zero as~$\alpha$ increases.
Also, it is supported by the result in Theorem~\ref{DFTAN-conv}.

However, note that as much as the precision parameter~$\alpha$ increases, the two curves for computed and estimated relative error does not match anymore.
It shows that above formalism for estimating the DFT approximation error needs more careful examination.

\section{Architecture Representation}
\label{dft-approx-arch}
\begin{figure*}
\centering
\input{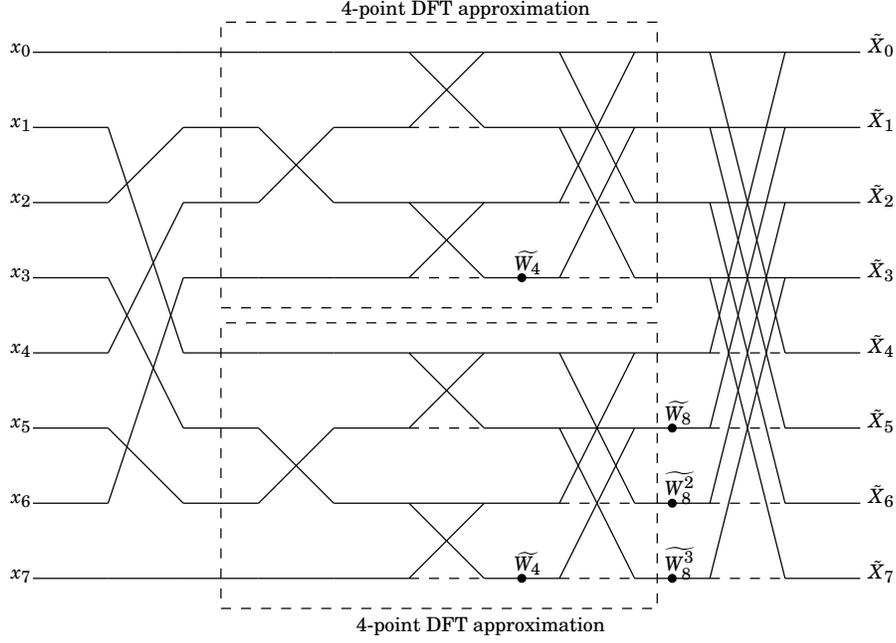}\\
\caption{SFG for~$8$-point DFT approximation for an arbitrary precision parameter~$\alpha$.
Dashed lines represents multiplication by~$-1$.}
\label{fig:FFTA8-SFG}
\end{figure*}

The introduced class of DFT approximations can be given familiar graphical interpretation.
Since it is built from the DIT  version of Cooley-Tukey radix-2 algorithm, it inherent the same structure.
For the specific case of~$N = 8$, Figure~\ref{fig:FFTA8-SFG} represents the signal flow graph (SFG) for the $8$-point DFT approximation for an arbitrary precision parameter~$\alpha$.
Approximate twiddle factors~$\widetilde{W_8}, \widetilde{W_8^2}$, and~$\widetilde{W_8^3}$ are obtained as defined in~\eqref{matrixmapping-dft}.
For this particular case where~$\alpha = 2$, we have that:
\begin{align*}
\widetilde{W_8} & = \frac{1}{2}-\frac{1}{2}j,\\
\widetilde{W^2_8} & = -j,\\
\widetilde{W^3_8} & = -\frac{1}{2}-\frac{1}{2}j.
\end{align*}
Note that we does not state~$\widetilde{W_4} = W_4 = -j$.
Using the complexity evaluation carried on the Section~\ref{complexity}, we have that~$8$-point DFT approximation built with a precision parameter~$\alpha = 2$ require only~$A_c(8) = 24$ complex additions for complex input.
In order to evaluate the number of real additions, note that for an arbitrary complex number~$a+bj$, we have that
\begin{equation*}
(a+bj)\cdot \frac{(\pm1-j)}{2} = \frac{(\pm a+b)}{2}+\frac{(\pm b-a)}{2}j.
\end{equation*}
Thus, a product of a complex number by~$\widetilde{W_8}$ or~$\widetilde{W^3_8}$ requires two real additions and two bit-shifting operations.
Therefore, we count~$2\cdot 24+4=52$ real additions and~$4$ bit-shifting operations for the complex input case.
After computational verification, we can note that the $8$-point DFT approximation proposed in~\cite{Suarez2014} is a particular case of class of DFT approximation proposed in this paper for precision parameter~$\alpha = 2$, as example furnished above.

\begin{figure*}
\centering
\input{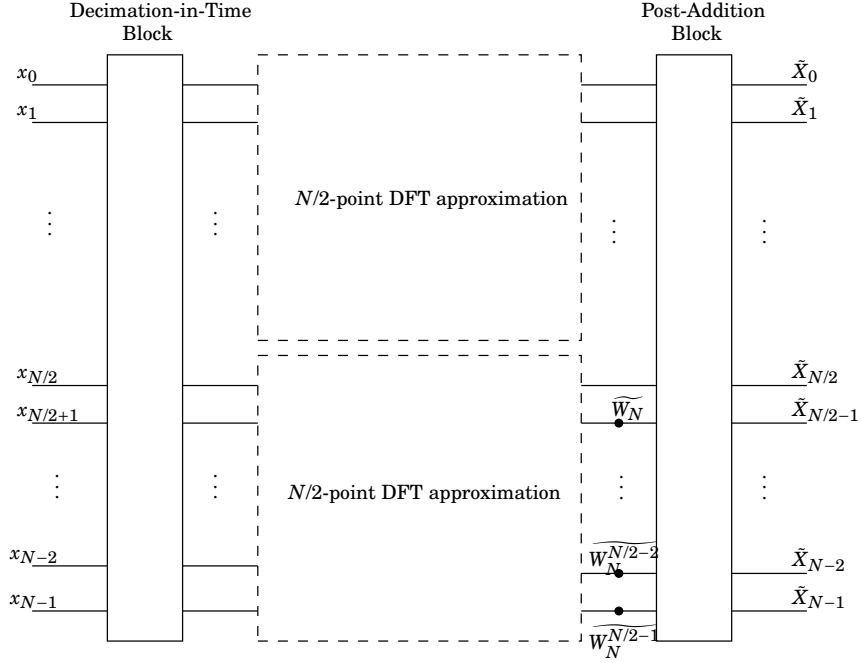}\\
\caption{SFG for an arbitrary~$N$-point DFT approximation for an arbitrary precision parameter~$\alpha$.}
\label{fig:FFTAN-SFG}
\end{figure*}

Figure~\ref{fig:FFTAN-SFG} shows the general case for an arbitrary~$N$-point DFT approximation architecture.
Also, approximate twiddle factors~$\widetilde{W_N^{k}}$ for~$k = 0, 1, \ldots, N/2-1$ are obtained by~\eqref{matrixmapping-dft}.
In the matrix factorization in Definition~\ref{DFT-approx-def}, the decimation-in-time block is mathematically represented by the matrix~$\mathbf{B}_{N}$, as the post addition block is represented by~$\mathbf{A}_N$.

\section{Application in Multi-Beam Forming}
\label{dft-approx-application-beamform}
In this section we furnish an example of application for the DFT approximations proposed in this paper.
The application is made in the field of multi-beam antennas.
We provide the pattern arrays for various lengths for DFT approximations.
All approximations  in this section furnish approximate twiddle factors with real and imaginary parts that are in the set~$0,\pm1,\pm1/2$.
This is done by setting the precision parameter~$\alpha = 1, 2$.

In this section we follows the formalization in~\cite{Suarez2014}.
Consider the exact~$N$-point DFT matrix~$\mathbf{F}_{N}$  and its elements~$f_{i,k}$ with~$i,k = 0, 1, \ldots, N-1$.
The rows of~$\mathbf{F}_{N}$ can be interpreted as a linear filter with transfer function denoted by~$H_i(\omega, \mathbf{F}_{N}) = \sum_{0}^{N-1} f_{i,k} \cdot e^{-jk\omega}$, where~$i = 0, 1, \ldots, N-1$ for~$\omega \in [-\pi, \pi)$.
The variable~$\omega$ represents the spatial frequency across the ULA of antenna~\cite{Suarez2014}.
For the scenario settled in~\cite{Suarez2014}, we have~$\omega = -\omega_n \sin(\psi)$ where we choose~$\omega_n = \pi$ and~$\psi \in [-\pi/2, \pi/2]$.
Therefore, the array patterns are given by
\begin{equation*}
P_i(\psi, \mathbf{F}_{N}) = \frac{|H_i(-\omega_n\sin(\psi), \mathbf{F}_{N})|}{\operatorname{max}(H_i(-\omega_n\sin(\psi), \mathbf{F}_{N}))},
\end{equation*}
for~$i = 0, 1, \ldots, N-1$.
Similarly, we can define the array pattern based on a~$N$-point DFT approximation as
\begin{equation*}
P_i(\psi, \tilde{\mathbf{F}}_{N}) = \frac{|H_i(-\omega_n\sin(\psi), \tilde{\mathbf{F}}_{N})|}{\operatorname{max}(H_i(-\omega_n\sin(\psi), \tilde{\mathbf{F}}_{N}))},
\end{equation*}
where~$i = 0, 1, \ldots, N-1$.

\begin{figure*}
\centering
\subfigure[Exact $8$-point DFT.]{\label{exact8}\includegraphics{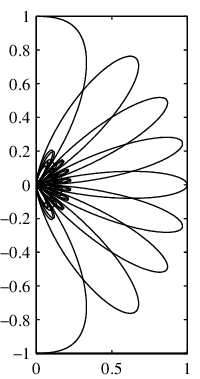}}
\subfigure[Approximate $8$-point DFT.]{\label{approx8}\includegraphics{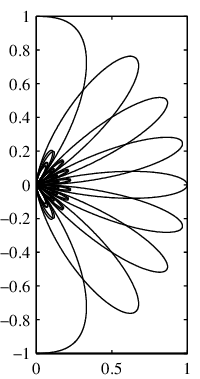}}
\subfigure[Error.]{\label{diff8}\includegraphics{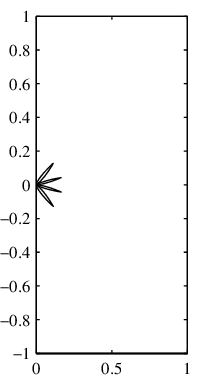}}
\caption{Visual representation of multi-beam pattern for exact and approximate~$8$-point DFT for precision parameter~$\alpha = 8$.}
\label{fig:multibeam8}
\end{figure*}
Figure~\ref{fig:multibeam8} shows the multi-beam pattern for exact and approximate~$8$-point DFT built with precision parameter~$\alpha = 2$.
This particular transform was already proposed in~\cite{Suarez2014}.
This $8$-point DFT approximation furnish beams pointed to~$\psi_k = 0.00, \pm 14.47, \pm 30.00, \pm 48.59, -90.00$ in degrees measured from array broadside direction.
These are values very close to exact~$8$-point DFT computation.
Figure~\ref{diff8} represents the error pattern associated with the $8$-point DFT approximation for precision parameter~$\alpha = 2$.

\begin{figure*}
\centering
\subfigure[Exact $16$-point DFT.]{\label{exact16}\includegraphics{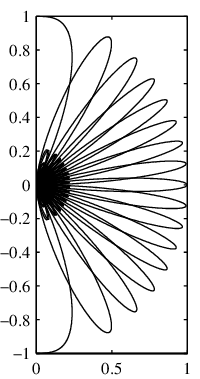}}
\subfigure[Approximate $16$-point DFT.]{\label{approx16}\includegraphics{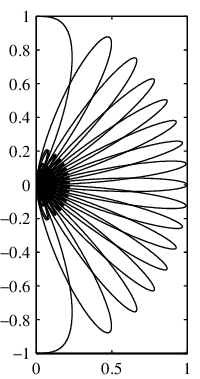}}
\subfigure[Error.]{\label{diff16}\includegraphics{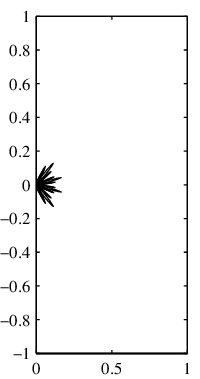}}
\caption{Visual representation of multi-beam pattern for exact and approximate~$16$-point DFT for precision parameter~$\alpha = 2$.}
\label{fig:multibeam16}
\end{figure*}
Figure~\ref{fig:multibeam16} shows the multi-beam pattern for exact and approximate~$16$-point DFT built with precision parameter~$\alpha = 2$.
This $16$-point DFT approximation furnish accurate beam point angles compared to the exact~$16$-point DFT.
It furnish beam angle  errors below Matlab numerical precision for all beams except for  absolute deviation  of~$0.0573$ in degrees for~$9$, $11$, and~$13$th beams when compared to the exact DFT.
Figure~\ref{diff16} represents the error pattern associated with the $16$-point DFT approximation for precision parameter~$\alpha = 2$.

\begin{figure*}
\centering
\subfigure[Exact $32$-point DFT.]{\label{exact32}\includegraphics{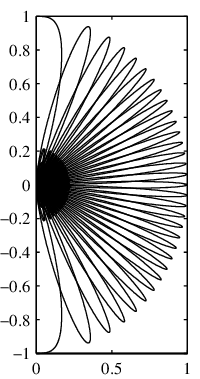}}
\subfigure[Approximate $32$-point DFT.]{\label{approx32}\includegraphics{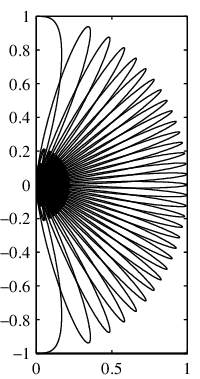}}
\subfigure[Error.]{\label{diff32}\includegraphics{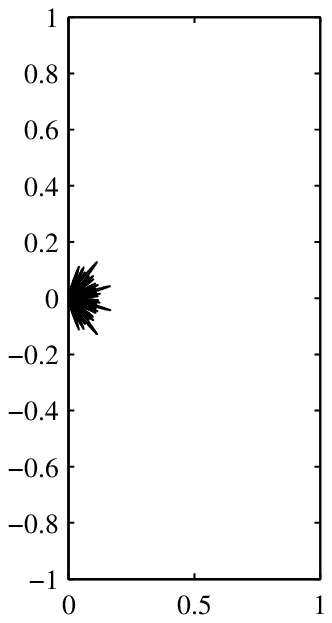}}
\caption{Visual representation of multi-beam pattern for exact and approximate~$32$-point DFT for precision parameter~$\alpha = 2$.}
\label{fig:multibeam32}
\end{figure*}
Figure~\ref{fig:multibeam32} shows the multi-beam pattern for exact and approximate~$32$-point DFT built with precision parameter~$\alpha = 2$.
Similar as the~$16$-point DFT approximation case described above, this~$32$-point DFT approximation furnish beam angle  errors below Matlab numerical precision for all beams except for  absolute deviation of~$0.0573$ in degrees for~$12$ and~$14$th beams when compared to the exact DFT.
Figure~\ref{diff32} represents the error pattern associated with the $32$-point DFT approximation for precision parameter~$\alpha = 2$.

Aforementioned DFT approximations for small blocklentghs.
In the following, we furnish the measures for larger blocklengths.
Hereafter we does not plot anymore the multi-beam pattern for exact and approximate DFT.
It is due the fact that there are a big number of beams (exactly $N$), thus the resulting image is just a black blurred image.
Consider the case where~$N = 512$ with precision parameter~$\alpha=2$.
This $512$-point DFT approximation furnish accurate beam point angles compared to the exact~$512$-point DFT.
This~$512$-point DFT approximation furnish beam angle  errors below Matlab numerical precision for all beams except for  absolute deviation of~$0.0573$ in degrees for~$46$,~$332$, and~$334$th beams when compared to the exact DFT.

For the case of~$1024$-point DFT approximation with~$\alpha=2$ we also obtain close values for beam angles.
The considered~$1024$-point DFT approximation furnish beam angle  errors below Matlab numerical precision for all beams except for  an absolute deviation of~$0.0573$ in degrees for~$54$,~$438$,~$514$,~$550$,~$876$, and~$960$th beams when compared to the exact DFT.

For the~$2048$-point DFT approximation with precision parameter~$\alpha = 2$, again we obtain very close values for the beam angles.
The $2048$-point DFT approximation with precision parameter~$\alpha = 2$ furnish beam angle  errors below Matlab numerical precision for all beams except for  and absolute deviation of of~$0.0573$ in degrees for~$1027$,~$1099$, and~$1919$th beams when compared to the exact DFT.

\section{Estimation Theory for DFT Approximations}
\label{dft-approx-estimation}
In this section we furnish background for the estimation and detection theory for the proposed class of DFT approximations.
Estimation and detection theory are remarks in DFT applications.
Indeed, most application of DFT in real world engineering problems are due its usefulness in estimation and detection in harmonic process.
An harmonic process can be defined as
\begin{equation}
\label{harmonic-process2}
x_n = \sum_{i=1}^{k}A_i \cos(\omega_in+\phi_i)+\epsilon_n,
\end{equation}
where~$\epsilon_n$ is a independent random variable of each~$\phi_i$ with~$E(\epsilon_n) = 0$ and~$E(\epsilon_n^2) = \sigma_\epsilon$, where~$\sigma_\epsilon$ is a known parameter.

In general, we have a set of~$N$ observations of a time series~$x_n$, denoted by~$x_0, x_1, \ldots, x_{N-1}$.
This time series is modeled as in~\eqref{harmonic-process2}.
The interest is estimate the quantities defined by~$k$, $\pm \omega_i$ and~$A_i$ for~$i = 1, 2, \ldots, k$ and~$\sigma_\epsilon^2$.

In order to simplify the analysis carried out here, consider the case that the number of harmonics~$k$ and each frequency~$\omega_i$ is already known.
We can rewrite~\eqref{harmonic-process2} as
\begin{equation}
\label{harmonic-process3}
x_n = \sum_{i=1}^k(A_i'\cos(\omega_in)+B_i'\sin(\omega_in))+\epsilon_n,
\end{equation}
where~$A_i' = A_i\cos(\phi_i)$ and~$B_i' = -A_i\sin(\phi_i)$, since~$A_i^2 = A_i'^2+B_i'^2$ and~$\phi_i = \tan^{-1}(-B_i'/A_i')$.

Note that~\eqref{harmonic-process3} can be seen as multiple linear regression model with regressors~$\cos(\omega_in)$ and~$\sin(\omega_in)$ and~$A_i'$ and~$B_i'$ being the coefficients to be estimated.
A widespread technique for multiple regression analysis that can be applied to this problem is minimize the squared error~\cite{Shumway2005}.
Thus, a way to estimate the coefficients~$A_i'$ and~$B_i'$  is obtained by minimizing the following quantity
\begin{equation*}
\operatorname{MSE} = \sum_{n = 0}^{N-1} \left[x_n-\sum_{i = 1}^k(A_i'\cos(\omega_in)+B_i'\sin(\omega_in))\right]^2.
\end{equation*}
Minimize above quantity require its differentiation with respect the coefficients of interest.
Doing so, we obtain the following prediction equations
\begin{equation}
\label{pred1}
\sum_{n = 0}^{N-1} x_n\cos(\omega_in) = \sum_{j=1}^k\hat{A}_i'c_{ij}+\sum_{j=1}^k\hat{B}_i'd_{ij}
\end{equation}
where
\begin{equation}
\label{pred2}
\sum_{n = 0}^{N-1} x_n\sin(\omega_in) = \sum_{j=1}^k\hat{A}_i'd_{ij}+\sum_{j=1}^k\hat{B}_i's_{ij},
\end{equation}
for~$j = 1, 2, \ldots, k$, where~$\hat{A}_i'$ and~$\hat{B}_i'$ represents the minimum squared error estimates for~$A_i'$ and~$B_i'$ and
\begin{align}
\label{csd}
c_{ij} = & \sum_{n=1}^{N-1}\cos(\omega_in)\cos(\omega_jn),\nonumber \\
s_{ij} = & \sum_{n=1}^{N-1}\sin(\omega_in)\sin(\omega_jn), \\
d_{ij} = & \sum_{n=1}^{N-1}\sin(\omega_in)\cos(\omega_jn).\nonumber
\end{align}

We aim at solve the linear system induced by~\eqref{pred1} and~\eqref{pred2} in order to find the estimates for~$A_i'$ and~$B_i'$.
For such a purpose, we identify properties for the quantities~$c_{ij}$, $s_{ij}$ and~$d_{ij}$ from the set of equations~\eqref{csd} for a particular case of frequencies~$\omega_i$.
A relevant case for spectral estimation occur when the sequence length~$N$ is a integer multiple of periods of each term~$\cos(\omega_in)$ and~$\sin(\omega_in)$.
This happens when the smallest frequency, say~$\omega_i$, obeys the rule~$2\pi p_i/\omega_i = N$.
As consequence
\begin{equation}
\label{freq}
\omega_i = \frac{2\pi p_i}{N},
\end{equation}
for~$i = 1, 2, \ldots, k$.
In such a conditions, we can use Euler equation~\cite{Folland2000} in order to show that
\begin{align}
\label{csd2}
c_{ij} =  s_{ij} = 0 & \qquad i\neq j \nonumber \\
d_{ij} = 0 & \qquad \text{any }~i,j \\
c_{ii} =  s_{ii} = \frac{N}{2} & \qquad \text{any }~i.\nonumber
\end{align}

Plugging the results in~\eqref{csd2} into~\eqref{pred1} and~\eqref{pred2}, we have that the minimum squared error estimators for ~$A_i'$ and~$B_i'$ for~$i = 1, 2, \ldots, k$ are given by
\begin{equation}
\label{estimate-Ai}
\hat{A}_i' = \frac{2}{N}\sum_{n = 0}^{N-1} x_n\cos(\omega_in)
\end{equation}
and
\begin{equation}
\label{estimate-Bi}
\hat{B}_i' = \frac{2}{N}\sum_{n = 0}^{N-1} x_n\sin(\omega_in),
\end{equation}
where each frequency~$\omega_i$ as in~\eqref{freq} is considered to be known.

We can evaluate above estimators for each~$A_i'$ e~$B_i'$.
Applying the expectation operator~\cite{Casella2002} we have that
\begin{equation*}
E(\hat{A}_i') = \frac{2}{N}\sum_{n = 0}^{N-1} E(x_n)\cos(\omega_in).
\end{equation*}
Using~\eqref{harmonic-process3} and noting that~$E(\epsilon_n) = 0$, we obtain
\begin{equation*}
E(\hat{A}_i') = \frac{2}{N}\left(A_i' c_{ij}+B_i'd_{ij}\right).
\end{equation*}
Using orthogonality relations in~\eqref{csd2}, we have
\begin{equation*}
E(\hat{A}_i') = A_i',
\end{equation*}
for~$i = 1, 2, \ldots, k$.
Similarly, we have
\begin{equation*}
E(\hat{B}_i') = B_i',
\end{equation*}
for~$i = 1, 2, \ldots, k$.

The variance for the above estimators are as follows.
We have that,
\begin{align}
\label{var-Ai}
\operatorname{Var}(\hat{A}_i') & = \left(\frac{2}{N}\right)^2\sum_{t=1}^{N-1} \sigma_\epsilon^2 \cos^2(\omega_in)\\\nonumber
 & = \left(\frac{2}{N}\right)^2\sigma_\epsilon^2\sum_{t=1}^{N-1}  \cos^2(\omega_in)\\\nonumber
& = \frac{2}{N}\sigma_\epsilon^2,
\end{align}
where we used the identity~$\sum_{t=1}^{N-1}  \cos^2(\omega_in) = N/2$.
Similarly, we have that
\begin{align}
\label{var-Bi}
\operatorname{Var}(\hat{B}_i') & = \frac{2}{N}\sigma_\epsilon^2,
\end{align}
where now we used the identity~$\sum_{t=1}^{N-1}  \sin^2(\omega_in) = N/2$.

Using~\eqref{estimate-Ai} and~\eqref{estimate-Bi}, we can show that
\begin{equation*}
\operatorname{Cov}(\hat{A}_i, \hat{B}_i) = 0,
\end{equation*}

Note that the estimators in~\eqref{estimate-Ai} and~\eqref{estimate-Bi} are obtained by a multiple linear regression based on the harmonic model in~\eqref{harmonic-process3}.
This estimation procedure return a total of~$2k$ parameters.
Thus, the estimator for the white noise variance~$\hat{\sigma}_\epsilon^2$ supposed in the model~\eqref{harmonic-process3} is based in the ordinary residue for the adjusted model, what results in
\begin{equation*}
\frac{1}{N-2k}\sum_{n=0}^{N-1} \left[x_n - \sum_{i=1}^k(\hat{A}_i'\cos(\omega_in)+\hat{B}_i'\sin(\omega_in))\right]^2.
\end{equation*}

The development provided above is furnished under the assumption that the frequencies~$\omega_i$ are known and they are in the format as in~\eqref{freq}.
When this condition required by~\eqref{freq} is not met, it is yet possible provide approximate properties for~$c_{ij}$, $s_{ij}$ and~$d_{ij}$ in~\eqref{csd2}.
Thus, we have
\begin{align*}
c_{ij} =  s_{ij} = \operatorname{O}(1) & \qquad i\neq j  \\
d_{ij} = 0 & \qquad \text{any }~i,j \\
c_{ii} =  s_{ii} \approx \frac{N}{2} & \qquad \text{any }~i,
\end{align*}
since~$|\omega_i\pm \omega_j| \gg 2\pi/N$ and none of frequencies are farther than~$\pi/N$ from~$0$ or~$\pi$ and where the operator~$\operatorname{O}(\cdot)$ represents asymptotic terms with the same order as its argument~\cite{Pristley1983}.
It is possible demonstrate that
\begin{equation*}
E(\hat{A}_i') = A_i'+\operatorname{O}(1/N)
\end{equation*}
and
\begin{equation*}
E(\hat{B}_i') = B_i'+\operatorname{O}(1/N).
\end{equation*}

The development above shows us how to estimate the coefficients~$A_i'$ and~$B_i'$ for~$i = 1, 2, \ldots, k$ for a harmonic process with~$k$ known frequencies.
However, this conditions of previous knowledge for the values assumed by the frequencies is in general not possible~\cite{Brandwood2003}.
Furthermore, the frequencies are not always in the format as in~\eqref{freq}.
Besides that, the number of harmonics~$k$ is not known.
A way to attack this problem is by making use of the periodogram as follows~\cite{Oppenheim1999}.

\subsection{The Periodogram}
\label{periodogramo}

In order to estimate~$A_i'$ and~$B_i'$ it is required previous knowledge of frequencies~$\omega_i$.
In general, the frequencies~$\omega_i$ and the number of harmonic terms are not known.
In such a case the approach based on minimum squared error is not anymore suitable.
Aiming at solving this problem, a tool named periodogram is used in order to furnish frequencies location in the harmonic process under analysis.
This tool was first proposed by Schuster in 1988 in~\cite{Schuster18989} with an application in meteorology.

\begin{definition}
\label{def1}
Let~$x_0$, $x_1$, \ldots, $x_{N-1}$ be~$N$ observations of an harmonic process.
The periodogram is defined for~\mbox{$-\pi\leq \omega \leq \pi$} by
\begin{equation*}
I_N(\omega) = A^2(\omega)+B^2(\omega),
\end{equation*}
where
\begin{equation*}
A(\omega) = \sqrt{\frac{2}{N}}\sum_{t=1}^{N-1}x_n\cos(\omega t)
\end{equation*}
and
\begin{equation*}
B(\omega) = \sqrt{\frac{2}{N}}\sum_{t=1}^{N-1}x_n\sin(\omega t).
\end{equation*}

\end{definition}

Note that if we use Euler equation~$e^{j\omega t} = \cos(\omega t)+j\sin(\omega t)$~\cite{Folland2000}, we can furnish a more concise definition as below.

\begin{definition}
\label{def1-alt}
Let~$x_0$, $x_1$, \ldots, $x_{N-1}$ be~$N$ observations of an harmonic process.
The periodogram is defined for~\mbox{$-\pi\leq \omega \leq \pi$} as
\begin{equation*}
I_N(\omega) = \frac{2}{N}\left|\sum_{t=1}^{N-1}x_n e^{-j\omega t}\right|^2.
\end{equation*}
\end{definition}
Hereafter we use both definitions for periodogram indistinguishably.
Note that the periodogram is nothing more than the absolute value of DFT components for the sequence~$x_n$~\cite{Oppenheim1999, Blahut2010}.

The frequencies in which we are interested are presented in the format as in~\eqref{freq}.
Note that the periodogram ordinates are nothing more than the sum of squared estimates in~\eqref{estimate-Ai} and~\eqref{estimate-Bi} when the frequency at which the periodogram is being evaluated in as in~\eqref{freq}.

If the estimates~$A_i'$ and~$B_i'$ assumes small values, there is no indication that the frequency~$\omega_i$ is in the harmonic process under analysis.
In this sense the periodogram is used to find the frequencies present in the harmonic process.
Therefore, we say that the periodogram~\emph{detects} the presence or not of a particular frequency~$\omega_i$ in the harmonic process under study.
Moreover, since the periodogram is able to detect the frequencies present in the harmonic process it also estimate the quantity~$k$ of harmonics in the process.

Since the support for~$\omega$ in the definition of periodogram is the interval~$[-\pi, \pi]$, the indexes for the frequencies in the form of~\eqref{freq} shall be~$i = 1, 2, \ldots, [N/2]$, where~$[\cdot]$  is the function that returns the largest integer smaller than its argument~\cite{Folland2000}.
In above conditions, we have
\begin{equation*}
A(\omega_i) = \sqrt{\frac{N}{2}}\hat{A}_i
\end{equation*}
and
\begin{equation*}
B(\omega_i) = \sqrt{\frac{N}{2}}\hat{B}_i.
\end{equation*}

For the sake of notational purposes, we represent~$I_N(\omega_i)$ by~$I_i$, for~$i = 0, 1, \ldots, [N/2]-1$, i.e.,~$I_i \equiv I_N(\omega_i)$.
As consequence from the Definition~\ref{def1}, we have that
\begin{equation*}
E(I_i) = \frac{N}{2} \left[E(\hat{A}_i'^2)+E(\hat{B}_i'^2)\right].
\end{equation*}
Note that~$E(\hat{A}_i'^2) = E^2(\hat{A}_i')+\operatorname{Var}(\hat{A}_i')$ and~$E(\hat{B}_i'^2) = E^2(\hat{B}_i')+\operatorname{Var}(\hat{B}_i')$.
From~\eqref{var-Ai} and~\eqref{var-Bi}, we have that
\begin{equation}
\label{EIi}
E(I_i) = \frac{N}{2} \left(A_i'^2+B_i'^2\right)+2\sigma_\epsilon^2 = \frac{N}{2}A_i^2+2\sigma_\epsilon^2.
\end{equation}

The values for the adjusted coefficients~$\hat{A}_i'$ and~$\hat{B}_i'$ will be nearly zero for periodogram evaluation in frequencies~$\omega = \omega_i$ not present in the harmonic process under study.
It is due the fact that~$A_i = 0$ for such a situation.
In above equation~\eqref{EIi}, we take~$A_i = 0$ what returns~$E(I_i) = 2\sigma_\epsilon^2$.
When the frequency~$\omega$ gets near of frequencies present in the harmonic process, the value of~$I_N(\omega)$ increases showing peaks with~$\operatorname{O}(N)$ and decreases to small values as the frequency gets distant from observed frequencies in the harmonic process.
Therefore, a way to estimate the frequencies present in the harmonic process is by making observation to the peaks of periodogram~\cite{Pristley1983}.
These results are exact if the frequencies in the harmonic process under analysis are as in~\eqref{freq}, but are only approximated otherwise.

\subsection{Periodogram Sampling Properties}
\label{amostragem-periodogramo}

As an initial step toward the analysis of periodogram sampling properties, we consider the harmonic process compounded only by the noise~$\epsilon_n$.
For this scenario, we assume that the noise is a white gaussian noise~\cite{Hayes1996}.
As consequence, all coefficients of terms~$\cos(\cdot)$ and~$\sin(\cdot)$ are null and~$k = 0$.

Such an assumption allow us to find exact probability distribution of~$I_N(\omega_i)$ for~$i = 1, 2, \ldots, [N/2]$.
Using the assumption of normality and independence for white gaussian noise, we have that~$x_n$ is null mean and possess variance~$\sigma_x^2 = \sigma_\epsilon^2$, what yields~$x_n \sim N(0, \sigma_x^2)$.
The knowledge about the distribution of~$x_n$ allows us to find the variance for~$A(\omega_i)$ and~$B(\omega_i)$ as the Definition~\ref{def1}.
Since~$A(\omega_i)$ and~$B(\omega_i)$ are linear combination of~$x_n$, we have that
\begin{align*}
\operatorname{Var}(A(\omega_i)) &  = \frac{2\sigma_x^2}{N}\left(\sum_{n=0}^{N-1}\cos^2(\omega_i)\right) \\
& =
\begin{cases}
\sigma_x^2,\text{ for~$i \neq 0$ and~$N/2$},\\
2\sigma_x^2,\text{ if $\omega_i = 0$,}\\
\end{cases}
\end{align*}
where~$\omega_i$ is the of the form in~\eqref{freq} and~$N$ is even.
As consequence of above result, we have that
\begin{align*}
A(\omega_i) =
\begin{cases}
N(0,\sigma_x^2), \text{ for~$i \neq 0$ and~$N/2$,}\\
N(0,2\sigma_x^2), \text{ if~$i = 0$ or~$N/2$,}\\
\end{cases}
\end{align*}
where~$\omega_i$ is the of the form in~\eqref{freq} and~$N$ is even.
Similarly, we can derive above results for~$B(\omega_i)$ with
\begin{align*}
\operatorname{Var}(B(\omega_i)) &  = \frac{2\sigma_x^2}{N}\left(\sum_{n=0}^{N-1}\sin^2(\omega_i)\right) \\
& =
\begin{cases}
\sigma_x^2,\text{  for~$i \neq 0$ and~$N/2$},\\
0,\text{ if $\omega_i = 0$},\\
\end{cases}
\end{align*}
where~$\omega_i$ is the of the form in~\eqref{freq} and~$N$ is even, what leads to
\begin{align*}
B(\omega_i) =
\begin{cases}
N(0,\sigma_x^2), \text{ for~$i \neq 0$ and~$N/2$ with~$N$ even,}\\
0, \text{ if~$i = 0$ or~$N/2$.}\\
\end{cases}
\end{align*}

Moreover, we have that~$A(\omega_i)$ and~$B(\omega_j)$ are not correlated since
\begin{align*}
\operatorname{Cov}(A(\omega_i), B(\omega_j)) & = \frac{2\sigma_x^2}{N}\left(\sum_{n=0}^{N-1} \cos(\omega_in)\sin(\omega_jn)\right) \\
& = \frac{2\sigma_x^2}{N}d_{ij} = 0,\\
\end{align*}
for any~$i,j = 1, 2, \ldots, k$ where we used~$d_{ij}$ in~\eqref{csd2}.
Using~\eqref{csd2} we can derive
\begin{equation*}
\operatorname{Cov}(A(\omega_i), A(\omega_j)) = \operatorname{Cov}(B(\omega_i), B(\omega_j)) = 0,
\end{equation*}
where~$i\neq j$.

In the following, we derive the probability distribution for the ordinates of periodogram~\mbox{$I_i \equiv I_N(\omega_i)$}.
Since~$A(\omega_i)$ and~$B(\omega_j)$ follows normal distribution with null mean and are not correlated, by the Theorem of  Cochran-Fisher~\cite{Montgomery2012} we have that~$I_i$ follows an~$\chi^2$ distribution scaled by the variance~$\sigma_x^2$ with two degrees of freedom for the case~$i \neq 0$ and~$N/2$ and one degree of freedom when~$i = 0$ or~$N/2$. Thus, we have that
\begin{align*}
I_i = A^2(\omega_i)+B^2(\omega_i) =
\begin{cases}
\sigma_x^2 \chi_2^2, \text{ for~$i \neq 0$ and~$N/2$,}\\
2\sigma_x^2 \chi_1^2, \text{ if~$i = 0$ or~$N/2$,}\\
\end{cases}
\end{align*}
where~$\omega_i$ is as in~\eqref{freq} and~$N$ is even.

The ordinates~$I_i$ of periodogram are not correlated since~$A(\omega_i)$ and~$B(\omega_j)$ are not correlated for~$i\neq j$.
In particular, it is possible demonstrate that the ordinates~$I_i$ of periodogram are independent.

\begin{theorem}
\label{teoremaDistI_i}
For the harmonic process~$x_n$ compounded only by the white gaussian noise with variance~$\sigma_x^2$, the ordinates~$I_i$ of periodogram for~$i = 1, 2, \ldots, [N/2]$ are independents and for each~$i$ we have that
\begin{align*}
I_i =
\begin{cases}
\sigma_x^2 \chi_2^2, \text{ for~$i \neq 0$ and~$N/2$,}\\
2\sigma_x^2 \chi_1^2, \text{ if~$i = 0$ or~$N/2$,}\\
\end{cases}
\end{align*}
where~$\omega_i$ is as in~\eqref{freq} and~$N$ is even.
\end{theorem}

Above theorem allows us to calculate the periodogram expectation and variance easily.
Remembering that the~$\chi^2_\nu$ distribution has mean value equals to~$\nu$ and variance~$2\nu$, we have that
\begin{align*}
E(I_i) = 2\sigma_x^2,
\end{align*}
for~$i = 0, 1 \ldots, [N/2]$, and
\begin{align}
\label{varI_i}
\operatorname{Var}(I_i) =
\begin{cases}
4\sigma_x^2, \text{ for~$i \neq 0$ and~$N/2$,}\\
8\sigma_x^2, \text{ if~$i = 0$ or~$N/2$,}\\
\end{cases}
\end{align}
where~$\omega_i$ is as in~\eqref{freq} and~$N$ is even.

In the following we furnish some important results for spectral estimation.
We derive expressions for~$\operatorname{Cov}(I_N(\omega), I_N(\omega'))$ where~$\omega$ and~$\omega'$ are not as in~\eqref{freq}.
Furthermore, it is furnished an expression for expected value~$E(I_N(\omega))$ when the process~$x_n$ is not only a white gaussian noise but contain sinusoidal parcels, i.e., it possess~$A_i \neq 0$.

Before providing the above mentioned results, we introduce an alternative representation to the periodogram in order to ease the understanding.
\begin{theorem}
Let~$I_N(\omega)$ be as in Definition~\ref{def1}.
Thus, for each~$\omega$ we can write the alternative formulation
\begin{align*}
I_N(\omega) = 2 \sum_{s = -(N-1)}^{(N-1)}\hat{R}(s)\cos(s\omega),
\end{align*}
where
\begin{align*}
\hat{R}(s) = \frac{1}{N} \sum_{t = 1}^{N-|s|}x_n X_{t+|s|}.
\end{align*}

\end{theorem}

A mathematical demonstration of above theorem can be conveyed by using the fact that if~$z$ is a complex number, so its norm is obtained by the product~$z\bar{z}$, where~$\bar{z}$ represents the complex conjugate of~$z$~\cite{Ahlfors1966}.
Applying this property to the periodogram, we have that
\begin{align*}
I_N(\omega) & = \frac{2}{N}\left|\sum_{n=0}^{N-1}x_n e^{-j\omega t}\right|^2\\
& =\frac{2}{N}\sum_{n=0}^{N-1} \sum_{r=1}^Nx_nx_r \cos((n-r)\omega)\\
& = 2 \sum_{s=-(N-1)}^{N-1} \left( \frac{1}{N} \sum_{n=0}^{N-|s|-1}x_nx_{n+|s|} \right) \cos(s\omega)\\
& = 2 \sum_{s = -(N-1)}^{(N-1)}\hat{R}(s)\cos(s\omega).
\end{align*}

In the following it is furnished a manner to evaluate the expected value for the periodogram when the harmonic process under consideration possess frequencies that does not follows~\eqref{freq}.

\begin{theorem}
\label{teoremaEI_i}
Let~$x_n$ be as in~\eqref{harmonic-process2}.
Thus, for any~$\omega$ we have that expected value for the periodogram of~$x_n$, $E(I_N(\omega))$ is given by
\begin{equation*}
2\sigma_\epsilon^2+\pi\sum_{i=1}^kA_i^2\Bigg[F_N(\omega+\omega_i)+F_N(\omega-\omega_i)\Bigg],
\end{equation*}
where each~$\omega_i$ follows~\eqref{freq} and~$F_N(\cdot)$ represents the Fejér kernel given by
\begin{equation}
\label{nucleofejer}
F_N(\theta) = \frac{1}{2\pi N}\frac{\sin^2(N\theta/2)}{\sin^2(\theta/2)}.
\end{equation}
\end{theorem}

\begin{figure}
\centering
\psfrag{FN}{$F_N(\theta)$}
\psfrag{theta}{$\theta$}
\includegraphics[]{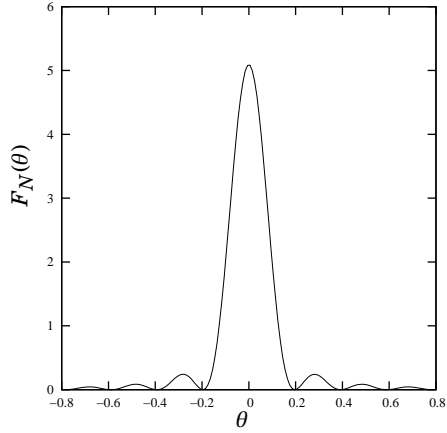}
\caption{Plot of Fejér kernel~$F_N(\theta)$ for~$ \theta \in \lbrack-8\pi/N, 8\pi/N\rbrack$ with~$N=32$.}
\label{fejer}
\end{figure}

Figure~\ref{fejer} shows the plot for the Fejér kernel for~$\theta$ up to~$8\pi/N$.
Note that Fejér kernel possess a peak in the origin and possess oscillations in its tails as~$|\theta|$ increases.

\begin{figure}
\centering
\psfrag{EI}{$E(I_N(\omega))$}
\psfrag{omega}{$\omega$}
\includegraphics[]{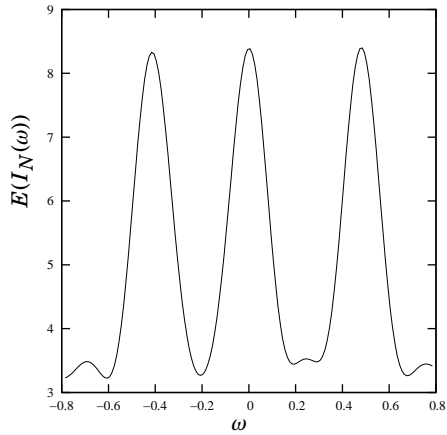}
\caption{Plot for the expected value for the periodogram~$I_i$ with~$N=32$ and frequencies~$\omega_1 = \pm 4.2\pi/N$ and~$\pm 4.9\pi/N$.}
\label{mfejer}
\end{figure}

According Theorem~\ref{teoremaEI_i}, we have that the expected value of~$I_i$ is a superposition of scaled Fejér kernels  centered in frequencies~$\pm \omega_i$ plus the parcel~$2\sigma_\epsilon^2$.
Figure~\ref{mfejer} illustrate this situation.

In Figure~\ref{mfejer}, the peaks are located at frequencies~$\omega$ present in the harmonic process.
Suppose there is in the process a frequency~$\omega_p$ that does not follows~\eqref{freq}.
In this situation, the peak centered in this frequency undergo heavy attenuation.
The worst case is when this frequency is exactly in the middle between two neighbors frequencies of the form in~\eqref{freq}.
For such a case, Whittle demonstrated in~\cite{Whittle1952} that there is an attenuation of~$4\pi^2$ and that it depends on the magnitude of~$A_i^2$.
The final effect of this consequent attenuation can affect the ability to detect the presence of such harmonic in the process.

A important assumption for detection theory development is that the periodogram ordinates shall be considered independents.
This condition is necessary for the proposition of statistical tests for the periodogram ordinates.
This condition is verified only in the very special case where all frequencies are as in~\eqref{freq}.
Not surprisingly, when the frequencies gets far away from the frequencies of the form~\eqref{freq}, the periodogram ordinates loose the independence condition.
The following theorem presents an way to compute the periodogram ordinates covariance when the frequencies does not match~\eqref{freq} and the harmonic process~$x_n$ is assumed to be compounded only by the noise, i.e.,~$A_i \equiv 0$ for~$i = 1, 2, \ldots, k$ as in~\eqref{harmonic-process2}.
Moreover, the noise is generalized to a white noise, thus not necessarily following an white gaussian noise.
As consequence, the harmonic process observations~$x_n$ are independent.

\begin{theorem}
\label{teoremavarI_i}
Let~$x_n = \epsilon_n$ be an harmonic process with independent observations and finite fourth cumulant~$\kappa_4$~\cite{Casella2002}, where~$\epsilon_n$ is a white noise.
The~$x_n$ periodogram covariance evaluated in~$\omega_1$ and~$\omega_2$, $\operatorname{Cov}(I_N(\omega_1), I_N(\omega_2))$, is given by
\begin{equation*}
\frac{8\pi\kappa_4}{N}+\frac{4\sigma_x^2}{N^2}\Bigg[F_N(\omega_1+\omega_2)+F_N(\omega_1-\omega_2)\Bigg],
\end{equation*}
where~$F_N(\cdot)$ is the Fejér kernel as in~\eqref{nucleofejer}.
\end{theorem}

Above result in Theorem~\ref{teoremavarI_i} was proved by Bartlett~\cite{Bartlett1966}.
An particular case for above theorem is when~$\omega_1=\omega_2=\omega$, what yields
\begin{equation}
\label{varI_igeneral}
\operatorname{Var}(I_N(\omega)) =
\begin{cases}
4\sigma_x^2+\frac{4\kappa_4}{N}+\operatorname{O}\left(\frac{1}{N^2}\right), \text{ $\omega \neq 0, \pm\pi$,}\\
8\sigma_x^2+\frac{4\kappa_4}{N}, \text{ $\omega = 0, \pm\pi$.}
\end{cases}
\end{equation}

For the special case where~$\epsilon_n$ is normal and~$\omega$ is as in~\eqref{freq}, we have that~\eqref{varI_igeneral} reduces to~\eqref{varI_i}.
The general case~\eqref{varI_igeneral} shows us that~$\operatorname{Var}(I_N(\omega))$ has asymptotic behavior with order~$\operatorname{O}(1/N)$ when the noise~$\epsilon_n$ is not normal and~$\omega$ does not match~\eqref{freq}.

Above result in Theorem~\ref{teoremavarI_i} will be useful for statistical tests development for harmonic detection in the following section.

\subsection{Statistical Tests for Periodogram Ordinates}
\label{statistical-test}

A periodogram of a harmonic process~$x_n$ may present various peaks, from which some of them can be due to the presence of harmonic parcels and only the mere consequence of random effects of physical process under study.
Thus, in order to verify if an particular frequency is part of harmonic process under analysis, we shall apply statistical tests~\cite{Casella2002}.

Consider the statistical test with null hypothesis~$H_0: A_i = 0$ for all~$i = 1, 2, \ldots, k$ as in~\eqref{harmonic-process2} and~$H_1:A_i\neq 0$ for some~$i$.
Note that these hypothesis are equivalent to test if~$k=0$.

We are interested in test the presence of any harmonic.
Initially, consider the case where~$\sigma_x^2$ is known.
We define
\begin{equation*}
\gamma = \frac{\lbrack \operatorname{max}_{1\leq i \leq [N/2]}(I_i) \rbrack}{\sigma_x^2}.
\end{equation*}

In the following development of statistical test we consider~$x_n$ with a normal distribution.
For such a case, as described in the previous section and shows the Theorem~\ref{teoremaDistI_i}, we have that under~$H_0$ the periodogram ordinates~$I_i$ follows an scaled~$\chi_2^2$ for~$i = 1, 2, \ldots, [N/2]$.
In particular, for~$i\neq 0$ we have that
\begin{equation*}
I_i/\sigma_x^2 \sim \chi_2^2.
\end{equation*}
Note that~$\chi_2^2$ is equivalent to exponential distribution~\cite{Casella2002}.
Therefore, we have that
\begin{equation*}
p[(I_i/\sigma_x^2)\leq z] = 1-\exp(-z/2).
\end{equation*}
Under~$H_0$, the periodogram ordinates are independent and identically distributed.
For any~$z$, we have that
\begin{align*}
p[\gamma > z] & = 1-p[\gamma \leq z]\\
& = 1-p[(I_i/\sigma_x^2)\leq z], \text{ for all~$i$}\\
& = 1-(1-\exp(-z/2))^{[N/2]}.
\end{align*}

Above equation furnish the distribution of~$\gamma$ when~$\sigma_x^2$ is known.
This distribution can be used in order to build the one-sided exact hypothesis test with length~$\alpha$ in order to verify if any~$A_i = 0$ for any~$i = 1, 2, \ldots, k$.

However, usually the value of~$\sigma_x^2$ is not known.
For such, we substitute~$\sigma_x^2$ by an unbiased estimator given by
\begin{equation*}
v = \frac{1}{2[N/2]} \sum_{i = 1}^{[N/2]} I_i.
\end{equation*}
Applying Theorem~\ref{teoremaEI_i} we have that under~$H_0$, the estimator~$v$ is unbiased for~$\sigma_x^2$, where~$\sigma_\epsilon^2$ represents~$\sigma_x^2$ in Theorem~\ref{teoremaEI_i}.
Thus, we define the statistic
\begin{equation*}
g^\ast = \frac{\operatorname{max}_{1\leq i \leq [N/2]}(I_i)}{\{1/(2[N/2])\}\sum_{i = 1}^{[N/2]} I_i}.
\end{equation*}

For high values of~$N$, we can ignore the effects of fluctuations assumed by~$v$ and thus use the same distribution as the statistic~$\gamma$,
what yields
\begin{equation*}
p[g^\ast>z]\approx 1-(1-\exp(-z/2))^{[N/2]}.
\end{equation*}

Fisher~\cite{Fisher1929} derived an exact test for~$\operatorname{max}_{1\leq i \leq [N/2]}(I_i)$ based on the statistics
\begin{equation*}
g = \frac{\operatorname{max}_{1\leq i \leq [N/2]}(I_i)}{\sum_{i = 1}^{[N/2]} I_i}.
\end{equation*}
Fisher demonstrated that the distribution of~$g$ can be exactly given by the series expansion
\begin{align*}
p[g>z] & = n(1-z)^{n-1}-\frac{n(n-1)}{2}(1-2z)^{n-1}\\
 & \phantom{=} + \ldots + (-1)^a\frac{n!}{a!(n-a)!}(1-az)^{n-1},
\end{align*}
where~$n=[N/2]$ and~$a$ is the largest integer less than~$1/z$.
Compute all the terms series expansion above can be unnecessary.
Fisher proposed to use only the first term of above series expansion, what result in~$n\exp(-z/2)$.
This test is known as test~$g$~\cite{Pristley1983}.

In case of Fisher test return a significant result, everything we can do is say that null hypothesis~$H_0$ must be rejected and as consequence we can accept the alternative hypothesis that there is an harmonic present one of the frequencies~\eqref{freq} in the process under study.
At this point we will use a result we own to Hartley\footnote{Although similar names, he is not the  R. V. L. Hartley, the proposer of so well known DHT.} in~\cite{Hartley1949}.

Suppose that~$\operatorname{max}_{1\leq i \leq [N/2]}(I_i)$ occur in~$i = p$.
In~\cite{Hartley1949}, it was demonstrated that the significance probability of~$\operatorname{max}_{1\leq i \leq [N/2]}(I_i) = I_p$ due to an existing harmonic component other than~$\omega_p = 2\pi p/N$ is less than the test significance level~$\alpha$.
Thus, if~$\operatorname{max}_{1\leq i \leq [N/2]}(I_i) = I_p$ is significant, we can estimate~$\hat{\omega}_p$ as the occurrence of an harmonic in this frequency.
Since the frequency of occurrence of an harmonic is estimated as~$\hat{\omega}_p$, we can use this in order to estimate the respective coefficients in Definition~\ref{def1},
\begin{equation*}
A(\hat{\omega}_p) = \sqrt{\frac{2}{N}}\sum_{n=0}^{N-1}x_n\cos(\hat{\omega}_p n)
\end{equation*}
and
\begin{equation*}
B(\hat{\omega}_p) = \sqrt{\frac{2}{N}}\sum_{n=0}^{N-1}x_n\sin(\hat{\omega}_p n).
\end{equation*}

The Fisher test allows us to verify only the significance where the periodogram peak occurs.
Whittle in~\cite{Whittle1952} extended the Fisher test in order to verify the significance of non-principal periodogram peaks.
This procedure can be easily implemented by taking off from the denominator of statistic~$g$ the largest periodogram ordinate and substituting the~$\operatorname{max}_{1\leq i \leq [N/2]}(I_i)$ by the second largest periodogram ordinate, here denoted by~$I_{p''}$.
Note that this procedure is nothing more than consider a fictional periodogram where the largest ordinate is taken off.
This way, we have that we can verify the significance of the second largest periodogram ordinate by setting
\begin{equation*}
g' = \frac{I_{p''}}{\big(\sum_{i = 1}^{[N/2]} I_i \big)- I_p},
\end{equation*}
and using the distribution of~$g$ and substituting~$n$ by~$n-1$.

If the second largest periodogram ordinate is then significant, we perform one more time the above procedure taking off the second largest periodogram ordinate.
If the third periodogram ordinate is then significant, we repeat the process for the third periodogram ordinate.
The process is then repeated until we find a non-significant periodogram ordinate.
All periodogram ordinate~$I_{p}, I_{p''}, I_{p'''}, \ldots$ found to be significant allows the estimation of existing harmonics in the frequencies~$\omega_{p}, \omega_{p''}, \omega_{p'''}, \ldots$ that follows~\eqref{freq}.
The number of significant periodogram ordinates is an estimate for~$k$ in the model of~\eqref{harmonic-process3}.
By this procedure proposed by Hartley~\cite{Hartley1949}, we have estimation for each coefficient of harmonic process, the number of harmonics and the frequencies of its location.

\subsection{The Approximate Periodogram}

The periodogram, hereafter called the exact periodogram, can be given a matrix format, in particular, at the sample frequencies as in~\eqref{freq}.
Consider an~$N$ power-of-two.
If we adjoin exact periodogram ordinates as a column vector, we have that its values will be given by
\begin{equation*}
\mathbf{I}_{N} = \frac{2}{N}\big|\mathbf{F}_{N}\cdot \mathbf{x}\big|^2,
\end{equation*}
where~$\mathbf{x} = \begin{bmatrix} x_0, x_1, \ldots, x_{N-1} \end{bmatrix}^\top$,~$\mathbf{I}_{N} = \begin{bmatrix} I_0, I_1, \ldots, I_{N-1} \end{bmatrix}^\top$ and for each~$i = 0, 1, \ldots, N-1$ we have that~$I_i \equiv I_N(\omega_i)$ with~$\omega_i$ as in~\eqref{freq} and~$\mathbf{F}_{N}$ is just the exact $N$-point DFT matrix.
In particular, since the input signal is real, we does not need to compute all coefficients.
Indeed, we need to compute only the ordinates~$i = 0, 1, \ldots, N/2$~\cite{Oppenheim2009}, since~$I_i = I_{N-i}$ for~$i = 0, 1, \ldots, N/2-1$.
Above formalism paves the way for defining the approximate periodogram as follows.
\begin{definition}
\label{periodogram-approx}
Let~$\tilde{\mathbf{F}}_{N}$ be an $N$-point DFT approximation matrix for an arbitrary precision parameter~$\alpha$.
Also, let~$\mathbf{x} = \begin{bmatrix} x_0, x_1, \ldots, x_{N-1} \end{bmatrix}^\top$ be the vector representation of observations of harmonic process under analysis.
We define the approximate periodogram as
\begin{equation*}
\tilde{\mathbf{I}}_\mathrm{N} = \frac{2}{N}\big|\tilde{\mathbf{F}}_{N}\cdot \mathbf{x}\big|^2,
\end{equation*}
where~$\mathbf{I}_{N} = \begin{bmatrix} \tilde{I}_0, \tilde{I}_1, \ldots, \tilde{I}_{N-1} \end{bmatrix}^\top$ for each~$i = 0, 1, \ldots, N-1$.%
\end{definition}
Similarly to the exact periodogram, we are not required to compute all periodogram ordinates.
Therefore, we compute only the approximate periodogram ordinates for~$i = 0, 1, \ldots, N/2$~\cite{Oppenheim2009}, since~$\tilde{I}_i = \tilde{I}_{N-i}$ for~$i = 0, 1, \ldots, N/2-1$.

Here we use the approximation developed in Section~\ref{sec:ERR_ANA_III} for the matrix representing the DFT approximation.
We use~\eqref{DFTA-approx} that states
\begin{equation*}
\tilde{\mathbf{F}}_{N} \approx a_1^{\operatorname{log}_2(N/4)}\mathbf{F}_{N},
\end{equation*}
where the sign~$\approx$ represents  an element-wise approximation.
In Appendix~\ref{a1-prop}, we derived upper and lower bounds for~$a_1$ and showed it goes to unity as the precision parameter~$\alpha$ is made arbitrarily large.
Thus, as larger is the precision parameter~$\alpha$, closer are~$\tilde{\mathbf{F}}_{N}$ and~$\mathbf{F}_{N}$.
It was formally demonstrated by Theorem~\ref{DFTAN-conv}.

Using above approximation to the matrix of $N$-point DFT approximation, we are induced to state the following theorem.
\begin{theorem}
Let~$\mathbf{I}_{N}$ be the exact periodogram associated with an harmonic process with observations~$\mathbf{x} = \begin{bmatrix} x_0, x_1, \ldots, x_{N-1} \end{bmatrix}^\top$.
Let~$\tilde{\mathbf{I}}_\mathrm{N}$ be its approximated periodogram built with a precision parameter~$\alpha$.
We have that
\begin{align*}
\tilde{\mathbf{I}}_\mathrm{N} & \approx a_1^{2\operatorname{log}_2(N/4)} \mathbf{I}_{N}.
\end{align*}

\begin{proof}
Take the~\eqref{DFTA-approx} into the Definition~\ref{periodogram-approx}.
We have that
\begin{align*}
\tilde{\mathbf{I}}_\mathrm{N} & \approx \frac{2}{N} \big| a_1^{\operatorname{log}_2(N/4)}\mathbf{F}_{N}\cdot \mathbf{x}\big|^2\\
& \approx  a_1^{2\operatorname{log}_2(N/4)} \frac{2}{N}\big|\mathbf{F}_{N}\cdot \mathbf{x}\big|^2\\
& \approx a_1^{2\operatorname{log}_2(N/4)} \mathbf{I}_{N}.
\end{align*}
\end{proof}
\end{theorem}

Furthermore, we furnish an approximate periodogram representation based on Definition~\ref{def1-alt} for the exact periodogram.
For frequencies~$\omega_i$ as in~\ref{freq} for~$i = 0, 1, \ldots N/2$, the exponential terms~$e^{-\omega_in}$ in the exact periodogram as in Definition~\ref{def1-alt} are the elements of exact $N$-point DFT matrix for~$i = 0, 1, \ldots N/2$.
It is clear form the above representation for th exact periodogram in terms of exact $n$-point DFT matrix.
Therefore, we introduce the representation of approximate periodogram as the following.
Let~$\tilde{\mathbf{I}}_\mathrm{N}$ be the approximate periodogram built with the $N$-point DFT approximation matrix with precision parameter~$\alpha$ as in Definition~\ref{periodogram-approx}.
We have that the approximate periodogram ordinates can be expressed as
\begin{equation*}
\tilde{I}_i  = \frac{2}{N}\Bigg|\sum_{n = 0}^{N-1}x_n \tilde{f}_{n,i}\Bigg|^2,
\end{equation*}
where~$i = 0, 1, \ldots, N/2$ and~$\tilde{f}_{n,i}$ are the~$(n,i)$ entries of $N$-point DFT approximation matrix with precision parameter~$\alpha$.

From Definition~\ref{def1} and the representation of exact periodogram based in the exact DFT matrix, we have that
\begin{equation*}
A(\omega_i) = \sqrt{\frac{2}{N}}\mathfrak{Re}\left(\sum_{n = 0}^{N-1}x_n e^{-j\omega_in}\right)
\end{equation*}
and
\begin{equation*}
B(\omega_i) = \sqrt{\frac{2}{N}}\mathfrak{Im}\left(\sum_{n = 0}^{N-1}x_n e^{-j\omega_in}\right),
\end{equation*}
where~$i = 0, 1, \ldots, N/2$.
In a likely manner, we can furnish an alternative definition for the approximate periodogram as follows.
\begin{definition}
\label{periodogram-approx-alt}
Let~$x_0, x_1, \ldots, x_{N-1}$ be~$N$ observations of an harmonic process.
Also, let~$\tilde{f}_{n,i}$ be the~$(n,i)$ matrix entry of~$N$-point DFT approximation~$\tilde{\mathbf{F}}_{N}$ for the precision parameter~$\alpha$.
The approximate periodogram for frequencies~$\omega_i$ as in~\eqref{freq} is defined as
\begin{equation*}
I_i = \tilde{A}^2(\omega_i)+\tilde{B}^2(\omega_i),
\end{equation*}
where
\begin{equation*}
\tilde{A}(\omega_i)  \sqrt{\frac{2}{N}}\sum_{n=0}^{N-1}x_n\mathfrak{Re}\left(\tilde{f}_{n,i}\right)
\end{equation*}
and
\begin{equation*}
\tilde{B}(\omega_i) = \sqrt{\frac{2}{N}}\sum_{n=0}^{N-1}x_n\mathfrak{Im}\left(\tilde{f}_{n,i}\right),
\end{equation*}
where~$i = 0, 1, \ldots, N/2$ and~$\mathfrak{Re}(\cdot)$ and~$\mathfrak{Im}(\cdot)$ returns the real and imaginary parts of its complex argument.
\end{definition}

Using the approximation in~\eqref{DFTA-approx}, we have that for each matrix entry~$\tilde{f}_{n,i}$ for~$i,n = 0, 1, \ldots, N-1$ of~$\tilde{\mathbf{F}}_{N}$, we have that
\begin{equation*}
\tilde{f}_{n,i} \approx a_1^{\operatorname{log}_2(N/4)} e^{-j\omega_in},
\end{equation*}
thus we have that
\begin{equation}
\label{Re-f}
\mathfrak{Re}(\tilde{f}_{n,i}) \approx a_1^{\operatorname{log}_2(N/4)} \cos(\omega_in)
\end{equation}
and
\begin{equation}
\label{Im-f}
\mathfrak{Im}(\tilde{f}_{n,i}) \approx a_1^{\operatorname{log}_2(N/4)} \sin(\omega_in).
\end{equation}
Using above relations in~$\tilde{A}(\omega_i)$ and~$\tilde{B}(\omega_i)$ as in Definition~\ref{periodogram-approx-alt}, we obtain
\begin{equation}
\label{A-approx}
\tilde{A}(\omega_i) \approx a_1^{\operatorname{log}_2(N/4)} A(\omega_i)
\end{equation}
and
\begin{equation}
\label{B-approx}
\tilde{B}(\omega_i) \approx a_1^{\operatorname{log}_2(N/4)} B(\omega_i).
\end{equation}

\subsection{Statistical Properties of Approximated Periodogram}

In this section we present some important results that paves the way for detection and estimation using DFT approximations.
We develop results for~$\tilde{A}(\omega_i)$ and~$\tilde{B}(\omega_i)$ based on those for the exact counterparts~$A(\omega_i)$ and~$B(\omega_i)$ in subsection~\ref{amostragem-periodogramo}.

Again, consider the case where the harmonic process is compounded only by the white gaussian noise~$\epsilon_n$ with variance~$\sigma_\epsilon^2$, for~$i = 0, 1, \ldots, N-1$.
From Definition~\ref{periodogram-approx-alt}, we have that~$\tilde{A}(\omega_i)$ is a linear combination of~$x_n = \epsilon_n$.
Thus, we have that
\begin{align*}
\operatorname{Var}(\tilde{A}(\omega_i)) &  = \frac{2\sigma_x^2}{N}\left(\sum_{n=0}^{N-1}\mathfrak{Re}(\tilde{f}_{n,i})^2\right),
\end{align*}
where~$\omega_i$ is the of the form in~\eqref{freq} and~$N$ is even.
We use the result in~\eqref{A-approx}, what enable us to state
\begin{align*}
\operatorname{Var}(\tilde{A}(\omega_i)) &  = \frac{2\sigma_x^2}{N}\left(\sum_{n=0}^{N-1}\mathfrak{Re}(\tilde{f}_{n,i})^2\right)\\
&  \approx \frac{2\sigma_x^2}{N}\left(\sum_{n=0}^{N-1}a_1^{2\operatorname{log}_2(N/4)} \cos^2(\omega_i)\right)\\
& \approx
\begin{cases}
a_1^{2\operatorname{log}_2(N/4)}\sigma_x^2,\text{ for~$i \neq 0$ and~$N/2$},\\
2a_1^{2\operatorname{log}_2(N/4)}\sigma_x^2,\text{ if $\omega_i = 0$,}\\
\end{cases}
\end{align*}
where~$\omega_i$ is the of the form in~\eqref{freq} and~$N$ is even.
Note as far as the precision parameter~$\alpha$ is made larger, the coefficient~$a_1$ goes to unity as in Appendix~\ref{a1-prop}.
It means that the variance of~$\tilde{A}(\omega_i)$ approximates the variance of~$A(\omega_i)$ as~$\alpha$ is made arbitrarily large.

Since~$x_n$ is a white noise with variance~$\sigma_\epsilon^2$, we have that
\begin{align*}
\tilde{A}(\omega_i) & \approx
\begin{cases}
N(0,a_1^{2\operatorname{log}_2(N/4)}\sigma_x^2), \text{ for~$i \neq 0$ and~$N/2$,}\\
N(0,2a_1^{2\operatorname{log}_2(N/4)}\sigma_x^2), \text{ if~$i = 0$ or~$N/2$,}\\
\end{cases}
\end{align*}
where~$\omega_i$ is the of the form in~\eqref{freq} and~$N$ is even.

A similar result can be established for~$\tilde{B}(\omega_i)$.
For the variance of~$\tilde{B}(\omega_i)$, we have
\begin{align*}
\operatorname{Var}(\tilde{B}(\omega_i)) &  = \frac{2\sigma_x^2}{N}\left(\sum_{n=0}^{N-1}\mathfrak{Im}(\tilde{f}_{n,i})^2\right),
\end{align*}
where~$\omega_i$ is the of the form in~\eqref{freq} and~$N$ is even.
Using  the result in~\eqref{B-approx}, we obtain
\begin{align*}
\operatorname{Var}(\tilde{B}(\omega_i)) &  = \frac{2\sigma_x^2}{N}\left(\sum_{n=0}^{N-1}\mathfrak{Im}(\tilde{f}_{n,i})^2\right)\\
&  \approx \frac{2\sigma_x^2}{N}\left(\sum_{n=0}^{N-1}a_1^{2\operatorname{log}_2(N/4)} \sin^2(\omega_i)\right)\\
& \approx
\begin{cases}
a_1^{2\operatorname{log}_2(N/4)}\sigma_x^2,\text{ for~$i \neq 0$ and~$N/2$},\\
0,\text{ if $\omega_i = 0$,}\\
\end{cases}
\end{align*}
where~$\omega_i$ is the of the form in~\eqref{freq} and~$N$ is even.
Thus we state
\begin{align*}
\tilde{B}(\omega_i) & \approx
\begin{cases}
N(0,a_1^{2\operatorname{log}_2(N/4)}\sigma_x^2), \text{ for~$i \neq 0$ and~$N/2$,}\\
0, \text{ if~$i = 0$ or~$N/2$,}\\
\end{cases}
\end{align*}
where~$\omega_i$ is the of the form in~\eqref{freq} and~$N$ is even.

Another thing of interest is the correlation between quantities~$\tilde{A}(\omega_i)$ and~$\tilde{B}(\omega_i)$.
In subsection~\ref{amostragem-periodogramo} we showed that~$A(\omega_i)$ and~$B(\omega_j)$ are not correlated since
\begin{align*}
\operatorname{Cov}(A(\omega_i), B(\omega_j)) & = \frac{2\sigma_x^2}{N}\left(\sum_{n=0}^{N-1} \cos(\omega_in)\sin(\omega_jn)\right) \\
& = \frac{2\sigma_x^2}{N}d_{ij} = 0,\\
\end{align*}
for any~$i,j = 0, 1, \ldots, N-1$ where we used~$d_{ij}$ in~\eqref{csd2}.
Using Definition~\ref{periodogram-approx-alt}, we obtain
\begin{align*}
\operatorname{Cov}(\tilde{A}(\omega_i), \tilde{B}(\omega_j)) & = \frac{2\sigma_x^2}{N}\left(\sum_{n=0}^{N-1} \mathfrak{Re}(\tilde{f}_{n,i})\mathfrak{Im}(\tilde{f}_{n,j})\right) %
\end{align*}
for any~$i,j = 0, 1, \ldots, N-1$.
We use the relations in~\eqref{Re-f} and~\eqref{Im-f}, what leads to
\begin{align*}
\operatorname{Cov}(\tilde{A}(\omega_i), \tilde{B}(\omega_j)) & \approx \frac{2\sigma_x^2}{N}\left(\sum_{n=0}^{N-1} a_1^{\operatorname{log}_2(N/4)}\cos(\omega_in) a_1^{\operatorname{log}_2(N/4)}\sin(\omega_jn)\right) \nonumber\\
& \approx a_1^{2\operatorname{log}_2(N/4)} \frac{2\sigma_x^2}{N}d_{ij} \nonumber\\
&  \approx 0,
\end{align*}
for any~$i,j = 0, 1, \ldots, N-1$.

Above results sets bases for further development and the proposition of an~\emph{approximated} probability distribution for the approximated periodogram.
For an sufficient large precision parameter~$\alpha$, we can make~$\operatorname{Cov}(\tilde{A}(\omega_i), \tilde{B}(\omega_j))$ arbitrarily close to zero.
Thus, the quantities~$\tilde{A}(\omega_i)$ and~$\tilde{B}(\omega_j)$ are~\emph{nearly} uncorrelated.
Therefore, we assume that~$\tilde{A}(\omega_i)$ and~$\tilde{B}(\omega_j))$ are uncorrelated.
Also, we have settled that~$\tilde{A}(\omega_i)$ and~$\tilde{B}(\omega_j))$ follows normal distributions.
Therefore, we apply Cochran-Fisher Theorem~\cite{Montgomery2012} and set the following result.
\begin{theorem}
\label{teoremaDistI_i-approx}
For the harmonic process~$x_n$ compounded only by the white gaussian noise with variance~$\sigma_x^2$,  the ordinates~$I_i$ of periodogram for~$i = 1, 2, \ldots, [N/2]$ are independents and for each~$i$ we have that
\begin{align*}
\tilde{I}_i = \tilde{A}^2(\omega_i)+\tilde{B}^2(\omega_i) \approx
\begin{cases}
a_1^{2\operatorname{log}_2(N/4)}\sigma_x^2 \chi_2^2, \text{ for~$i \neq 0$ and~$N/2$,}\\
2a_1^{2\operatorname{log}_2(N/4)}\sigma_x^2 \chi_1^2, \text{ if~$i = 0$ or~$N/2$,}\\
\end{cases}
\end{align*}
where~$\omega_i$ is as in~\eqref{freq} and~$N$ is even.
\end{theorem}

\subsection{Statistical Test for Approximate Periodogram Ordinates}

In this subsection we propose an approximate statistical test for approximate periodogram ordinates.
For such, we use~$g$-test~\cite{Pristley1983} as defined in subsection~\ref{statistical-test} where we substitute the exact periodogram ordinates by the approximate periodogram ordinates.
The~$g$-test establishes null hypothesis~$H_0: A_i = 0$ for all~$i = 1, 2, \ldots, k$ as in~\eqref{harmonic-process2} and~$H_1:A_i\neq 0$ for some~$i$.
Note that these hypothesis are equivalent to test if~$k=0$~\cite{Pristley1983}.
Thus, we have that
\begin{equation*}
g = \frac{\operatorname{max}_{1\leq i \leq [N/2]}(\tilde{I}_i)}{\sum_{i = 1}^{[N/2]} \tilde{I}_i},
\end{equation*}
where we use the approximate distribution
\begin{align*}
p[g>z] & \approx [N/2]\exp(-z/2).
\end{align*}

If we set a significance level~$\zeta$~\footnote{Standard manuscripts in statistical tests uses the Greek letter~$\alpha$ to represent significance level. However, this particular letter is already in use throughout this draft, that is why we resort to use~$\zeta$ in order to represent the test significance level.} we obtain
\begin{equation*}
p[g>z_{\zeta}] = \zeta.
\end{equation*}
Thus, if the calculated~$g$ statistic exceed the significance level quantile~$z_\zeta$, then~$\tilde{I}_p = \operatorname{max}_{1\leq i \leq [N/2]}(\tilde{I}_i)$ is significant at~$100\cdot\zeta\%$ and we reject the hypothesis~$H_0$ what leads us to conclude that~$x_n$ has an harmonic and we estimate its frequency location at~$\hat{\omega}_p = \omega_p$.

Since the frequency of occurrence of an harmonic is estimated as~$\hat{\omega}_p$, we can use this in order to estimate the respective coefficients in Definition~\ref{periodogram-approx-alt},
\begin{equation*}
\tilde{A}(\hat{\omega}_p) = \sqrt{\frac{2}{N}}\sum_{n=0}^{N-1}x_n\mathfrak{Re}(\tilde{f}_{n,p})
\end{equation*}
and
\begin{equation*}
\tilde{B}(\hat{\omega}_p) = \sqrt{\frac{2}{N}}\sum_{n=0}^{N-1}x_n\mathfrak{Im}(\tilde{f}_{n,p}).
\end{equation*}
Above estimation correspond to the simple task of taking the real and imaginary parts of~$p$th component of approximated DFT of~$\mathbf{x}$, that is already calculated during the computation of its approximate periodogram.

In a likely manner, we can test the significance of the remaining approximate periodogram ordinates by making use of the scheme proposed by Whittle~\cite{Whittle1952}.
We do this by taking off from the denominator of statistic~$g$ the largest approximate periodogram ordinate and substituting the~$\operatorname{max}_{1\leq i \leq [N/2]}(I_i)$ by the second largest periodogram ordinate, here denoted by~$\tilde{I}_{p''}$.
Note that this procedure is nothing more than consider a fictional approximate periodogram where the largest ordinate is taken off.
This way, we have that we can verify the significance of the second largest approximate periodogram ordinate by setting
\begin{equation*}
g' = \frac{\tilde{I}_{p''}}{\big(\sum_{i = 1}^{[N/2]} \tilde{I}_i \big)- \tilde{I}_p},
\end{equation*}
and using the distribution of~$g$ and substituting~$[N/2]$ by~$[N/2]-1$.

If the second largest approximate periodogram ordinate is then significant, we perform one more time the above procedure taking off the second largest approximate periodogram ordinate.
If the third approximate periodogram ordinate is then significant, we repeat the process for the third approximate periodogram ordinate.
The process is then repeated until we find a non-significant approximate periodogram ordinate.
All approximate periodogram ordinate~$\tilde{I}_{p}, \tilde{I}_{p''}, \tilde{I}_{p'''}, \ldots$ found to be significant allows the estimation of existing harmonics in the frequencies~$\omega_{p}, \omega_{p''}, \omega_{p'''}, \ldots$ that follows~\eqref{freq}.
The number of significant approximate periodogram ordinates is an estimate for~$k$ in the model in~\eqref{harmonic-process3}.
By this procedure originally proposed by Hartley~\cite{Hartley1949} for the of test exact periodogram ordinates, we have an estimation for each coefficient of harmonic process, the number of harmonics and the frequencies of its location.

\section{Appendix}
\label{dft-approx-appendix}

\subsection{Matrix Product Convergence in Frobenius Norm}
\label{prod-conv}
Let~$\mathbf{A}$ and~$\mathbf{B}$ be two square complex valued matrices.
Let the series of matrices~$\mathbf{A}_1, \mathbf{A}_2, \ldots$ and $\mathbf{B}_1, \mathbf{B}_2, \ldots$ be respectively convergent to~$\mathbf{A}$ and~$\mathbf{B}$ in Frobenius norm, i.e., $\|\mathbf{A}-\mathbf{A}_k\|_\mathrm{F} \to 0$ and~$\|\mathbf{B}-\mathbf{B}_k\|_\mathrm{F} \to 0$ as~$k \to \infty$.
Also, assume there exist a real number~$M_A$ such that~$\|\mathbf{A}_k\|_\mathrm{F} \leq M_A$ for~$k = 1, 2 \ldots$.
Thus, we have that~$\|\mathbf{A}\mathbf{B}-\mathbf{A}_k\mathbf{B}_k\|_\mathrm{F} \to 0$ as~$k \to \infty$.

\begin{proof}
Calculating the Frobenius norm of~$\mathbf{A}\mathbf{B}-\mathbf{A}_k\mathbf{B}_k$, we have
\begin{align*}
\|\mathbf{A}\mathbf{B}-\mathbf{A}_k\mathbf{B}_k\|_\mathrm{F} & = \|\mathbf{A}\mathbf{B}-\mathbf{A}_k\mathbf{B}+\mathbf{A}_k\mathbf{B}-\mathbf{A}_k\mathbf{B}_k\|_\mathrm{F}\\
& = \|\left(\mathbf{A}-\mathbf{A}_k\right)\mathbf{B}+\mathbf{A}_k\left(\mathbf{B}-\mathbf{B}_k\right)\|_\mathrm{F}\\
& \leq \|\left(\mathbf{A}-\mathbf{A}_k\right)\mathbf{B}\|_\mathrm{F}+\|\mathbf{A}_k\left(\mathbf{B}-\mathbf{B}_k\right)\|_\mathrm{F}\\
& \leq \|\left(\mathbf{A}-\mathbf{A}_k\right)\|_\mathrm{F}\|\mathbf{B}\|_\mathrm{F}+\|\mathbf{A}_k\|_\mathrm{F}\|\left(\mathbf{B}-\mathbf{B}_k\right)\|_\mathrm{F}
\end{align*}
Consider now that~$\|\mathbf{A}_k\|_\mathrm{F} \leq M_A$ for~$k = 1, 2 \ldots$.
Thus, we have that
\begin{align*}
\|\mathbf{A}\mathbf{B}-\mathbf{A}_k\mathbf{B}_k\|_\mathrm{F} %
& \leq \|\left(\mathbf{A}-\mathbf{A}_k\right)\|_\mathrm{F}\|\mathbf{B}\|_\mathrm{F}+M_A\|\left(\mathbf{B}-\mathbf{B}_k\right)\|_\mathrm{F}.
\end{align*}
Considering the assumption that~$\|\mathbf{A}-\mathbf{A}_k\|_\mathrm{F} \to 0$ and~$\|\mathbf{B}-\mathbf{B}_k\|_\mathrm{F} \to 0$ as~$k \to \infty$, we have that
\begin{align*}
\|\mathbf{A}\mathbf{B}-\mathbf{A}_k\mathbf{B}_k\|_\mathrm{F} \to 0,
\end{align*}
as~$k \to \infty$.
\end{proof}

\subsection{Frobenius Norm of Particular Matrix Product Involving Block Diagonal Matrix}
\label{prod-frob}
Let~$\mathbf{A}_N$ and~$\mathbf{B}_{N}$ be as described in Section~\ref{review} in the Cooley-Tukey decimation-in-time algorithm for the DFT~\cite{Oppenheim2009}.
Let~$\mathbf{C}$ be a~$N\times N$ matrix such as
\begin{equation*}
\mathbf{C} =
\left[
\begin{rsmallmatrix}
\mathbf{C}_1 & \\
& \mathbf{C}_2
\end{rsmallmatrix}
\right],
\end{equation*}
where the blocks~$\mathbf{C}_1$ and~$\mathbf{C}_2$ are~$N/2 \times N/2$ matrices.
Thus, we have that
\begin{equation*}
\|\mathbf{A}_N \mathbf{C} \mathbf{B}_{N}\|_\mathrm{F} = \sqrt{2\|\mathbf{C}_1\|^2+2\|\mathbf{C}_2\|^2}.
\end{equation*}

\begin{proof}
Remember that
\begin{equation*}
\mathbf{A}_N =  \left[
\begin{rsmallmatrix}
1 & 1\\
1 & -1
\end{rsmallmatrix}\right] \otimes
\mathbf{I}_{N/2}
=
\left[
\begin{rsmallmatrix}
\mathbf{I}_{N/2} & \mathbf{I}_{N/2}\\
\mathbf{I}_{N/2} & -\mathbf{I}_{N/2}
\end{rsmallmatrix}\right],
\end{equation*}
where~$\mathbf{I}_{N/2}$ is the identity matrix of order~$N/2$ and~$\otimes$ denotes the Kronecker tensor product~\cite{Manassah2001}.
Thus, we have that
\begin{equation*}
\mathbf{A}_N \mathbf{C} =
\left[
\begin{rsmallmatrix}
\mathbf{I}_{N/2} & \mathbf{I}_{N/2}\\
\mathbf{I}_{N/2} & -\mathbf{I}_{N/2}
\end{rsmallmatrix}
\right]
\cdot
\left[
\begin{rsmallmatrix}
\mathbf{C}_1 & \\
& \mathbf{C}_2
\end{rsmallmatrix}
\right]
=
\left[
\begin{rsmallmatrix}
\mathbf{C}_1 & -\mathbf{C}_2\\
\mathbf{C}_1 & \mathbf{C}_2
\end{rsmallmatrix}
\right].
\end{equation*}
Above matrix has Frobenius norm equals to
\begin{equation*}
\|\mathbf{A}_N \mathbf{C}\|_\mathrm{F} = \sqrt{2\|\mathbf{C}_1\|^2+2\|\mathbf{C}_2\|^2}.
\end{equation*}
Note that~$\mathbf{B}_{N}$ is a permutation matrix, thus why it does not alter the Frobenius norm of its multiplicands matrices.
Therefore, we have that
\begin{equation*}
\|\mathbf{A}_N \mathbf{C}\mathbf{B}_{N}\|_\mathrm{F} = \sqrt{2\|\mathbf{C}_1\|^2+2\|\mathbf{C}_2\|^2}.
\end{equation*}

\end{proof}

\subsection{Visual Inspection of DFT Approximation}
\label{matrices-images}
\begin{figure*}
\centering
\subfigure[Real part of $8$-point DFT approximation.]{\includegraphics{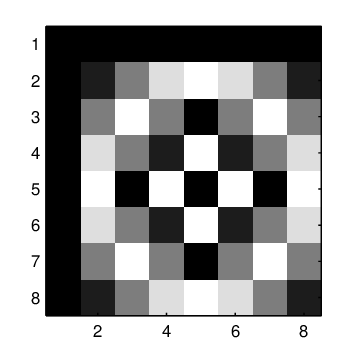}}
\subfigure[Imaginary part of $8$-point DFT approximation.]{\includegraphics{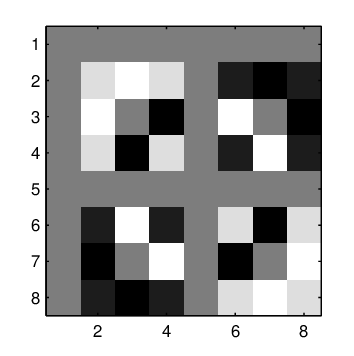}}\\
\subfigure[Real part of $8$-point exact DFT.]{\includegraphics{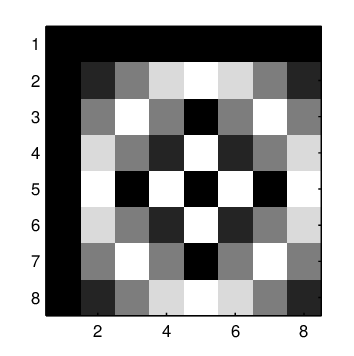}}
\subfigure[Imaginary part of $8$-point exact DFT.]{\includegraphics{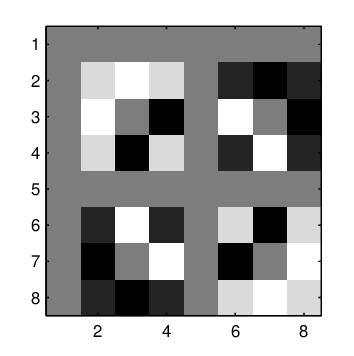}}
\caption{Visual representation of real and imaginary parts of $8$-point exact and approximated DFT for precision parameter~$\alpha = 8$.}
\label{fig:FFTA8_alpha8}
\end{figure*}

\begin{figure*}
\centering
\subfigure[Real part of $16$-point DFT approximation.]{\includegraphics{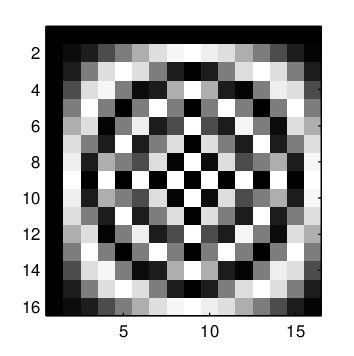}}
\subfigure[Imaginary part of $16$-point DFT approximation.]{\includegraphics{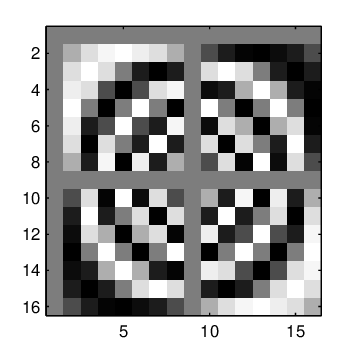}}\\
\subfigure[Real part of $16$-point exact DFT.]{\includegraphics{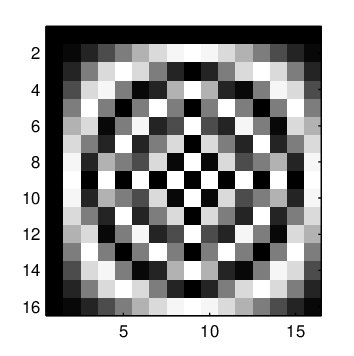}}
\subfigure[Imaginary part of $16$-point exact DFT.]{\includegraphics{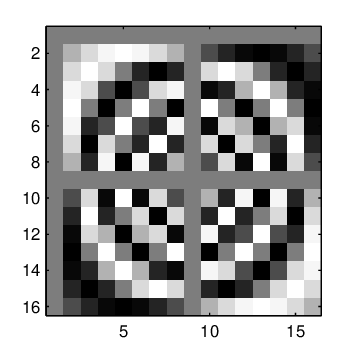}}
\caption{Visual representation of real and imaginary parts of $16$-point exact and approximated DFT for precision parameter~$\alpha = 8$.}
\label{fig:FFTA16_alpha8}
\end{figure*}

\begin{figure*}
\centering
\subfigure[Real part of $32$-point DFT approximation.]{\includegraphics{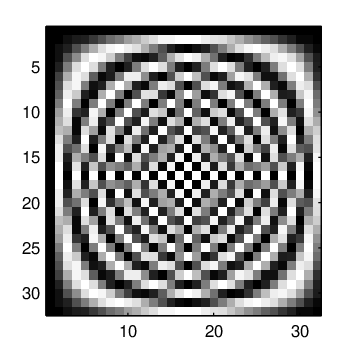}}
\subfigure[Imaginary part of $32$-point DFT approximation.]{\includegraphics{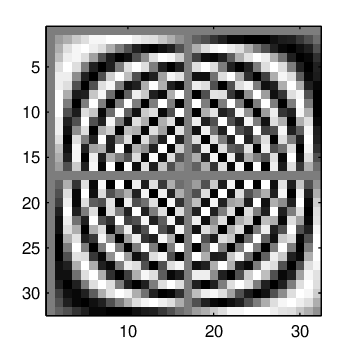}}\\
\subfigure[Real part of $32$-point exact DFT.]{\includegraphics{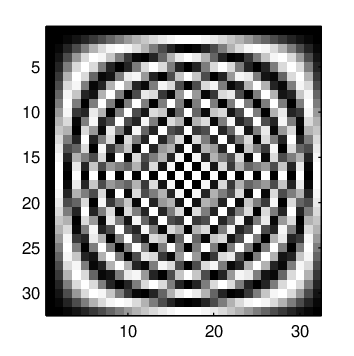}}
\subfigure[Imaginary part of $32$-point exact DFT.]{\includegraphics{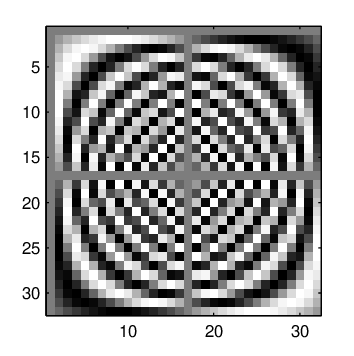}}
\caption{Visual representation of real and imaginary parts of $32$-point exact and approximated DFT for precision parameter~$\alpha = 8$.}
\label{fig:FFTA32_alpha8}
\end{figure*}

\begin{figure*}
\centering
\subfigure[Real part of $64$-point DFT approximation.]{\includegraphics{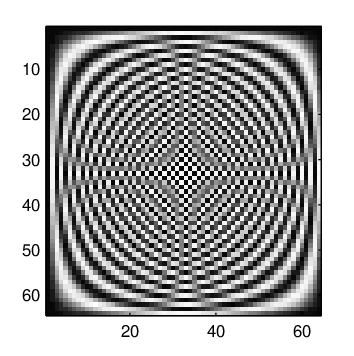}}
\subfigure[Imaginary part of $64$-point DFT approximation.]{\includegraphics{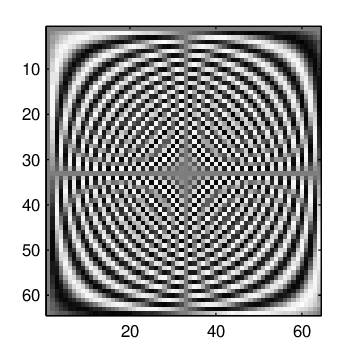}}\\
\subfigure[Real part of $64$-point exact DFT.]{\includegraphics{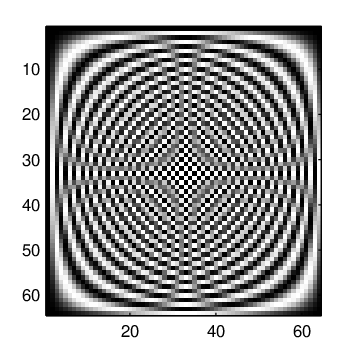}}
\subfigure[Imaginary part of $64$-point exact DFT.]{\includegraphics{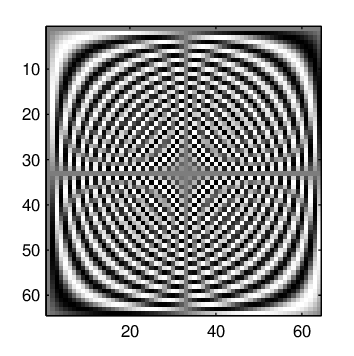}}
\caption{Visual representation of real and imaginary parts of $64$-point exact and approximated DFT for precision parameter~$\alpha = 8$.}
\label{fig:FFTA64_alpha8}
\end{figure*}

In this appendix we provide visual inspection to the class of DFT approximation for particular lengths and parameter precision~$\alpha$.
Taking the case where~$\alpha = 8$ we plot the visual representation of real and imaginary parts of~$8, 16, 32$ and~$64$-point DFT approximation in Figures~\ref{fig:FFTA8_alpha8},~\ref{fig:FFTA16_alpha8},~\ref{fig:FFTA32_alpha8} and~\ref{fig:FFTA64_alpha8}, respectively.
The exact~$8, 16, 32$ and~$64$-point DFT are also plotted for comparisons.

\subsection{A Relevant Summation Formula Simplification for Approximate Twiddle Factor Fourier Series}
\label{summation-simplification}
This appendix furnish a summation formula simplification that is useful in Fourier series representation for the approximate twiddle factor.
Consider the following summation formula
\begin{equation*}
S_{N-1} = \sum_{k = 1}^{N-1} k(f_{k+1}-f_k),
\end{equation*}
where~$f_{k}$ is a arbitrary real valued function for~$k = 1, 2, \ldots, N$.
Opening the expression above we have
\begin{align*}
S_{N-1} & = (f_2-f_1)+2(f_3-f_2)+3(f_4-f_3)+4(f_5-f_4)+\ldots\\
&\phantom{=} \ldots+(N-2)(f_{N-1}-f_{N-2})+(N-1)(f_{N}-f_{N-1}).
\end{align*}
We combine each first term in each parcel inside a brackets.
Thus, we take the terms~$f_2$ in the first parcel and~$-2f_2$ in the second parcel, which result in~$-f_2$.
Take the term~$2f_3$ in the second parcel and~$-3f_3$ in the third parcel, which result in~$-f_3$.
Take the term~$3f_4$ in the third parcel and~$-4f_4$ in their fourth parcel, which result in~$-f_4$.
Doing so, we have
\begin{align*}
S_{N-1} & = -f_1+\overbrace{f_2-2f_2}^{-f_2}+\overbrace{2f_3-3f_3}^{-f_3}+\overbrace{3f_4-4f_4}^{-f_4}+4f_5+\ldots\\
&\phantom{=} \ldots-(N-2)f_{N-2}+\underbrace{(N-2)f_{N-1}-(N-1)f_{N-1}}_{-f_{N-1}}+(N-1)f_{N},
\end{align*}
what yields
\begin{align}
\label{S_N-1}
S_{N-1} & = -f_1-f_2-f_3-\ldots-f_{N-1}+(N-1)f_N\nonumber\\
& = (N-1)f_N-\sum_{k = 1}^{N-1}f_k.
\end{align}

\subsection{Development of Approximate Twiddle Factor Fourier Series and Its Representation based on Chebyshev Polynomials}
\label{sf-round}

This appendix furnish the Fourier series representation for the approximate twiddle factor.
The Fourier series representation is a central tool in signal processing and analysis~\cite{Papoulis1997}.
We aim at furnishing Fourier series representation that will provide means for DFT approximation analysis and error investigation.
Yet, in this appendix we make a connection between the Fourier series coefficients of approximate twiddle factor and Chebyshev polynomials.

In general, unity root arguments are unitary complex numbers in the form of~$e^{j\theta} = \cos(\theta)+j\sin(\theta)$, where~$\theta = 2\pi/N$ for a positive integer.
In order to facilitate the Fourier series expansion we resort to provide Fourier coefficients for the general case of complex inputs~$e^{j\theta} = \cos(\theta)+j\sin(\theta)$ where~$\theta \in [0, 2\pi)$.
Thus, we have that
\begin{equation*}
\frac{1}{\alpha}\operatorname{round}(\alpha e^{j\theta}) = \frac{1}{\alpha}\operatorname{round}(\alpha \cos(\theta))+j\frac{1}{\alpha}\operatorname{round}(\alpha \sin(\theta)),
\end{equation*}
where~$\theta \in [0, 2\pi)$.

We begin our analysis remembering the fact that the scaled rounding operator for complete periods of sine and cosine arguments is symmetric along~\emph{ordinates} axis.
As consequence, the mean term of Fourier series expansion is always zero for complete periods of sine and cosine arguments.
Moreover, sine and cosine are odds and even functions, respectively.
Since the scaled rounding operator is an odd function, the aforementioned properties of sine and cosine functions are preserved.
Therefore, the scaled rounded sine function and the scaled rounded cosine function posses the following form for Fourier series:
\begin{equation*}
\frac{1}{\alpha}\operatorname{round}(\alpha \cos(\theta)) = \sum_{n = 0}^\infty a_n \cos(n\theta)
\end{equation*}
and
\begin{equation*}
\frac{1}{\alpha}\operatorname{round}(\alpha \sin(\theta)) = \sum_{n = 0}^\infty b_n \sin(n\theta).
\end{equation*}

Let us initiate the analysis through the Fourier coefficients of scaled rounded sine function.
Both sine and the scaled rounded sine are odd functions with period~$2\pi$, therefore their product is a even function.
As consequence, the integral for obtaining the~$b_n$ Fourier coefficient is twice the integral over the interval~$[0,\pi)$.
Thus, we have that
\begin{align*}
b_n = & \frac{2}{2\pi}\int_{0}^{2\pi}\frac{1}{\alpha} \operatorname{round}(\alpha \sin(\theta))\sin(n\theta)d \theta\\
= & \frac{2}{\pi}\int_{0}^{\pi}\frac{1}{\alpha} \operatorname{round}(\alpha \sin(\theta))\sin(n\theta)d \theta\\
=& \frac{2}{\pi\alpha} \left[ \sum_{i=1}^{\alpha-1}i\int_{\sin^{-1}(\frac{2i-1}{2\alpha})}^{\sin^{-1}(\frac{2i+1}{2\alpha})}\sin(n\theta)d\theta + \alpha\int_{\sin^{-1}(1-\frac{1}{2\alpha})}^{\pi-\sin^{-1}(1-\frac{1}{2\alpha})}\sin(n\theta)d\theta \right.\\
&\left.\phantom{\frac{1}{\pi\alpha}} +\sum_{i=1}^{\alpha-1}i\int_{\pi-\sin^{-1}(\frac{2i+1}{2\alpha})}^{\pi-\sin^{-1}(\frac{2i-1}{2\alpha})}\sin(n\theta)d\theta \right]\\
=& -\frac{2}{\pi\alpha} \left[ \alpha\frac{\cos(n\theta)}{n}\Big|_{\sin^{-1}(1-\frac{1}{2\alpha})}^{\pi-\sin^{-1}(1-\frac{1}{2\alpha})} \right.\\
& \left. \phantom{\frac{2}{\pi\alpha}}+ \sum_{i=1}^{\alpha-1}i\left( \frac{\cos(n\theta)}{n}\Big|_{\sin^{-1}(\frac{2i-1}{2\alpha})}^{\sin^{-1}(\frac{2i+1}{2\alpha})} -\frac{\cos(n\theta)}{n}\Big|_{\pi-\sin^{-1}(\frac{2i-1}{2\alpha})}^{\pi-\sin^{-1}(\frac{2i+1}{2\alpha})}\right) \right].
\end{align*}
In above expression we note that~$\cos(n(\pi-\theta)) = (-1)^n\cos(n\theta)$ for~$\theta \in [0, 2\pi)$.
Therefore, we have that
\begin{align*}
b_n = &-\frac{2}{\pi\alpha} \left[ \alpha\frac{\cos(n\theta)}{n}\Big|_{\sin^{-1}(1-\frac{1}{2\alpha})}^{\pi-\sin^{-1}(1-\frac{1}{2\alpha})} \right.\\
& \left. \phantom{\frac{2}{\pi\alpha}}+ \sum_{i=1}^{\alpha-1}i\left( \frac{\cos(n\theta)}{n}\Big|_{\sin^{-1}(\frac{2i-1}{2\alpha})}^{\sin^{-1}(\frac{2i+1}{2\alpha})} -\frac{\cos(n\theta)}{n}\Big|_{\pi-\sin^{-1}(\frac{2i-1}{2\alpha})}^{\pi-\sin^{-1}(\frac{2i+1}{2\alpha})}\right) \right]\\
=& -\frac{2}{\pi\alpha} \left[ \alpha\frac{\cos(n\theta)}{n}\Big|_{\sin^{-1}(1-\frac{1}{2\alpha})}^{\pi-\sin^{-1}(1-\frac{1}{2\alpha})} \right.\\
& \left. \phantom{\frac{2}{\pi\alpha}}+ \sum_{i=1}^{\alpha-1}i\left( \frac{\cos(n\theta)}{n}\Big|_{\sin^{-1}(\frac{2i-1}{2\alpha})}^{\sin^{-1}(\frac{2i+1}{2\alpha})} -(-1)^n\frac{\cos(n\theta)}{n}\Big|_{\sin^{-1}(\frac{2i-1}{2\alpha})}^{\sin^{-1}(\frac{2i+1}{2\alpha})}\right) \right]\\
=& -\frac{2}{n\pi\alpha} \left[ \alpha\cos(n\theta)\Big|_{\sin^{-1}(1-\frac{1}{2\alpha})}^{\pi-\sin^{-1}(1-\frac{1}{2\alpha})} \right.\\
& \left. \phantom{\frac{2}{\pi\alpha}}+ (1 +(-1)^{n+1})\sum_{i=1}^{\alpha-1}i\cos(n\theta)\Big|_{\sin^{-1}(\frac{2i-1}{2\alpha})}^{\sin^{-1}(\frac{2i+1}{2\alpha})} \right].
\end{align*}
Applying above interval limits, we obtain
\begin{align*}
b_n =& -\frac{2}{n\pi\alpha} \left[ \alpha\cos(n\theta)\Big|_{\sin^{-1}(1-\frac{1}{2\alpha})}^{\pi-\sin^{-1}(1-\frac{1}{2\alpha})} \right.\\
& \left. \phantom{\frac{2}{\pi\alpha}}+ (1 +(-1)^{n+1})\sum_{i=1}^{\alpha-1}i\cos(n\theta)\Big|_{\sin^{-1}(\frac{2i-1}{2\alpha})}^{\sin^{-1}(\frac{2i+1}{2\alpha})} \right]\\
=& -\frac{2}{n\pi\alpha} \left[ \alpha((-1)^n -1)\left(\cos\left(n\sin^{-1}\left(1-\frac{1}{2\alpha}\right)  \right)\right) \right.\\
& \left. \phantom{\frac{2}{\pi\alpha}}+ (1+(-1)^{n+1})\sum_{i=1}^{\alpha-1}i\left( \cos\left(n\sin^{-1}\left(\frac{2i+1}{2\alpha}\right)\right) \right. \right.\\
& \left. \left. \phantom{\frac{2}{\pi\alpha}(1+(-1)^{n+1})\sum_{i=1}^{\alpha-1}} -\cos\left(n\sin^{-1}\left(\frac{2i-1}{2\alpha}\right)\right)\right) \right]\\
= & -\frac{2}{n\pi\alpha} (1+(-1)^{n+1}) \left[ -\alpha\left(\cos\left(n\sin^{-1}\left(1-\frac{1}{2\alpha}\right)  \right)\right) \right.\\
& \left. \phantom{\frac{2}{\pi\alpha}}+ \sum_{i=1}^{\alpha-1}i\left( \cos\left(n\sin^{-1}\left(\frac{2i+1}{2\alpha}\right)\right) \right. \right.\\
& \left. \left. \phantom{\frac{2}{\pi\alpha}(1+(-1)^{n+1})\sum_{i=1}^{\alpha-1}} -\cos\left(n\sin^{-1}\left(\frac{2i-1}{2\alpha}\right)\right)\right) \right].
\end{align*}

Above expression shows us that~$b_n = 0$ for all~$n$ even, i.e., $b_0, b_2, b_4 , \ldots$ are zero.
Also, in above expression we note the connection with Chebyshev polynomials.
We have that~$\cos(n\theta) = T_{n}(\cos(\theta))$ for~$\theta \in (0,2\pi]$, where~$T_{n}(\cdot)$ represents the Chebyshev polynomial of first kind of degree~$n$.
In particular, since~$\sin^{-1}\left(1-1/2\alpha\right)$ is in the first quadrant, we have that
\begin{equation*}
\cos\left(n\sin^{-1}\left(1-\frac{1}{2\alpha}\right)\right)=T_{n}\left(\sqrt{1-\left(1-\frac{1}{2\alpha}\right)^2}\right).
\end{equation*}
Similarly, angles~$\sin^{-1}\left((2i\pm1)/2\alpha\right)$ for~$i = 1, 2, \ldots, \alpha-1$ are in the first quadrant.
Thus, we have that
\begin{equation*}
 \cos\left(n\sin^{-1}\left(\frac{2i+1}{2\alpha}\right)\right)=T_n\left(\sqrt{1-\left(\frac{2i+1}{2\alpha}\right)^2}\right)
\end{equation*}
and
\begin{equation*}
 \cos\left(n\sin^{-1}\left(\frac{2i-1}{2\alpha}\right)\right)=T_n\left(\sqrt{1-\left(\frac{2i-1}{2\alpha}\right)^2}\right),
\end{equation*}
where~$i = 1, 2, \ldots, \alpha-1$.
Applying above equations in terms of first kind Chebyshev polynomials into the expression for Fourier coefficient~$b_n$, we obtain
\begin{align*}
b_n =& -\frac{2}{n\pi\alpha} (1+(-1)^{n+1}) \left[ -\alpha T_{n}\left(\sqrt{1-\left(1-\frac{1}{2\alpha}\right)^2}\right) \right.\\
& \left. \phantom{\frac{2}{\pi\alpha}}+ \sum_{i=1}^{\alpha-1}i\left( T_n\left(\sqrt{1-\left(\frac{2i+1}{2\alpha}\right)^2}\right) \right. \right.\\
& \left. \left. \phantom{\frac{2}{\pi\alpha}(1+(-1)^n)\sum_{i=1}^{\alpha-1}} -T_n\left(\sqrt{1-\left(\frac{2i-1}{2\alpha}\right)^2}\right)\right) \right].
\end{align*}
Note in above expression that~$b_n = 0$ for even~$n$.
It is due the existence of~$(1+(-1)^{n+1})$ parcel multiplying the complete expression for~$b_n$.
Also, above expression for~$b_n$ can be simplified.
Here we apply the result in Appendix~\ref{summation-simplification}.
In particular, we apply~\eqref{S_N-1} what allow us substitute
\begin{align*}
\sum_{i=1}^{\alpha-1} i\left( T_n\left(\sqrt{1-\left(\frac{2i+1}{2\alpha}\right)^2}\right) -T_n\left(\sqrt{1-\left(\frac{2i-1}{2\alpha}\right)^2}\right)\right)
\end{align*}
by
\begin{align*}
(\alpha-1)T_n\left(\sqrt{1-\left(1-\frac{1}{2\alpha}\right)^2}\right)- \sum_{i=1}^{\alpha-1} T_n\left(\sqrt{1-\left(\frac{2i-1}{2\alpha}\right)^2}\right).
\end{align*}
Above substitution leads to
\begin{align*}
b_n =& -\frac{2}{n\pi\alpha} (1+(-1)^{n+1}) \left[ -\alpha T_{n}\left(\sqrt{1-\left(1-\frac{1}{2\alpha}\right)^2}\right) \right.\\
& \left. \phantom{\frac{2}{\pi\alpha}}+ (\alpha-1)T_n\left(\sqrt{1-\left(1-\frac{1}{2\alpha}\right)^2}\right) \right.\\
& \left. \phantom{\frac{2}{\pi\alpha}(1+(-1)^n)\sum_{i=1}^{\alpha-1}} - \sum_{i=1}^{\alpha-1} T_n\left(\sqrt{1-\left(\frac{2i-1}{2\alpha}\right)^2}\right) \right],
\end{align*}
what result in
\begin{align}
\label{sf-bn}
b_n =& \frac{2}{n\pi\alpha} (1+(-1)^{n+1})  \sum_{i=1}^{\alpha} T_n\left(\sqrt{1-\left(\frac{2i-1}{2\alpha}\right)^2}\right) .
\end{align}

Strictly, we should calculate the expression for each~$a_n$ for Fourier series coefficients for the scaled rounded cosine function.
However, we can exploit sine and cosine symmetries in order to evaluate~$a_n$.
Since~$\sin(\theta+\pi/2) = \cos(\theta)$ for~$\theta \in [0, 2\pi)$, we have that
\begin{equation*}
\frac{1}{\alpha}\operatorname{round}(\alpha \cos(\theta)) = \frac{1}{\alpha}\operatorname{round}(\alpha \sin(\theta+\pi/2)).
\end{equation*}
Applying the Fourier series representation for scaled rounded sine function and using that~$\sin(n(\theta+\pi/2)) = \sin(n\theta)\cos(n\pi/2)+\sin(n\pi/2)\cos(n\theta)$, we have that
\begin{align*}
\sum_{n = 0}^{\infty}a_n \cos(n\theta) =& \sum_{n = 0}^{\infty} b_n \sin(n(\theta+\pi/2))\\
=& \sum_{n = 0}^{\infty} b_n
\begin{cases}
\phantom{-}\cos(n\theta), \quad 1\equiv n\pmod{4},\\
-\sin(n\theta),\quad 2\equiv n\pmod{4},\\
-\cos(n\theta),\quad 3\equiv n\pmod{4},\\
\phantom{-}\sin(n\theta), \quad 0\equiv n\pmod{4}.
\end{cases}
\end{align*}
Since~$b_n = 0$ for~$n$ even, we have that
\begin{align*}
a_n  =&
\begin{cases}
\phantom{-}b_n, \quad 1\equiv n\pmod{4},\\
-b_n,\quad 3\equiv n\pmod{4},\\
\phantom{-}0, \quad 0,2\equiv n\pmod{4}.
\end{cases}
\end{align*}
Above equation in form of cases can be given a compact form as
\begin{equation*}
a_n = (-1)^{\frac{n-1}{2}}b_n.
\end{equation*}
Applying~\eqref{sf-bn} into above equation, we have
\begin{align}
\label{sf-an}
a_n =& \frac{2}{n\pi\alpha}(-1)^{\frac{n-1}{2}} (1+(-1)^{n+1})  \sum_{i=1}^{\alpha} T_n\left(\sqrt{1-\left(\frac{2i-1}{2\alpha}\right)^2}\right) .
\end{align}

\subsection{Properties of Fourier Series First Harmonic Coefficient of Approximate Twiddle Factor}
\label{a1-prop}

This appendix provide properties for the first harmonic coefficient for the Fourier series of approximate twiddle factor.
Consider the inequality in~\eqref{z-round-im-ineq} for~$z = e^{j\theta}$, what results in
\begin{equation*}
\sin(\theta)-\frac{1}{2\alpha} \leq  \frac{1}{\alpha} \operatorname{round}(\alpha \sin(\theta)) \leq \sin(\theta)+\frac{1}{2\alpha},
\end{equation*}
where~$\theta \in [0,2\pi)$.
Taking twice the integral of function in above inequality  multiplied by~$\sin(\theta)$ and~$2/\pi$, we have
\begin{align*}
\frac{2}{\pi}\int_{0}^{\pi}\left(\sin(\theta)-\frac{1}{2\alpha}\right)\sin(\theta)d\theta & \leq  \frac{2}{\pi}\int_{0}^{\pi}\left(\frac{1}{\alpha} \operatorname{round}(\alpha \sin(\theta))\right)\sin(\theta)d\theta
\end{align*}
and
\begin{align*}
\frac{2}{\pi}\int_{0}^{\pi}\left(\frac{1}{\alpha} \operatorname{round}(\alpha \sin(\theta))\right)\sin(\theta)d\theta & \leq \frac{2}{\pi}\int_{0}^{\pi}\left(\sin(\theta)+\frac{1}{2\alpha}\right)\sin(\theta)d\theta.
\end{align*}

In Appendix~\ref{sf-round} we settled that~$a_n = (-1)^{(n-1)/2}b_n$, what yields~$a_1 = b_1$.
Doing so, we recognize the expression
\begin{equation*}
a_1 = b_1 = \frac{2}{\pi}\int_{0}^{\pi}\left(\frac{1}{\alpha} \operatorname{round}(\alpha \sin(\theta))\right)\sin(\theta)d\theta.
\end{equation*}
Therefore, above inequality becomes
\begin{align*}
\frac{2}{\pi}\int_{0}^{\pi}\left(\sin(\theta)-\frac{1}{2\alpha}\right)\sin(\theta)d\theta & \leq  a_1  \leq \frac{2}{\pi}\int_{0}^{\pi}\left(\sin(\theta)+\frac{1}{2\alpha}\right)\sin(\theta)d\theta\\
\frac{2}{\pi}\int_{0}^{\pi}\left(\sin(\theta)^2-\frac{\sin(\theta)}{2\alpha}\right)d\theta & \leq  a_1  \leq \frac{2}{\pi}\int_{0}^{\pi}\left(\sin(\theta)^2+\frac{\sin(\theta)}{2\alpha}\right)d\theta\\
\frac{2}{\pi}\int_{0}^{\pi}\left(\frac{1-\cos(2\theta)}{2}-\frac{\sin(\theta)}{2\alpha}\right)d\theta & \leq  a_1  \leq \frac{2}{\pi}\int_{0}^{\pi}\left(\frac{1-\cos(2\theta)}{2}+\frac{\sin(\theta)}{2\alpha}\right)d\theta\\
\frac{2}{\pi}\left(\frac{\theta}{2}-\frac{\sin(2\theta)}{4}+\frac{\cos(\theta)}{2\alpha}\right)\Big|_{0}^{\pi} & \leq  a_1  \leq \frac{2}{\pi}\left(\frac{\theta}{2}-\frac{\sin(2\theta)}{4}-\frac{\cos(\theta)}{2\alpha}\right)\Big|_{0}^{\pi}\\
\frac{2}{\pi}\left(\frac{\pi}{2}-\frac{1}{\alpha}\right) & \leq  a_1  \leq \frac{2}{\pi}\left(\frac{\pi}{2}+\frac{1}{\alpha}\right)\\
1-\frac{2}{\pi\alpha} & \leq  a_1  \leq 1+\frac{2}{\pi\alpha}.
\end{align*}
From above sandwich for the first Fourier coefficient of approximate twiddle factor~$a_1$ we can state that
\begin{equation}
\label{lim-a1}
\lim_{\alpha \to \infty} a_1 = 1.
\end{equation}
\begin{figure*}
\centering
\psfrag{a1}{$a_1$}
\psfrag{logALPHA}{$\operatorname{log}_2(\alpha)$}
\includegraphics{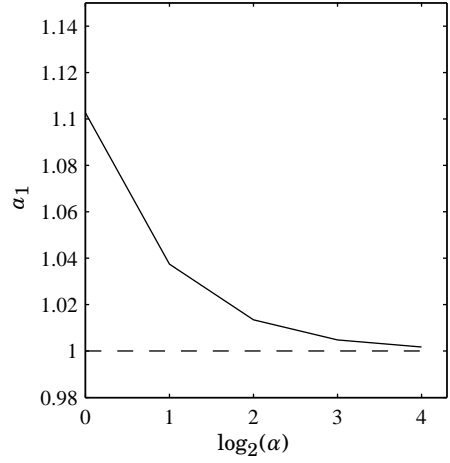}
\caption{First Fourier coefficient for approximate twiddle factor against the logarithm of precision parameter~$\alpha$.}
\label{fig:a1}
\end{figure*}
Using~\eqref{a1b1} we can find algebraic expression for~$a_1$ for the simplest precision parameter~$\alpha = 1$, what leads to
\begin{equation*}
a_1 = \frac{2}{\pi}\sqrt{3}.
\end{equation*}
For this particular precision parameter~$\alpha = 1$ we have that~$a_1 > 1$.
Therefore, we hope the curve for the first Fourier coefficient~$a_1$ as a function of~$\alpha$ to be decreasing curve as far as~$\alpha$ increases.
That is testified by Figure~\ref{fig:a1} that plots the first Fourier coefficient~$a_1$ against the logarithm of precision parameter~$\alpha$.
The first Fourier coefficient fast approximate the unity for high values of~$\alpha$.

\onecolumn

{\small
\singlespacing
\bibliographystyle{siam}
\bibliography{bibcleanoutput}
}

\end{document}